\documentclass[prx,twocolumn,amsmath,amssymb]{revtex4-1}
\usepackage[T1]{fontenc} % for Rzazewski
\usepackage{bm}
\usepackage{graphicx}
\usepackage[bookmarks=false]{hyperref}

\newcommand{\ket}[1]{|#1\rangle}
\newcommand{\braket}[1]{\langle #1 \rangle}

\def\dd{\mathrm{d}}
\def\ee{\mathrm{e}}
\def\ii{\mathrm{i}}

\def\Re{\mathrm{Re}}
\def\Im{\mathrm{Im}}

\def\Tr{\mathrm{Tr}}
\def\Re{\mathrm{Re}}
\def\Im{\mathrm{Im}}
\def\const{\text{const.}}
\def\ddt#1{\frac{\partial #1}{\partial t}}
\def\ddtin{(\partial/\partial t)}
\def\vzero{\bm{0}}

\def\half{\frac{1}{2}}

\def\Erion{\mathrm{Er}^{3+}}
\def\Yion{\mathrm{Y}^{3+}}
\def\Feion{\mathrm{Fe}^{3+}}
\def\ErFeO{\mathrm{ErFeO_3}}
\def\ErYFeO{\mathrm{Er}_x\mathrm{Y}_{1-x}\mathrm{FeO_3}}
\def\ErEr{$\Erion$--$\Erion$}
\def\ErFe{$\Erion$--$\Feion$}
\def\FeFe{$\Feion$--$\Feion$}

\def\sumnn{\sum_{\mathrm{n.n.}}}

\def\EEr{E_{\mathrm{Er}}}

\def\wEr{\omega_{\mathrm{Er}}}
\def\wex{\omega_{\text{ex}}}
\def\wph{\omega_{\text{ph}}}
\def\rabi{g}

\def\aa{\bar{a}}
\def\ba{\bar{b}}

\def\JFe{J_{\text{Fe}}}
\def\JEr{J_{\text{Er}}}
\def\DFe{D^{\text{Fe}}}
\def\Ax{A_x}
\def\Ay{A_y}
\def\Az{A_z}
\def\Axy{A_{xy}}
\def\Axz{A_{xz}}
\def\NUC{N_{0}}

\def\rabi{g}

\def\muB{\mu_{\text{B}}}
\def\kB{k_{\text{B}}}
\def\meV{\mathrm{meV}}
\def\THz{\mathrm{THz}}

\def\ZZ{\mathcal{Z}}
\def\ZZb{\bar{\mathcal{Z}}}

\def\FFb{\bar{\mathcal{F}}}
\def\SSb{\bar{\mathcal{S}}}

\def\HH{\mathcal{H}}

\def\HHFe{\mathcal{H}_{\text{Fe}}}
\def\HHFeEr{\mathcal{H}_{\text{Fe--Er}}}
\def\HHEr{\mathcal{H}_{\text{Er}}}

\def\oHH{\hat{\mathcal{H}}}

\def\oHHDicke{\oHH_{\text{Dicke}}}

\def\oHHeff{\hat{\mathcal{H}}^{\text{eff}}}
\def\oHHeffDicke{\hat{\mathcal{H}}^{\text{eff}}_{\mathrm{Dicke}}}
\def\oHHa{\oHH^{\text{a}}}
\def\oHHaDicke{\oHH^{\text{a}}_{\mathrm{Dicke}}}

\def\oHHFe{\oHH_{\text{Fe}}}

\def\oHHFeEr{\oHH_{\text{Er--Fe}}}
\def\oHHEr{\oHH_{\text{Er}}}
\def\oHHstat{\oHH_{\text{Er--Fe}}^{\varSigma}}

\def\oHHdyn{\oHH_{\text{Er--Fe}}^{\text{coupling}}}
\def\oHHMFFe#1{\oHH_{\text{Fe}}^{#1}}
\def\oHHMFEr#1{\oHH_{\text{Er}}^{#1}}

\def\oa{\hat{a}}
\def\oad{\hat{a}^{\dagger}}
\def\ob{\hat{b}}
\def\obd{\hat{b}^{\dagger}}

\def\oS{\hat{S}}
\def\oT{\hat{T}}
\def\oY{\hat{Y}}
\def\ovmu{\hat{\bm{\mu}}}
\def\oR{\hat{R}}
\def\ovR{\hat{\bm{R}}}
\def\ovS{\hat{\bm{S}}}
\def\ovsigma{\hat{\bm{\sigma}}}
\def\ovSigma{\hat{\bm{\varSigma}}}

\def\osigma{\hat{\sigma}}
\def\oSigma{\hat{\varSigma}}

\def\RR#1#2{R^{#1}_{#2}}
\def\SS#1#2{S^{#1}_{#2}}

\def\RRR{\mathcal{R}}
\def\SSS{\mathcal{S}}

\def\vuFe#1{\bm{u}_{\text{Fe}}^{#1}}
\def\vuEr#1{\bm{u}_{\text{Er}}^{#1}}

\def\Bext{B^{\text{DC}}}

\def\vBstat{\bm{B}^{\text{DC}}}
\def\vBext{\bm{B}^{\text{DC}}}

\def\vBMFEr#1{\bm{B}_{\text{Er}}^{#1}}
\def\vBMFFe#1{\bm{B}_{\text{Fe}}^{#1}}
\def\vBMFEra#1{\bar{\bm{B}}_{\text{Er}}^{#1}}
\def\vBMFFea#1{\bar{\bm{B}}_{\text{Fe}}^{#1}}

\def\vD#1{\bm{D}^{#1}}
\def\vsigmaa#1{\bar{\bm{\sigma}}^{#1}}

\def\vSa#1{\bar{\bm{S}}^{#1}}

\def\Sigmaa#1{\bar{\varSigma}^{#1}}
\def\sigmaa#1{\bar{\sigma}^{#1}}
\def\Sa#1{\bar{S}^{#1}}

\def\vsigma#1{\bm{\sigma}^{#1}}
\def\vS#1{\bm{S}^{#1}}
\def\vR#1{\bm{R}^{#1}}

\def\ZFe#1{Z_{\text{Fe}}^{#1}}
\def\ZEr#1{Z_{\text{Er}}^{#1}}

\def\gf{\mathfrak{g}}
\def\gfFe{\mathfrak{g}^{\mathrm{Fe}}}
\def\gfEr{\mathfrak{g}^{\mathrm{Er}}}

\def\mgfFe{\bm{\gf}^{\mathrm{Fe}}}
\def\mgfEr{\bm{\gf}^{\mathrm{Er}}}

\def\Tc{T_{\text{c}}}

\def\zz{z_{\mathrm{Fe}}}
\def\zEr{z_{\mathrm{Er}}}

\begin{document}

%\preprint{}

% \title{$\bm{\ErFeO}$ shows magnonic superradiant phase transition}
\title{Magnonic Superradiant Phase Transition}

\author{Motoaki Bamba}
\altaffiliation{E-mail: bamba.motoaki.y13@kyoto-u.jp}
\affiliation{Department of Physics, Kyoto University, Kyoto 606-8502, Japan}
\affiliation{PRESTO, Japan Science and Technology Agency, Kawaguchi 332-0012, Japan}
\author{Xinwei Li}
\affiliation{Department of Electrical and Computer Engineering, Rice University, Houston 77005, USA}
\author{Nicolas Marquez Peraca}
\affiliation{Department of Physics and Astronomy, Rice University, Houston 77005, USA}
\author{Junichiro Kono}
\affiliation{Department of Electrical and Computer Engineering, Rice University, Houston 77005, USA}
\affiliation{Department of Material Science and NanoEngineering, Rice University, Houston 77005, USA}
\affiliation{Department of Physics and Astronomy, Rice University, Houston 77005, USA}
\date{\today}

\begin{abstract}
We show that the low-temperature phase transition in $\ErFeO$ that occurs at a critical temperature of $\sim 4$\,K can be described as a magnonic version of the superradiant phase transition (SRPT).  The role of photons in the quantum-optical SRPT is played by $\Feion$ magnons, while that of two-level atoms is played by $\Erion$ spins.  Our spin model, which is reduced to an extended Dicke model, takes into account the short-range, direct exchange interactions between $\Erion$ spins in addition to the long-range $\Erion$--$\Erion$ interactions mediated by $\Feion$ magnons.  By using realistic parameters determined by recent terahertz magnetospectroscopy and magnetization experiments, we demonstrate that it is the cooperative, ultrastrong coupling between $\Erion$ spins and $\Feion$ magnons that causes the phase transition.  This work thus proves $\ErFeO$ to be a unique system that exhibits a SRPT in thermal equilibrium, in contrast to previous observations of laser-driven non-equilibrium SRPTs.
\end{abstract}

% \keywords{}

\maketitle
\section{Introduction}
In 1973, it was proposed \cite{Hepp1973AP,Wang1973PRA} that a static transverse electromagnetic field (a \emph{photon} field) and a static polarization (a \emph{matter} field) spontaneously appear in thermal equilibrium, when the photon--matter coupling strength exceeds a certain threshold, entering the so-called ultrastrong coupling regime  \cite{Ciuti2005PRB,Forn-Diaz2018,Kockum2018}. This phenomenon has come to be known as the superradiant phase transition (SRPT), or the Dicke phase transition, since the Dicke model (originally developed for the phenomenon of superradiance \cite{Dicke1954PR}) was used in the theoretical calculations \cite{Hepp1973AP,Wang1973PRA}.

While the focus of optical science has traditionally been on non-equilibrium excited-state dynamics, a unique aspect of the SRPT is that it is concerned with the thermal-equilibrium state of a light--matter coupled system.  Non-equilibrium SRPTs have been demonstrated in cold atom systems driven by laser light  \cite{Baumann2010N,Kirton2018a}, but realization of the SRPT in true thermal equilibrium has been challenging.  The existence of an analog of the SRPT has been theoretically shown for a superconducting circuit in thermal equilibrium  \cite{Bamba2016circuitSRPT}, but no experimental observations have been reported for this situation, either.

Early studies suggested no-go theorems against the SRPT  \cite{Rzazewski1975PRL,Knight1978,Bialynicki-Birula1979,Gawedzki1981PRA} suggesting that the thermal-equilibrium SRPT is impossible to realize in systems described by the minimal-coupling Hamiltonian, i.e., charged particles (without spins) interacting with electromagnetic fields.  Since the classical treatment of the electromagnetic fields used in proofs of such no-go theorems can be justified only in limited situations  \cite{Wang1973PRA,Hepp1973PRA,Bialynicki-Birula1979,Hemmen1980PLA,Gawedzki1981PRA,Bamba2017NogoCircuit}, proposals of counter-examples against the no-go theorems and criticisms against the counter-examples have been repeated in the research history of the SRPT  \cite{Keeling2007JPCM,Vukics2012PRA,Vukics2014PRL,Bamba2014SPT,Vukics2015PRA,Griesser2016PRA,Hagenmuller2012PRL,Chirolli2012PRL,Mazza2018,Andolina2019,Nataf2019a}.
% Eventually, the SRPT has not yet been observed until now.
%including its analogue,
%while an analogous thermal phase transition was theoretically proposed
%for a superconducting circuit  \cite{Bamba2016circuitSRPT}.

One way to evade the no-go theorems is by introducing another degree of freedom, such as spin  \cite{Knight1978}. For example, it has been shown that the Rashba spin--orbit coupling can cause a paramagnetic instability in an ultrastrongly coupled system between a cyclotron resonance and a cavity photon field, implying a SRPT  \cite{Nataf2019a}.   Another way is to utilize various types of  interactions in magnetic materials, which cannot be described by the minimal-coupling Hamiltonian. Ultrastrong photon--magnon coupling has been reported  \cite{Zhang2014PRLa,Goryachev2014PRA,Bourhill2016,Kostylev2016,Flower2019}, but evidence for a SRPT has not been achieved.  A variety of phase transitions exist in magnetic systems, and it is conceivable that some of the known phase transitions can be understood as a realization of the SRPT.  In this context, it is noteworthy that the problem of ultrastrong coupling between $\Erion$ spins and $\Feion$ magnons in $\ErFeO$ has been mapped to the Dicke model in a recent experimental study \cite{Li2018a}.  In this extraordinary situation of matter--matter ultrastrong coupling, the role of photons in the usual Dicke model is played by magnons.

In this paper, we theoretically show that the phase transition in $\ErFeO$ with a critical temperature ($\Tc$) of $\sim4\;\mathrm{K}$, known as the low-temperature phase transition (LTPT), is a magnonic SRPT, i.e., an analog of the SRPT where $\Erion$ spins cooperatively couple with a $\Feion$ magnonic field, instead of a photonic field as in the originally proposed SRPT.

We determined the parameters in our spin model from terahertz magnetospectroscopy \cite{Li2018a} and magnetization \cite{Zhang2019} experiments.  We derived an extended version of the Dicke model \cite{Dicke1954PR} from the spin model and clarified the correspondence between the LTPT and the SRPT.  We found that the LTPT can occur due to the $\Erion$--magnon coupling even in the absence of direct $\Erion$--$\Erion$ exchange interactions.  Also, we observed that the critical temperature $\Tc$ of the LTPT is enhanced by the $\Erion$--magnon coupling, compared to that obtained only by the direct $\Erion$--$\Erion$ interactions.  These results demonstrate that $\ErFeO$ is a unique physical system in which a SRPT can be experimentally realized in thermal equilibrium.  %The realization of the original SRPT would be accelerated by the experimental observation of the magnonic SRPT in magnetic materials.

This paper is organized as follows.  We first review the SRPT in the Dicke model and the LTPT in $\ErFeO$ in Secs.~\ref{sec:SRPT} and \ref{sec:LTPT}, respectively.  Our spin model of $\ErFeO$ is described in Sec.~\ref{sec:spin_model}.  Calculated phase diagrams are shown in Sec.~\ref{sec:phase_diagram}.  For discussing the analogy with the SRPT, an extended version of the Dicke model is derived from the spin model in Sec.~\ref{sec:spin-boson}.  The analogy is fully discussed in Sec.~\ref{sec:analogy}.  Section~\ref{sec:summary} summarizes our findings.

Appendix~\ref{app:meanfield} shows the details of our mean-field calculation. In Appendix \ref{app:symmetry}, we show how the number of parameters in the spin model can be reduced by considering the low-temperature spin configuration in $\ErFeO$.  In Appendix~\ref{sec:spin_resonance}, spin resonance frequencies are numerically calculated by the mean-field method and by the extended Dicke Hamiltonian for comparing these methods as well as for determining the parameters.  In Appendix~\ref{app:param}, the actual values of the parameters are listed. In Appendix~\ref{app:magnon}, the magnon quantization procedure for the $\Feion$ subsystem is described.  In Appendix~\ref{app:boundary}, we discuss small differences of the phase diagrams between that obtained by the mean-field method and that obtained by the extended Dicke Hamiltonian.

\section{Superradiant phase transition in the Dicke model} \label{sec:SRPT}
The SRPT was first suggested in 1973 by Hepp and Lieb \cite{Hepp1973AP} and has been extensively discussed based on the Dicke model \cite{Dicke1954PR} expressed as
%%%
\begin{equation} \label{eq:Dicke} % !!!!!!!!!
\frac{\oHHDicke}{\hbar}
\equiv \wph\oad\oa + \wex\oS_x
  + \frac{\ii2\rabi}{\sqrt{N}}(\oad-\oa)\oS_z.
\end{equation}
%%%
Here, $\oa$ is the annihilation operator of a photon in a photonic mode with a resonance frequency $\wph$, $\oS_{x,y,z}$ are the spin-$\frac{N}{2}$ operators representing an ensemble of two-level atoms with a transition frequency $\wex$, and $N$ is the number of atoms.  The last term represents the coupling between the photonic mode and the atomic ensemble with a strength of $\rabi$.  In the thermodynamic limit, i.e., $N\to\infty$, the SRPT arise when $4\rabi^2 > \wph\wex$, i.e., in the ultrastrong coupling regime, $\rabi \gtrsim \wph, \wex$  \cite{Ciuti2005PRB,Forn-Diaz2018,Kockum2018} Below $\Tc$, the expectation values of the photon annihilation operator $\braket{\oa}$ and spin operator $\braket{\oS_z}$ become nonzero, signaling the spontaneous appearance of a static electromagnetic field and a static polarization (or a persistent electric current) in thermal equilibrium.

A simpler calculation method for the SRPT was demonstrated by Wang and Hioe, also in 1973  \cite{Wang1973PRA}, and its validity for the Dicke model was confirmed by Hepp and Lieb  \cite{Hepp1973PRA}. The partition function at temperature $T$
\begin{equation}
\ZZ_{\mathrm{Dicke}}(T) \equiv \Tr[\ee^{-\oHHDicke/(\kB T)}]
\end{equation}
in the thermodynamic limit, $N\to\infty$, can be approximately evaluated by replacing the trace over the photonic variables with an integral over coherent states $\ket{\sqrt{N}\aa}$ ($\aa\in\mathbb{C}$; giving $\oa\ket{\sqrt{N}\aa}=\sqrt{N}\aa\ket{\sqrt{N}\aa}$) as
\begin{subequations}
\begin{align}
\ZZb_{\mathrm{Dicke}}(T)
& \equiv \int\frac{\dd^2\aa}{\pi/N}
  \Tr[\ee^{-\oHHeffDicke(\aa)/(\kB T)}] \\
& = \int\frac{\dd^2\aa}{\pi/N} \ee^{-\SSb_{\mathrm{Dicke}}(\aa, T)/(\kB T)},
\end{align}
\end{subequations}
where we defined an effective Hamiltonian
\begin{equation}
\frac{\oHHeffDicke(\aa)}{\hbar}
\equiv N\wph|\aa|^2 + \wex \oS_x
  + \ii2\rabi(\aa^*-\aa)\oS_z,
\end{equation}
an action
\begin{subequations}
\begin{align}
& \SSb_{\mathrm{Dicke}}(\aa,T) \nonumber \\
& \equiv -\kB T\ln\Tr[\ee^{-\oHHeffDicke(\aa)/(\kB T)}] \\
& = N\left\{ \hbar\wph|\aa|^2 - \kB T\ln\Tr[\ee^{-\oHHaDicke(\aa)/(\kB T)}] \right\},
\end{align}
\end{subequations}
and an effective Hamiltonian per atom
\begin{equation}
\frac{\oHHaDicke(\aa)}{\hbar}
\equiv \frac{\wex}{2} \osigma_x
  + \ii\rabi(\aa^*-\aa)\osigma_z.
\end{equation}
The normalized expectation value $\aa = \braket{\oa}/\sqrt{N}$ of the annihilation operator of a photon at temperature $T$ can determined for minimizing the action, i.e., $\partial \SSb/\partial \Re[\aa] = 0$ and $\partial \SSb/\partial \Im[\aa] = 0$.  We find that $\aa$ acquires a nonzero value below $\Tc$ when $4\rabi^2>\wph\wex$ is satisfied
%In a cavity of a vlume of $\vol$ with perfect mirrors,
%the vector potential $\oA$ and electric field $\oE$
($\sqrt{N}\aa$ gives a finite electric (displacement) field or vector potential even in the thermodynamic limit, $N\to\infty$, if the atomic density is fixed).  The above approximation is justified if the free energy $\FFb_{\mathrm{Dicke}}(T) \equiv - (\kB T/N)\ln\ZZb_{\mathrm{Dicke}}(T)$ per atom satisfies $\hbar\wph/N\ll|\FFb_{\mathrm{Dicke}}(T)|$ in the thermodynamic limit  \cite{Bialynicki-Birula1979,Hemmen1980PLA,Gawedzki1981PRA,Bamba2017NogoCircuit}

Based on the above semiclassical calculation scheme, Rz\k{a}\.zewski {\it et al.}\ derived no-go theorems starting from the minimal-coupling Hamiltonian in the long-wavelength approximation in 1979 \cite{Bialynicki-Birula1979} and in the general case in 1981  \cite{Gawedzki1981PRA}. However, since the proof had the above-mentioned limitation of validity due to the semiclassical treatment employed, the presence of the SRPT in the minimal-coupling Hamiltonian is still controversial  \cite{Keeling2007JPCM,Vukics2012PRA,Vukics2014PRL,Bamba2014SPT,Vukics2015PRA,Griesser2016PRA,Hagenmuller2012PRL,Chirolli2012PRL,Mazza2018,Andolina2019,Nataf2019a}.

%Below, we consider a spin model that describes the LTPT in $\ErFeO$.  We transform it into an extended version of the Dicke model, Eq.~\eqref{eq:Dicke}, and will discuss the analogy between the SRPT and LTPT.

\section{Low-temperature phase transition in $\bm{\ErFeO}$} \label{sec:LTPT}
Resonance frequencies of magnons, quanta of spin waves, in magnetic materials have provided rich information on the spin configurations of materials.  Softening (i.e., decrease of resonance frequency) of magnon modes has been discussed in connection with magnetic phase transitions.  Magnons also provide a platform for electrodynamics studies both in the classical and quantum regimes  \cite{Tabuchi2014PRL,Zhang2014PRLa,Goryachev2014PRA,Tabuchi2015,Tabuchi2016,Bourhill2016,Kostylev2016,Morris2017,Li2018a,Flower2019a,Macneill2019,Flower2019,Liensberger2019,Lachance-Quirion2019}.

% Especially, ultrastrong magnon--photon (quantum of electromagnetic wave) and spin--magnon  \cite{Li2018a} couplings are essential for realizing superradiant phase transitions (SRPTs)  \cite{Hepp1973AP,Wang1973PRA} and its magnon analogues, respectively, which originate from long-range interactions mediated by photons or magnons rather than the short-range exchange interactions in standard magnetic phase transitions.

Due to the coupling (amplitude exchange) between a magnon in magnetic materials and a photon (electromagnetic wave) in a cavity, which can be described by the last term in the Dicke model [Eq.~\eqref{eq:Dicke}], we can observe anticrossing on their resonance frequencies.  If the anticrossing frequency is higher than dephasing rates (broadening or linewidth), we can exchange the amplitude coherently between the magnon and photon modes.  Such a regime is called the strong coupling regime, and it attracts much attention for coherent transfer of quantum information between different media of quanta \cite{Tabuchi2014PRL,Zhang2014PRLa,Goryachev2014PRA,Tabuchi2015,Tabuchi2016,Morris2017} and for magnon detection  \cite{Flower2019a,Flower2019,Lachance-Quirion2019}

On the other hand, the anticrossing frequency ($2\rabi$) can be comparable to the original resonance frequency ($\wph$) of photons, magnons, or other material excitations ($\wex$), i.e., the ultrastrong coupling regime  \cite{Ciuti2005PRB,Forn-Diaz2018,Kockum2018}. Ultrastrong photon--magnon coupling has been reported for a yttrium-iron-garnet (YIG) sphere embedded in a cavity with a resonance frequency in the gigahertz (GHz) region  \cite{Zhang2014PRLa,Goryachev2014PRA,Bourhill2016,Kostylev2016,Flower2019}. Recently, $\rabi/\omega \sim 0.46$ has been achieved for the purpose of detecting dark matter (galactic axions)  \cite{Flower2019a,Flower2019}. Ultrastrong spin--magnon \cite{Li2018a} and magnon--magnon \cite{Macneill2019,Liensberger2019} coupling have also been observed.  Among such magnetic materials with ultrastrong coupling, $\ErFeO$ is a candidate material showing the magnonic SRPT as explained below.

\begin{figure}[tbp]
\centering
\includegraphics[width=\linewidth]{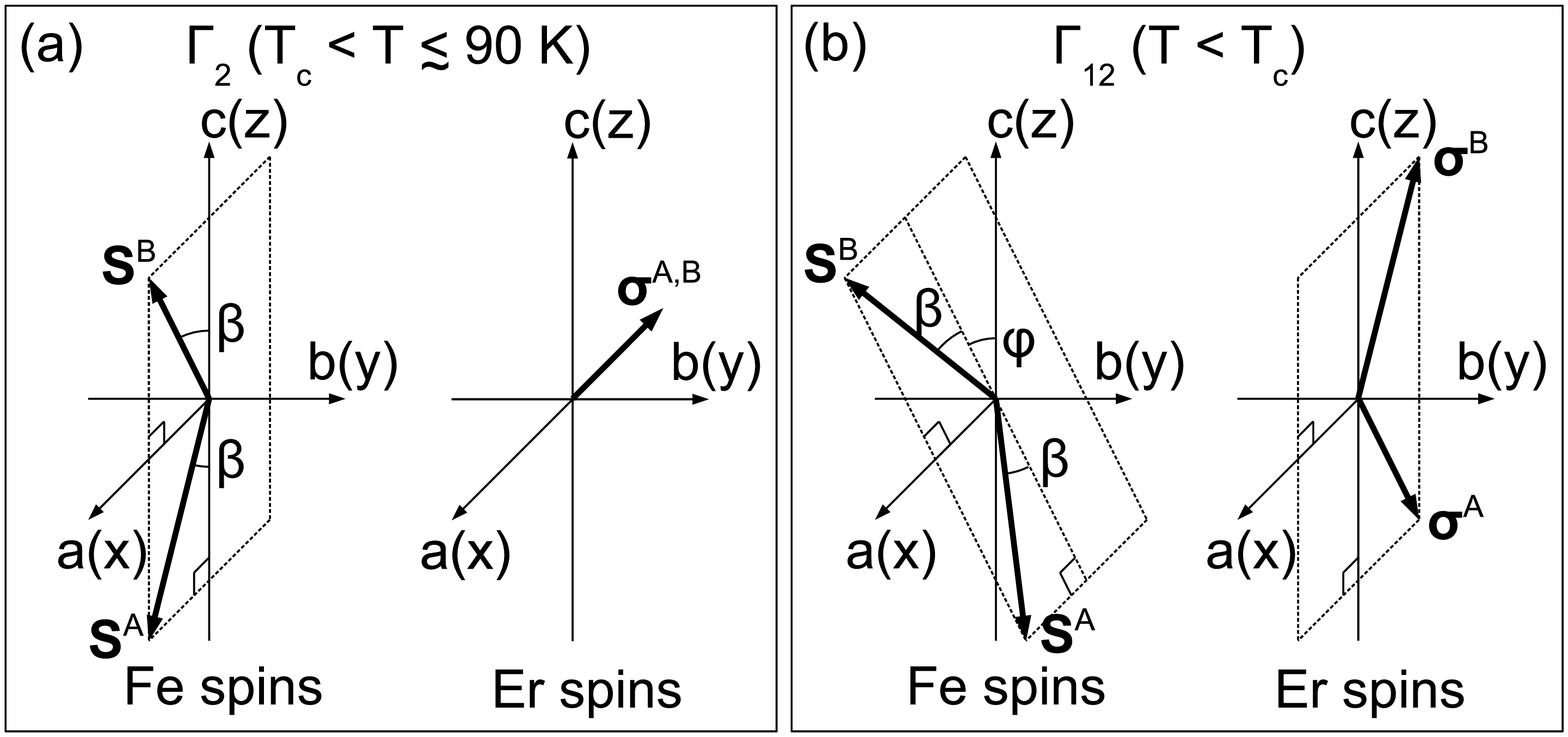}
\caption{Spin configurations in $\mathrm{ErFeO}_3$ below and above $\Tc \sim 4\;\mathrm{K}$.  In this paper, we consider two-sublattice models both for $\Erion$ and $\Feion$ spins.  (a)~In the high-temperature ($\Tc < T \lesssim 90\;\mathrm{K}$, $\Gamma_2$) phase, the $\Feion$ spins are ordered antiferromagnetically along the $c$ axis with a slight canting toward the $a$ axis.  The $\Erion$ spins are paramagnetic and directed to the $a$ axis by the weak $\Feion$ magnetization.  (b)~In the low-temperature ($T < \Tc$, $\Gamma_{12}$) phase, the $\Erion$ spins are ordered antiferromagnetically along the $c$ axis, and the AFM vector $\vS{\text{A}}-\vS{\text{B}}$ of the $\Feion$ spins rotates in the $bc$ plane.
}
\label{fig:ErFeO3_4.5K}
\end{figure}
As shown in Fig.~\ref{fig:ErFeO3_4.5K}, at $\Tc \sim 4\;\mathrm{K}$, $\ErFeO$ shows the LTPT  \cite{Gorodetsky1973,Klochan1975}, a second-order phase transition where $\Erion$ spins are ordered antiferromagnetically along the $c$ axis together with a rotation of the $\Feion$ antiferromagnetism (AFM) vector $\vS{\text{A}}-\vS{\text{B}}$ in the $bc$ plane due to the $\Erion$--$\Feion$ exchange interactions.

In the absence of those exchange interactions, as in Fig.~\ref{fig:ErFeO3_4.5K}(a), $\Feion$ spins are ordered antiferromagnetically just along the $c$ axis with a slight canting to the $a$ axis in the ground state of the $\Feion$ subsystem.  When we consider that the magnon excitation in this $\Feion$ subsystem corresponds to the photon excitation in the electromagnetic vacuum, the rotation of the $\Feion$ AFM vector (at $T<\Tc$ as shown in Fig.~\ref{fig:ErFeO3_4.5K}(b)) means a spontaneous appearance of magnons, which corresponds to the appearance of photons (a static electromagnetic field) in the ordinary SRPT, in thermal equilibrium. The ordering of $\Erion$ spins correspond to the spontaneous appearance of an atomic field (a polarization) in the SRPT. In this way, we can expect that there is an analogy between the LTPT in $\ErFeO$ and the SRPT in the Dicke model.

A theoretical model for describing the LTPT was proposed by Vitebskii and Yablonskii in 1978  \cite{Vitebskii1978}. The ratio between the \ErEr{} and \ErFe{} interaction strengths was theoretically investigated by Kadomtseva, Krynetskii, and Matveev in 1980  \cite{Kadomtseva1980}. They also mentioned the analogy between the LTPT and the cooperative Jahn--Teller transition  \cite{Gehring1975,Kugel1982}. The analogy between the cooperative Jahn--Teller transition and the SRPT was discussed by Loos in 1984 \cite{Loos1984} and also by Larson in 2008  \cite{Larson2008PRA}. Loos also suggested a magnetic system consisting of coupled ferromagnetic and paramagnetic spins, such as rare-earth iron garnets, as a candidate system for observing the above analogy. However, this analogy has not yet been verified experimentally.

$\ErFeO$ can be modeled as coupled antiferromagnetic and paramagnetic (or antiferromagnetic) spins.  In the above-mentioned studies, unfortunately, the analogy between the LTPT and the SRPT was not directly drawn either theoretically or experimentally.  In 2018, the $\sqrt{N}$-dependence ($N$ is the $\Erion$ density) of the anticrossing frequency, or vacuum Rabi splitting ($2\rabi$), between paramagnetic $\Erion$ spins and a $\Feion$ magnon mode was confirmed experimentally at $T>\Tc$ by Li \textit{et al}  \cite{Li2018a}. This $\sqrt{N}$-dependence, the Dicke cooperativity, can be taken as evidence that the coupling between the $\Erion$ spin ensemble and the $\Feion$ magnon mode is cooperative, well described by the Dicke model or its extension.

As pointed out in the early studies  \cite{Vitebskii1978,Kadomtseva1980}, it is important to take into account not only the $\Erion$--magnon coupling but also the antiferromagnetic \ErEr exchange interactions for discussing the LTPT in $\ErFeO$.  Therefore, we must extend the Dicke model to fully describe the LTPT, because Eq.~\eqref{eq:Dicke} does not include the atom--atom interactions that correspond to the $\Erion$--$\Erion$ exchange interactions.  In our experiments  \cite{Li2018a}, while the $\Erion$--magnon coupling was clearly observed through terahertz absorption spectroscopy, the influence of the \ErEr{} interactions remained unclear.

We determined the parameters in our spin model (Sec.~\ref{sec:spin_model}), including the \ErEr{} interactions, through terahertz spectra that we observed previously \cite{Li2018a} as well as the phase diagrams obtained in a recent magnetization study  \cite{Zhang2019}. The parameter estimation method is discussed in Appendices, and we focus on the analogy between the LTPT and SRPT in the following sections.

\section{Spin Model} \label{sec:spin_model} % !!!!!!!!!!!!!!!!!!!!!!!!!!!!!!!!
Each unit cell of $\ErFeO$ contains four $\Erion$ ions and four $\Feion$ ions.  The four $\Feion$ spins, each of which has an angular momentum of $\hbar S = (5/2)\hbar$, are oriented in different directions with each other even in the absence of an external DC magnetic field  \cite{Herrmann1964}. However, it is known that the $\Feion$ spin resonances (magnon modes) are well described by considering only two spins $\ovS^\text{A/B}$, each of which in fact consists of two real $\Feion$ spins but is usually treated as a single spin with $S = 5/2$.  In such a two-sublattice model of $\Feion$, as depicted in Fig.~\ref{fig:ErFeO3_4.5K}(a), at $\Tc < T \lesssim 90\;\mathrm{K}$, the two spins $\ovS^\text{A/B}$ are ordered antiferromagnetically along the $c$ axis, while they are slightly canted toward the $a$ axis and show a weak magnetization (the $\Feion$ spins show the so-called spin-reorientation transition at $90\;\mathrm{K} \lesssim T \lesssim 100\;\mathrm{K}$ \cite{Gorodetsky1973,Klochan1975,Zhang2019}).  On the other hand, $\Erion$ spins are paramagnetic at $T > \Tc$, and they are directed along the $a$ axis by the weak $\Feion$ magnetization.  This phase is called the $\Gamma_2$ phase  \cite{Kadomtseva1980}.

At $T < \Tc$, as shown in Fig.~\ref{fig:ErFeO3_4.5K}(b), when we use a two-sublattice model also for $\Erion$ spins, they are ordered antiferromagnetically along the $c$ axis, with a canting toward the $a$ axis due to the $\Feion$ magnetization.  Simultaneously, the $\Feion$ AFM vector gradually rotates in the $bc$ plane.  The rotation angle measured from the $c$ axis, $\varphi$, has been estimated to be $49^{\circ}$ at $T = 0\;\mathrm{K}$  \cite{Kadomtseva1980}. This low-temperature phase is called the $\Gamma_{12}$ phase  \cite{Kadomtseva1980}.

In the following, we describe our spin model for $\ErYFeO$ ($0 \le x \le 1$), which is consistent with our previous experimental study  \cite{Li2018a}. The $x$-dependence is described in more detail in Appendix~\ref{sec:spin_resonance}.  The replacement of $\Erion$ ions by non-magnetic $\Yion$ ones simply reduces the density of the rare-earth ($\Erion$) spins without changing the crystal structure or the magnetic configuration of $\Feion$ spins in the $\Gamma_2$ phase  \cite{Li2018a,Wood1969}.

Our Hamiltonian for the spins in $\ErYFeO$ consists of three parts:
\begin{equation} \label{eq:HH_ErYFeO3} % !!!!!!!!!!!!!!!!!!!!!!!!!!!!!!!!!!!!!
\HH = \HHFe + \HHEr + \HHFeEr,
\end{equation}
where $\HHFe$, $\HHEr$, and $\HHFeEr$ are the Hamiltonians of the $\Feion$ spins, $\Erion$ spins, and \ErFe{} interactions, respectively.

As explained above, we employ the two-sublattice model for $\Feion$ spins by following Herrmann's model \cite{Herrmann1963JPCS} and our previous studies  \cite{Li2018a,Bamba2019SPIE}. The Hamiltonian of $\Feion$ spins is described as
\begin{align} \label{eq:VFe} % !!!!!!!!!!!!!!!!!!!!!!!!!!!!!!!!!!!!!!!!!!
\oHHFe
& = \sum_{s=\text{A,B}} \sum_{i=1}^{\NUC} \muB \ovS_i^s\cdot\mgfFe\cdot\vBstat
  + \JFe \sumnn \ovS_i^\text{A} \cdot \ovS_{i'}^\text{B}
\nonumber \\ & \quad
- \DFe_y \sumnn \left( \oS_{i,z}^\text{A} \oS_{i',x}^\text{B} - \oS_{i',z}^\text{B} \oS_{i,x}^\text{A} \right)
\nonumber \\ & \quad
- \sum_{i=1}^{\NUC} \left( \Ax \oS_{i,x}^\text{A}{}^2 + \Az \oS_{i,z}^\text{A}{}^2 + \Axz \oS_{i,x}^\text{A} \oS_{i,z}^\text{A} \right)
\nonumber \\ & \quad
- \sum_{i=1}^{\NUC} \left( \Ax \oS_{i,x}^\text{B}{}^2 + \Az \oS_{i,z}^\text{B}{}^2 - \Axz \oS_{i,x}^\text{B} \oS_{i,z}^\text{B} \right).
\end{align}
Here, $\ovS_i^\text{A/B}$ is the operator of the $\Feion$ spin with $S=5/2$ at the $i$-th site in the A/B sublattice.  $\sumnn$ means a summation over all the nearest neighbor couplings.  The number of nearest neighbors is
\begin{equation}
\zz = 6.
\end{equation}
$\NUC$ is the number of $\Feion$ spins in each sublattice and is equal to the number of unit cells in $\ErFeO$.  Then, there are in total $2\NUC$ spins representing the $\Feion$ subsystem. $\muB$ is the Bohr magneton, and
\begin{equation}
\mgfFe \equiv \begin{pmatrix} \gfFe_x & 0 & 0 \\ 0 & \gfFe_y & 0 \\ 0 & 0 & \gfFe_z \end{pmatrix}
\end{equation}
is the $g$-factor tensor for the $\Feion$ spins.  In the following, the $g$-factor of free electron spin is expressed as $\gf$.  $\vBstat$ is an external DC magnetic flux density.  $\JFe$ and $\DFe_y$ are, respectively, the strengths of isotropic and Dzyaloshinkii--Moriya-type exchange interaction strengths between $\Feion$ spins.  $\Ax$, $\Az$, and $\Axz$ are the energies expressing the magnetic anisotropy of $\Feion$ spins.

While we expressed the $\Erion$ subsystem by a single spin lattice for the paramagnetic $\Erion$ spins ($T > \Tc$) in our previous studies  \cite{Li2018a,Bamba2019SPIE}, we use a two-sublattice model for $\Erion$ spins in this paper in order to describe the $\Erion$--$\Erion$ exchange interaction and the LTPT.  The Hamiltonian of $\Erion$ spins is expressed as
\begin{equation} \label{eq:VEr} % !!!!!!!!!!!!!!!!!!!!!!!!!!!!!!!!!!!!!!!!!!
\oHHEr = - \sum_{s=\text{A,B}}\sum_{i=1}^{\NUC} \ovmu_i^s \cdot \vBstat
+ \JEr \sum_{\text{n.n.}} \ovR_{i}^\text{A} \cdot \ovR_{i'}^\text{B}.
\end{equation}
Here, $\ovR_i^\text{A/B}$ is the operator of rare-earth ($\Erion$ or $\Yion$) spin at the $i$-site in the A/B sublattice.  For $\ErYFeO$, the rare-earth spins are represented randomly as ($s=\text{A,B}$)
\begin{equation}
\ovR_i^s = \begin{cases}
\ovsigma_i^s & \text{for $\Erion$}\\
\vzero & \text{for $\Yion$}
\end{cases}
\end{equation}
%While the real $\Erion$ spin in $\ErYFeO$ has an angular momentum of $(15/2)\hbar$,
We describe each $\Erion$ spin by a Pauli operator $\ovsigma_i^s$. The $\Yion$ ion is nonmagnetic and $\ovR_i^s$ is replaced by $\vzero$.  The first term in Eq.~\eqref{eq:VEr} represents the Zeeman effect, and the magnetic moment is expressed in terms of anisotropic $g$-factors, $\gfEr_{x,y,z}$, for the $\Erion$ spins as
\begin{equation} \label{eq:ovmu} % !!!!!!!!!!!!!!!!!!!!!!!!!!!!!!!!!!!!!!!!!!!
\ovmu_i^s
\equiv - \half\muB (\gfEr_x\oR_{i,x}^s, \gfEr_y\oR_{i,y}^s, \gfEr_z\oR_{i,z}^s)^{\text{t}}
= - \half\muB \mgfEr\cdot\ovR^s_i.
\end{equation}
The factor $1/2$ is added since $(1/2)\ovsigma_i^s$ corresponds to a spin-$\half$ operator theoretically.  We defined the $g$-factor tensor for $\Erion$ spins as
\begin{equation}
\mgfEr \equiv \begin{pmatrix} \gfEr_x & 0 & 0 \\ 0 & \gfEr_y & 0 \\ 0 & 0 & \gfEr_z \end{pmatrix}.
\end{equation}
The second term in Eq.~\eqref{eq:VEr} represents the \ErEr exchange interaction with a strength of $\JEr$.
% In our model, the two spins $\ovR_i^A$ and $\ovR_i^B$
% in the same unit cell interact with each other
% but do not interact with the spins in other unit cells.
Since $\Erion$ ions are diluted in $\ErYFeO$, the number of nearest neighbor $\Erion$ spins is effectively given by
\begin{equation}
\zEr = 6x.
\end{equation}

In a similar manner to our previous studies  \cite{Li2018a,Bamba2019SPIE}, we describe the \ErFe{} interaction Hamiltonian as
\begin{align} \label{eq:VFeEr} % !!!!!!!!!!!!!!!!!!!!!!!!!!!!!!!!!!!!!!!!!!
\oHHFeEr
& = \sum_{i=1}^{\NUC} \sum_{s,s'=\text{A,B}} \left[
      J \ovR_{i}^s \cdot \ovS_{i}^{s'}
    + \vD{s,s'}\cdot (\ovR_{i}^s \times \ovS_{i}^{s'})
    \right].
\end{align}
In our model, the \ErFe{} interaction is closed in each unit cell, i.e., the $\Erion$ and $\Feion$ spins in the same unit cell interact with each other but do not interact with the spins in other unit cells.  $J$ and $\vD{s,s'}$ are the strengths of the isotropic and antisymmetric exchange interactions, respectively.  Considering the spin configuration at $T < \Tc$ with no external DC magnetic field (see more details in Appendix~\ref{app:symmetry}), we assume that $\vD{s,s'}$ are expressed in terms of two values $D_x$ and $D_y$ as
\begin{subequations}
\begin{align}
\vD{\text{A,A}} & = (D_x, D_y, 0)^{\text{t}}, \\
\vD{\text{A,B}} & = (- D_x, -D_y, 0)^{\text{t}}, \\
\vD{\text{B,A}} & = (- D_x, D_y, 0)^{\text{t}}, \\
\vD{\text{B,B}} & = (D_x, -D_y, 0)^{\text{t}}.
\end{align}
\end{subequations}
Note that, as explained in Appendix~\ref{app:meanfield}, we assume that the $y$ components, $\oR^\mathrm{A/B}_{i,y}$, of the $\Erion$ spins are not influenced by the \ErFe{} interaction by implicitly considering a higher energy potential than the \ErFe{} interaction strengths $J$ and $\vD{s,s'}$ along the $b$ axis.  This assumption is required for properly describing the LTPT.

The actual values of the parameters that appears in our spin model are shown in Appendix~\ref{app:param}, together with a description of how we determined them.

\section{Phase diagrams} \label{sec:phase_diagram}
In this section, we show thermal-equilibrium (averaged) values of the $\Erion$ spins $\vsigmaa{\text{A/B}}$ and of the $\Feion$ spins $\vSa{\text{A/B}}$ in the zero-wavenumber (infinite-wavelength) limit by a mean-field method. Details of the mean-field method are given in Appendix~\ref{app:meanfield}. Since we simply consider a homogeneous external DC magnetic flux density, $\vBext$, in this paper, $\vsigmaa{\text{A/B}}$ and $\vSa{\text{A/B}}$ are independent of the site index $i$.

\begin{figure}[tbp]
\centering
\includegraphics[width=\linewidth]{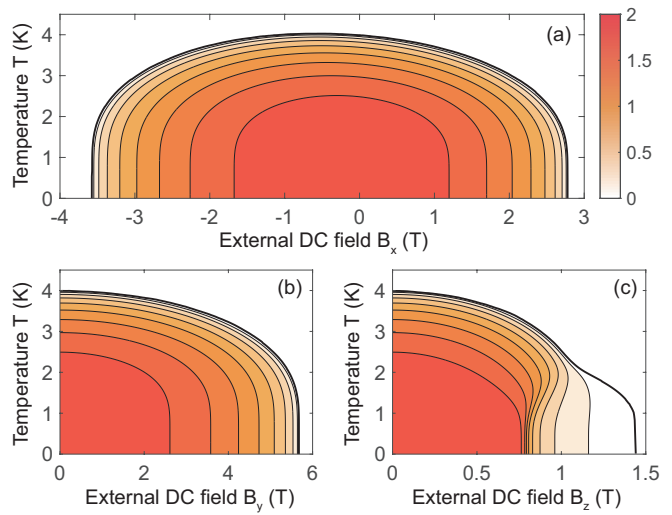}
\caption{Phase diagrams of spins in $\ErFeO$ calculated by the mean-field method.  An external DC magnetic field is applied along the (a)~$a$, (b)~$b$, and (c)~$c$ axes, respectively.  The difference $|\sigmaa{\text{A}}_z - \sigmaa{\text{B}}_z|$ of the $z$ components of the thermal-equilibrium values of $\Erion$ spins is mapped with red color.  The bold solid lines represent the phase boundaries.  The external DC magnetic field is varied from zero to positive or negative values at a fixed temperature.  Since $\ErFeO$ shows a weak magnetization along the $a$ axis, the critical field depends on whether the field is parallel or antiparallel to the magnetization in Fig.~\ref{fig:phase_diagram}(a).
}
\label{fig:phase_diagram}
\end{figure}

Figures~\ref{fig:phase_diagram}(a), (b), and (c) show calculated phase diagrams as a function of temperature, $T$, and external DC magnetic flux density, $\vBext$, applied along the $a$, $b$, and $c$ axes, respectively.  We plot the difference $|\sigmaa{\text{A}}_z - \sigmaa{\text{B}}_z|$ of the $z$ components of the thermal-equilibrium values of $\Erion$ spins (AFM vector) with red color.  It is the order parameter for the LTPT in the presence of an external DC magnetic field in general, while the rotation angle of the $\Feion$ AFM vector can be an alternative order parameter if the external DC field is zero or along the $a$ axis.  The bold solid lines represent the phase boundaries.  These phase diagrams well reproduce those observed by Zhang {\it et al.} \cite{Zhang2019}. As shown in Fig.~\ref{fig:phase_diagram}(a), since $\ErFeO$ possesses a weak magnetization along the $a$ axis, the critical field depends on whether the field is parallel or antiparallel to the magnetization.  The parameters used in the calculation are shown in Appendix~\ref{app:param}.

\begin{figure}[tbp]
\centering
\includegraphics[width=\linewidth]{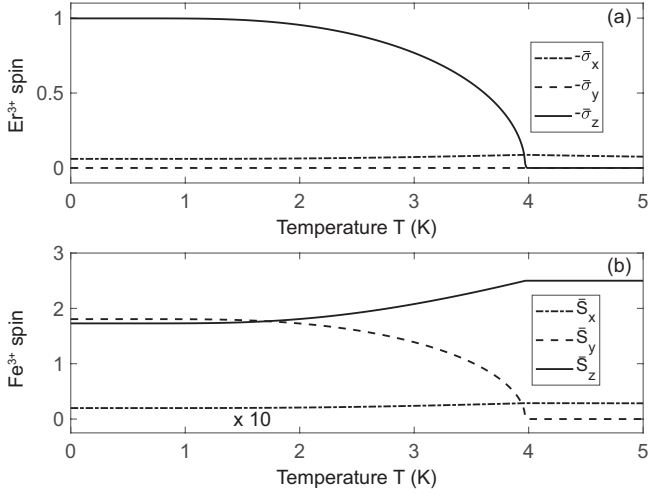}
\caption{The thermal-equilibrium values of (a)~$\Erion$ spin and (b)~$\Feion$ spin calculated by the mean-field method are plotted as a function of temperature $T$ in the case of zero external DC magnetic field.  As shown in Fig.~\ref{fig:RS_T}(a), $\sigmaa{}_z = \sigmaa{\text{A}}_z = - \sigmaa{\text{B}}_z$ spontaneously appears below the critical temperature $\Tc = 4.0\;\mathrm{K}$, i.e., the $\Erion$ spins are antiferromagnetically ordered along the $c$ axis.  They show a magnetization along the $a$ axis as $\sigmaa{}_x = \sigmaa{\text{A/B}}_x$ due to the $\Erion$--$\Feion$ exchange interaction with the weak $\Feion$ magnetization along the $a$ axis, while $\sigmaa{}_y = \sigmaa{\text{A/B}}_y = 0$.  As shown in Fig.~\ref{fig:RS_T}(b), above $\Tc$, the $\Feion$ spins are ordered antiferromagnetically along the $c$ axis as $\Sa{}_z = - \Sa{\text{A}}_z = \Sa{\text{B}}_z$, while they are slightly canted toward the $a$ axis as $\Sa{}_x = \Sa{\text{A/B}}_x$, and $\Sa{}_y = \Sa{\text{A}}_y = - \Sa{\text{B}}_y$ = 0.  Below $\Tc$, the $\Feion$ spins rotate in the $bc$ plane, and the rotation angle is $\varphi = \arctan(\Sa{}_y/\Sa{}_z) = 46^{\circ}$ at $T = 0\;\mathrm{K}$ with our parameters.
}
\label{fig:RS_T}
\end{figure}

In Fig.~\ref{fig:RS_T}, we plot the thermal-equilibrium values of the $\Erion$ and $\Feion$ spins in the absence of an external DC magnetic field as a function of temperature.  The LTPT, i.e., the antiferromagnetic ordering of $\Erion$ spins along the $c$ axis and the rotation of the $\Feion$ spins in the $bc$ plane \cite{Kadomtseva1980} are well reproduced in our spin model.  The rotation angle of the $\Feion$ AFM vector  is $\varphi = 46^{\circ}$ at $T = 0\;\mathrm{K}$ with our parameters.  This is approximately equal to the experimentally estimated value $\varphi = 49^{\circ}$  \cite{Kadomtseva1980}.

\section{Extended Dicke Hamiltonian} \label{sec:spin-boson}
In the previous sections, we discussed the LTPT of $\ErFeO$ through mean-field calculations based on our spin model.  It is a standard approach for analyzing magnetic phase transitions.  In this section, in order to discuss the analogy between the LTPT and the SRPT in the Dicke model, we transform the spin model, Eq.~\eqref{eq:HH_ErYFeO3}, into an extended version of the Dicke model, including direct $\Erion$--$\Erion$ exchange interactions, which were not considered in our previous studies  \cite{Li2018a,Bamba2019SPIE}.

We first rewrite the $\Feion$ subsystem in terms of the annihilation and creation operators of a magnon in Sec.~\ref{sec:Fe}.  The $\Erion$ subsystem is rewritten by large spin operators in Sec.~\ref{sec:Er}.  The $\Erion$--$\Feion$ exchange interactions are transformed into five $\Erion$--magnon couplings in Sec.~\ref{sec:ErFe}.  The total Hamiltonian is given in Sec.~\ref{sec:total}.

\subsection{$\Feion$ subsystem} \label{sec:Fe}
We assume that the most-stable values of the $\Feion$ spins at zero temperature, $\vSa{\text{A/B}}$, are unchanged when an external DC magnetic flux density $\vBext$ ($\lesssim10\;\mathrm{T}$) is applied, as we also assumed in our previous studies  \cite{Li2018a,Bamba2019SPIE}. Under this assumption, as depicted in Fig.~\ref{fig:ErFeO3_4.5K}(a), the most stable state (i.e., ground state) of the $\Feion$ subsystem $\oHHFe$, Eq.~\eqref{eq:VFe}, are expressed as
\begin{equation} \label{eq:vSa12} % !!!!!!!!!!!!!!!!!!!!!!!!!!!!!!!!!!!!!!!!!!
\vSa{\text{A}}_0 = \begin{pmatrix} S \sin\beta_0 \\ 0 \\ - S \cos\beta_0 \end{pmatrix}, \quad
\vSa{\text{B}}_0 = \begin{pmatrix} S \sin\beta_0 \\ 0 \\ S \cos\beta_0 \end{pmatrix}.
\end{equation}
Here, the canting angle $\beta_0$ is expressed as (see Appendix~\ref{app:magnon} or Refs.~\onlinecite{Herrmann1963JPCS,Li2018a,Bamba2019SPIE})
\begin{equation} \label{eq:beta0} % !!!!!!!!!!!!!!!!!!!!!!!!!!!!!!!!!!!!!!!!!!
\beta_0 = - \frac{1}{2}\arctan\frac{\Axy+\zz\DFe_y}{\zz\JFe-\Ax+\Az}.
\end{equation}

The magnon is the quantum of spin fluctuations from this stable state.  As shown in Appendix~\ref{app:magnon} as well as in Refs.~\onlinecite{Li2018a,Bamba2019SPIE}, in the long wavelength limit, the $\Feion$ Hamiltonian $\oHHFe$, Eq.~\eqref{eq:VFe}, can be rewritten in terms of the annihilation (creation) operators $\oa_K$ ($\oad_K$) of $\Feion$ magnons as
\begin{align} \label{eq:oHHFe2} % !!!!!!!!!!!!!!!!!!!!!!!!!!!!!!!!!!!!!!!!!!!!
\oHHFe
& \approx \sum_{K=0,\pi} \hbar\omega_K \oad_{K}\oa_{K}
  + \const
\end{align}
Here, $K=0$ and $\pi$ correspond to the quasi-ferromagnetic (qFM) and quasi-antiferromagnetic (qAFM) magnon modes  \cite{Herrmann1963JPCS}, respectively.  Their eigenfrequencies can be obtained as
\begin{equation}
\omega_K = \gamma\sqrt{(b\cos K-a)(d\cos K+c)},
\end{equation}
where we defined
\begin{subequations} \label{eq:abcd} % !!!!!!!!!!!!!!!!!!!!!!!!!!!!!!!!!!!!!!!
\begin{align}
a & = [S/(\gf\muB)] [ - \Az - \Ax - (\zz\JFe + \Az - \Ax)\cos(2\beta_0)
\nonumber \\ & \quad
    + (\Axz+\zz\DFe_y)\sin(2\beta_0)], \\
b & = [S/(\gf\muB)] (\zz\JFe), \\
c & = [S/(\gf\muB)] [(\zz\JFe+2\Az-2\Ax)\cos(2\beta_0)
\nonumber \\ & \quad
+ \zz\DFe_y\sin(2\beta_0)], \\
d & = [S/(\gf\muB)] [-\zz\JFe\cos(2\beta_0)
\nonumber \\ & \quad
 - (2\Axz+\zz\DFe_y)\sin(2\beta_0)].
\end{align}
\end{subequations}
The operators of the spin fluctuations $\delta\ovS^\mathrm{A/B}_i \equiv \ovS^\mathrm{A/B}_i - \vSa{\text{A/B}}_0$ are expressed as
\begin{subequations} \label{eq:delta_ovS_AB} % !!!!!!!!!!!!!!!!!!!!!!!!!!!!!!!
\begin{align}
\delta\ovS^\mathrm{\text{A}}_i
& = \sqrt{\frac{S}{2\NUC}} \begin{pmatrix}
    - ( \oT_0 - \oT_\pi ) \cos\beta_0 \\
      ( \oY_0 - \oY_\pi ) \\
    - ( \oT_0 - \oT_\pi ) \sin\beta_0
    \end{pmatrix}, \\
\delta\ovS^\mathrm{B}_i
& = \sqrt{\frac{S}{2\NUC}} \begin{pmatrix}
      ( \oT_0 + \oT_\pi ) \cos\beta_0 \\
      ( \oY_0 + \oY_\pi ) \\
    - ( \oT_0 + \oT_\pi ) \sin\beta_0
    \end{pmatrix},
\end{align}
\end{subequations}
where we defined
\begin{subequations} \label{eq:oTY_a} % !!!!!!!!!!!!!!!!!!!!!!!!!!!!!!!!!!!!!!
\begin{align}
\oT_K
& \equiv \left( \frac{b\cos K-a}{d\cos K+c} \right)^{1/4}
    \frac{(\oad_{-K}+\oa_K)}{\sqrt{2}}, \\
\oY_K
& \equiv \left( \frac{d\cos K+c}{b\cos K-a} \right)^{1/4}
    \frac{\ii(\oad_{-K}-\oa_K)}{\sqrt{2}}.
\end{align}
\end{subequations}
For the discussion in the next subsections, we define the sum and difference of the spins as
\begin{equation}
\ovS^{\pm}_i \equiv \ovS^\mathrm{A}_{i} \pm \ovS^\mathrm{B}_{i}.
\end{equation}
Their equilibrium (most stable) values are
\begin{subequations} \label{eq:vSa+-} % !!!!!!!!!!!!!!!!!!!!!!!!!!!!!!!!!!!!!!!!!
\begin{align}
\vSa{+}_0 \equiv \vSa{\text{A}}_0 + \vSa{\text{B}}_0
& = (2S\sin\beta_0, 0, 0)^{\text{t}}, \\
\vSa{-}_0 \equiv \vSa{\text{A}}_0 - \vSa{\text{B}}_0
& = (0, 0, -2S\cos\beta_0)^{\text{t}},
\end{align}
\end{subequations}
and their fluctuations are expressed as
\begin{subequations} \label{eq:delta_ovS+-} % !!!!!!!!!!!!!!!!!!!!!!!!!!!!!!!
\begin{align}
\delta\ovS^+
& \equiv \delta\ovS^\mathrm{A}_{i} + \delta\ovS^\mathrm{B}_{i}
  = \sqrt{\frac{2S}{\NUC}} \begin{pmatrix}
      \oT_\pi \cos\beta_0 \\
      \oY_0 \\
    - \oT_0 \sin\beta_0
    \end{pmatrix}, \\
\delta\ovS^-
& \equiv \delta\ovS^\mathrm{A}_{i} - \delta\ovS^\mathrm{B}_{i}
  = \sqrt{\frac{2S}{\NUC}} \begin{pmatrix}
    - \oT_0 \cos\beta_0 \\
    - \oY_\pi \\
      \oT_\pi \sin\beta_0
    \end{pmatrix}.
\end{align}
\end{subequations}

\subsection{$\Erion$ subsystem} \label{sec:Er}

We define following new operators:
\begin{equation}
\ovSigma^\mathrm{A/B} \equiv \frac{1}{2}\sum_{i=1}^{\NUC} \ovR^\mathrm{A/B}_{i}.
\end{equation}
For an $\Erion$ ion, $(1/2)\ovR^\mathrm{A/B}_i$ is a spin-$\half$ operator.  The total number of spin-$\half$ spins ($\Erion$ spins) in the two sublattices is
\begin{equation}
N \equiv 2x\NUC.
\end{equation}
Then, $\ovSigma^\mathrm{A/B}$ is a spin-$\frac{N}{4}$ operator representing the rare-earth spins in the A/B sublattice.  We also define the sum and difference of the two sublattice spins as
\begin{equation}
\ovSigma^{\pm} \equiv \ovSigma^\text{A} \pm \ovSigma^\text{B}.
\end{equation}
In the long-wavelength limit, all the spins in each sublattice have the same values in both static and dynamical situations.  Then, the $\Erion$ Hamiltonian in Eq.~\eqref{eq:VEr} can be rewritten as
\begin{subequations} \label{eq:oHHEr_R+-} % !!!!!!!!!!!!!!!!!!!!!!!!!!!!!!!!!!!
\begin{align}
& \oHHEr
\nonumber \\
& \approx \sum_{\xi=x,y,z}\gfEr_{\xi}\muB\oSigma^+_{\xi}\Bext_{\xi}
+ \zEr\JEr \sum_{i=1}^{\NUC}\ovR^\mathrm{A}_i \cdot\sum_{i'=1}^{\NUC}\frac{\ovR^\mathrm{B}_{i'}}{x\NUC} \\
& = \sum_{\xi=x,y,z}\gfEr_{\xi}\muB\oSigma^+_{\xi}\Bext_{\xi}
+ \frac{8\zEr\JEr}{N} \ovSigma^\mathrm{A}\cdot\ovSigma^\mathrm{B}.
\end{align}
\end{subequations}

\subsection{$\Erion$--$\Feion$ interactions}  \label{sec:ErFe}

In the same manner as in Refs.~\onlinecite{Li2018a,Bamba2019SPIE}, we rewrite the Hamiltonian of the $\Erion$--$\Feion$ exchange interactions, Eq.~\eqref{eq:VFeEr}, as
\begin{align} \label{eq:oHHFeEr_Sigma_S} % !!!!!!!!!!!!!!!!!!!!!!!!!!!!!!!!!!!
\oHHFeEr
& \approx 2J
    \left(
      \ovSigma^+ \cdot \vSa{+}_0
    + \ovSigma^+ \cdot \delta\ovS^+
    \right)
\nonumber \\ & \quad
  + \begin{pmatrix} 0 \\ 2D_y \\ 0 \end{pmatrix} \cdot
    \left(
      \ovSigma^+ \times \vSa{-}_0
    + \ovSigma^+ \times \delta\ovS^-
    \right)
\nonumber \\ & \quad
  + \begin{pmatrix} 2D_x \\ 0 \\ 0 \end{pmatrix} \cdot
    \left(
      \ovSigma^- \times \vSa{-}_0
    + \ovSigma^- \times \delta\ovS^-
    \right).
\end{align}
In each parenthesis, the first terms represent the influence of the static components (equilibrium values) $\vSa{\text{A/B}}_0$ of $\Feion$ spins to $\Erion$ spins $\ovSigma^{\pm}$, and the second terms represent the coupling between the $\Feion$ fluctuation $\delta\ovS^{\pm}$ and the $\Erion$ spins $\ovSigma^{\pm}$.  We divide these terms into the two Hamiltonians as
\begin{equation} \label{eq:oHHFeEr} % !!!!!!!!!!!!!!!!!!!!!!!!!!!!!!!!!!!!!!!!
\oHHFeEr = \oHHstat + \oHHdyn.
\end{equation}
The first term gives a part of the $\Erion$ spin resonance frequency, and it is expressed as
\begin{align} \label{eq:oHHstat} % !!!!!!!!!!!!!!!!!!!!!!!!!!!!!!!!!!!!!!!!
\oHHstat
& = E_x\oSigma^+_{x},
\end{align}
where we used Eqs.~\eqref{eq:vSa+-} and $E_{x}$ is defined as
\begin{equation} \label{eq:Exyz} % !!!!!!!!!!!!!!!!!!!!!!!!!!!!!!!!!!!!!!!
E_x \equiv 4S(J\sin\beta_0+D_y\cos\beta_0).
\end{equation}
Note that we neglected $(-4SD_x\cos\beta_0)\oSigma^-_y$ under the assumption explained at the end of Sec.~\ref{sec:spin_model}. The second term in Eq.~\eqref{eq:oHHFeEr} is rewritten in terms of the $\Feion$ fluctuations as
\begin{align} \label{eq:oHHdyn_approx} % !!!!!!!!!!!!!!!!!!!!!!!!!!!!!!!!!!!!!
\oHHdyn
& = \sqrt{\frac{8S}{\NUC}} \left[
    (J\cos\beta_0-D_y\sin\beta_0)\oT_\pi \oSigma^+_{x}
\right. \nonumber \\ & \quad
  + J\oY_0\oSigma^+_y
  + (D_x\sin\beta_0) \oT_{\pi} \oSigma^-_{y}
  + D_x \oY_\pi \oSigma^-_{z}
\nonumber \\ & \quad \left.
  + (-J\sin\beta_0-D_y\cos\beta_0)\oT_0 \oSigma^+_{z}
  \right].
\end{align}

\subsection{Total system}  \label{sec:total}

The total Hamiltonian derived from our spin model is finally expressed as
\begin{align} \label{eq:extended_Dicke} % !!!!!!!!!!!!!!!!!!!!!!!!!!!!!!!!!!!!
\oHH
& \approx \sum_{K=0,\pi} \hbar\omega_K\oad_K\oa_K
  + E_x \oSigma^+_{x}
  + \sum_{\xi=x,y,z} \gfEr_{\xi}\muB\Bext_{\xi} \oSigma^+_{\xi}
\nonumber \\ & \quad
  + \frac{8\zEr\JEr}{N} \ovSigma^\mathrm{A} \cdot \ovSigma^\mathrm{B}
+ \frac{2\hbar\rabi_x}{\sqrt{N}}(\oad_\pi+\oa_\pi)\oSigma^+_x
\nonumber \\ & \quad
+ \frac{\ii2\hbar\rabi_y}{\sqrt{N}}(\oad_0-\oa_0)\oSigma^+_y
+ \frac{2\hbar\rabi_y'}{\sqrt{N}}(\oad_\pi+\oa_\pi)\oSigma^-_y
\nonumber \\ & \quad
+ \frac{\ii2\hbar\rabi_z}{\sqrt{N}}(\oad_\pi-\oa_\pi)\oSigma^-_z
+ \frac{2\hbar\rabi_z'}{\sqrt{N}}(\oad_0+\oa_0)\oSigma^+_z.
\end{align}
Here, the five $\Erion$--magnon coupling terms in Eq.~\eqref{eq:oHHdyn_approx} were rewritten in terms of the annihilation (creation) operators $\oa_{K}$ ($\oad_{K}$) of a magnon. The five coupling strengths are defined as
\begin{subequations} \label{eq:rabi} % !!!!!!!!!!!!!!!!!!!!!!!!!!!!!!!!!!!!!!
\begin{align}
\hbar\rabi_x
& = \sqrt{2xS}(J\cos\beta_0-D_y\sin\beta_0)\left(\frac{b+a}{d-c}\right)^{1/4}
\nonumber \\ & = h\times\sqrt{x}\times 0.051\;\THz, \\
\hbar\rabi_y & = \sqrt{2xS}J\left(\frac{d+c}{b-a}\right)^{1/4}
\nonumber \\ & = h\times\sqrt{x}\times 0.041\;\THz, \\
\hbar\rabi_y' & = \sqrt{2xS}(D_x\sin\beta_0)\left(\frac{b+a}{d-c}\right)^{1/4}
\nonumber \\ & = h\times\sqrt{x}\times 3.1\times10^{-5}\;\THz, \\
\hbar\rabi_z & = \sqrt{2xS}D_x\left(\frac{d-c}{b+a}\right)^{1/4}
\nonumber \\ & = h\times\sqrt{x}\times 0.116\;\THz, \label{eq:rabi_z} \\
\hbar\rabi_z' & = \sqrt{2xS}(-J\sin\beta_0-D_y\cos\beta_0)\left(\frac{b-a}{d+c}\right)^{1/4}
\nonumber \\ & = h\times\sqrt{x}\times (-0.040\;\THz).
\end{align}
\end{subequations}
The actual values are evaluated by the parameters shown in Appendix~\ref{app:param}.  Note that, compared with the expression in our previous studies  \cite{Li2018a,Bamba2019SPIE}, the above coupling strengths have additional factors: $\sqrt{2}$ and $\sqrt{S}$.  The first of these factors, $\sqrt{2}$, originates from the number of $\Erion$ sublattices in the present study, while a single $\Erion$ lattice was considered in our previous studies  \cite{Li2018a,Bamba2019SPIE}.
%In other words, $J$, $D_x$, and $D_y$ in this paper
%correspond to those in our previous studies \cite{Li2018a,Bamba2019SPIE}
%divided by $\sqrt{2}$.
On the other hand, the second factor, $\sqrt{S}$, comes from the difference in the way of normalizing the $\Feion$ spins between the present
and previous studies  \cite{Li2018a,Bamba2019SPIE}.

\section{Analogy between the two phase transitions} \label{sec:analogy}

Based on the extended Dicke Hamiltonian, Eq.~\eqref{eq:extended_Dicke}, derived in the previous section, we show in this section that the LTPT in ErFeO$_3$ is a magnonic SRPT.

In Sec.~\ref{sec:thermal}, we show that the $\Erion$--qAFM magnon coupling with a strength of $\rabi_z$ corresponds to the matter--photon coupling in the SRPT case.  We also demonstrate that the thermal SRPT predicted by the extended Dicke Hamiltonian correctly reproduces the temperature-dependence of the $\Erion$ and $\Feion$ spins shown in Fig.~\ref{fig:RS_T}.  In Sec.~\ref{sec:quantum}, we quantitatively compare the contributions of the $\Erion$--magnon coupling and the $\Erion$--$\Erion$ exchange interactions in the LTPT.  We show that the LTPT can be caused solely by the $\Erion$--magnon coupling.  Furthermore, we demonstrate that the $\Erion$--magnon coupling enhances the critical temperature and critical magnetic field of the phase transition, compared with the case in which the phase transition is driven by the $\Erion$--$\Erion$ exchange interactions alone.

\subsection{Correspondence} \label{sec:thermal}
In this section, by using the semiclassical method described in Sec.~\ref{sec:SRPT} with the extended Dicke Hamiltonian, Eq.~\eqref{eq:extended_Dicke}, we calculate the thermal-equilibrium values of $\Erion$ and $\Feion$ spins and magnon amplitudes as a function of temperature.
% The results in Fig.~\ref{fig:RS_T} calculated by the meanfield method will be reproduced.

While the $\Erion$ spin ensemble is described by six operators, $\oSigma^+_{x,y,z}$ and $\oSigma^-_{x,y,z}$, in the extended Dicke Hamiltonian, only $\oSigma_x^+$ and $\oSigma_z^-$ are relevant to the LTPT depicted in Fig.~\ref{fig:ErFeO3_4.5K}. $\oSigma_x^+$ corresponds to the paramagnetic alignment by the $\Feion$ magnetization along the $a$ axis, and $\oSigma_z^-$ corresponds to the antiferromagnetic ordering along the $c$ axis.  Then, for analyzing the thermal-equilibrium values of the spins, we need to consider only the following two terms in the $\Erion$--$\Erion$ exchange interactions:
\begin{align}
\frac{8\zEr\JEr}{N} \ovSigma^\mathrm{A} \cdot \ovSigma^\mathrm{B} & = \frac{2\zEr\JEr}{N}\sum_{\xi=x,y,z} \left[ (\oSigma_{\xi}^+)^2 - (\oSigma_{\xi}^-)^2 \right]
\nonumber \\
& \to \frac{2\zEr\JEr}{N} \left[ (\oSigma_{x}^+)^2 - (\oSigma_{z}^-)^2 \right].
\end{align}

On the other hand, while $\Feion$ spins are described by the qFM and qAFM magnon modes in the extended Dicke Hamiltonian, only the qAFM mode is relevant to the LTPT.  As shown in Fig.~\ref{fig:ErFeO3_4.5K}, $\delta\oS_y^-$ and $\delta\oS_z^-$ are required for describing the rotation of the $\Feion$ AFM vector in the $bc$ plane, and $\delta\oS_x^+$ is required for the possible modulation of canting along the $a$ axis. As seen in Eqs.~\eqref{eq:delta_ovS+-}, they are related to the qAFM magnon mode ($K=\pi$), and the qFM mode ($K=0$) plays no role in the LTPT.

Consequently, among the terms in the total Hamiltonian given by Eq.~\eqref{eq:extended_Dicke}, we only need to consider the following terms for describing the LTPT (the other terms are required for fully reproducing the THz spectra as discussed in Appendix \ref{sec:spin_resonance}):
\begin{align} \label{eq:oHH_pi} % !!!!!!!!!!!!!!!!!!!!!!!!!!!!!!!!!!!!!!!!!!!!
{\oHH}/{\hbar}
& \to \omega_{\pi}\oad_{\pi}\oa_{\pi}
  + \wEr\oSigma_{x}
  + \frac{2\zEr\JEr}{N\hbar} \left( \oSigma_{x}{}^2 - \oSigma_{z}{}^2 \right)
\nonumber \\ & \quad
+ \frac{2\rabi_x}{\sqrt{N}}(\oad_\pi+\oa_\pi)\oSigma_x
+ \frac{\ii2\rabi_z}{\sqrt{N}}(\oad_\pi-\oa_\pi)\oSigma_z.
\end{align}
Here, the $\Erion$ resonance frequency is defined as
\begin{equation}
\wEr \equiv \frac{|E_x+\gfEr_{x}\muB\Bext_{x}|}{\hbar}.
\end{equation}
Note that we re-wrote the large spin operators representing the $\Erion$ spin ensemble as
\begin{equation}
\begin{cases}
\oSigma_{x}^{+} \to \oSigma_{x} \equiv \sum_{i=1}^N {\osigma_{i,x}}/{2} \\
\oSigma_{y}^{-} \to \oSigma_{y} \equiv \sum_{i=1}^N {\osigma_{i,y}}/{2} \\
\oSigma_{z}^{-} \to \oSigma_{z} \equiv \sum_{i=1}^N {\osigma_{i,z}}/{2}
\end{cases}
\end{equation}
where we re-indexed the Pauli operators representing the $\Erion$ spins in the two sublattices as
\begin{equation}
\begin{cases}
\osigma_{i,x}^\text{A} \to \osigma_{2i-1,x} \\
\osigma_{i,y}^\text{A} \to \osigma_{2i-1,y} \\
\osigma_{i,z}^\text{A} \to \osigma_{2i-1,z}
\end{cases} \quad
\begin{cases}
\osigma_{i,x}^\text{B} \to \osigma_{2i,x} \\
\osigma_{i,y}^\text{B} \to -\osigma_{2i,y} \\
\osigma_{i,z}^\text{B} \to -\osigma_{2i,z}
\end{cases}
\end{equation}
In Eq.~\eqref{eq:oHH_pi}, we assumed that the external DC magnetic field is applied along the $a$ axis for keeping the $\Gamma_{12}$ symmetry, where either $|\sigmaa{\text{A}}_z-\sigmaa{\text{B}}_z|$ or the rotation angle $\varphi$ of the $\Feion$ AFM vector from the $c$ axis can be the order parameter for the LTPT.  Among the five $\Erion$--magnon couplings in Eq.~\eqref{eq:extended_Dicke}, only the $\rabi_x$ and $\rabi_z$ terms are required for considering the coupling between $\oSigma_{x,z}$ and the qAFM magnons.  While the $\rabi_y$ term also couples $\oSigma_y$ and qAFM magnons, its coupling strength is negligible compared with $\rabi_{x,z}$, as shown in Eqs.~\eqref{eq:rabi}, consistent with the experimentally observed antiferromagnetic ordering of $\Erion$ spins along the $c$ axis ($\braket{\oSigma^-_y} = 0$).

Through comparison of Eq.~\eqref{eq:oHH_pi} with Eq.~\eqref{eq:Dicke} (the Dicke model), we can identify the $\rabi_z$ term to correspond to the matter--photon coupling (transverse coupling). Additionally, the $\rabi_x$ term represents longitudinal coupling and the $\JEr$ term describes the $\Erion$--$\Erion$ exchange interactions in Eq.~\eqref{eq:oHH_pi}.  The coupling strength $\rabi_z=2\pi\times0.116\;\THz$ puts the systems in the ultrastrong regime, since it is a significant fraction of the $\Erion$ resonance and qAFM magnon frequencies, $E_x = h\times0.023\;\THz$ and $\omega_{\pi} = 2\pi\times0.896\;\THz$.  When the $\rabi_z$ term causes a SRPT, $\braket{\oSigma_z}=\braket{\oSigma_z^-}$ spontaneously acquires a nonzero value in thermal equilibrium, corresponding to the antiferromagnetic ordering of $\Erion$ spins along the $c$ axis.  As will be discussed later, the spontaneous appearance of nonzero $\braket{\ii(\oad_{\pi}-\oa_{\pi})}$, which is coupled with $\oSigma_z$ in the $\rabi_z$ term, corresponds to the rotation of the $\Feion$ AFM vector.

Following the semiclassical treatment in Sec.~\ref{sec:SRPT}, we calculate the expectation values of the $\Erion$ spins and $\Feion$ qAFM magnon operators at a finite temperature.  In the thermodynamic limit, $N\to\infty$, the partition function $\ZZ(T) \equiv \Tr[\ee^{-\oHH/(\kB T)}]$ can be approximately evaluated by replacing the trace over the magnonic variables with an integral over c-numbers $\aa_r,\aa_i\in\mathbb{R}$, giving $\oa_{\pi}\to\sqrt{N}(\aa_r + \ii \aa_i)$ as
\begin{subequations}
\begin{align}
\ZZb(T)
& \equiv \int\frac{\dd\aa_r\dd\aa_i}{\pi/N}
  \Tr[\ee^{-\oHHeff(\aa_r,\aa_i)/(\kB T)}] \\
& = \int\frac{\dd\aa_r\dd\aa_i}{\pi/N} \ee^{-\SSb(\aa, T)/(\kB T)}, \label{eq:ZZb_SSb}
\end{align}
\end{subequations}
where we defined an effective Hamiltonian
\begin{align}
\oHHeff(\aa_r,\aa_i)/\hbar
& \equiv N\omega_{\pi}(\aa_r{}^2+\aa_i{}^2)
  + \wEr\oSigma_{x}
\nonumber \\ & \quad
  + \frac{4\zEr\JEr}{N\hbar} \left( \braket{\oSigma_{x}}\oSigma_{x}
    - \braket{\oSigma_{z}}\oSigma_{z} \right)
\nonumber \\ & \quad
  - \frac{2\zEr\JEr}{N\hbar} \left( \braket{\oSigma_{x}}^2 - \braket{\oSigma_{z}}^2 \right)
\nonumber \\ & \quad
+ 4\rabi_x\aa_r\oSigma_x
+ 4\rabi_z\aa_i\oSigma_z
\end{align}
by introducing the $\Erion$ components $\braket{\oSigma_{x,z}}$ of the mean-fields for the $\Erion$ ensemble.  The action appearing in Eq.~\eqref{eq:ZZb_SSb} is defined as
\begin{subequations}
\begin{align}
& \SSb(\aa_r,\aa_i,T)
\nonumber \\
& \equiv -\kB T\ln\Tr[\ee^{-\oHHeff(\aa_r,\aa_i)/(\kB T)}] \\
& = N\left\{ \hbar\omega_{\pi}(\aa_r{}^2+\aa_i{}^2)
  - \frac{2\zEr\JEr}{N^2} \left( \braket{\oSigma_{x}}^2
 - \braket{\oSigma_{z}}^2 \right) \right\}
\nonumber \\ & \quad
 - N \kB T\ln\Tr[\ee^{-\oHHa(\aa_r,\aa_i)/(\kB T)}],
\end{align}
\end{subequations}
where we defined an effective Hamiltonian per $\Erion$ spin as
\begin{align} \label{eq:oHHa} % !!!!!!!!!!!!!!!!!!!!!!!!!!!!!!!!!!!!!!!!!!!!!!
\frac{\oHHa(\aa_r,\aa_i)}{\hbar}
& \equiv \frac{\wEr}{2} \osigma_{x}
  + \frac{2\zEr\JEr}{N\hbar} \left( \braket{\oSigma_{x}^+}\osigma_{x}
    - \braket{\oSigma_{z}^-}\osigma_{z} \right)
\nonumber \\ & \quad
+ 2\rabi_x\aa_r\osigma_x
+ 2\rabi_z\aa_i\osigma_z.
\end{align}
We omitted the site index $i$ here, since all the spins are identical.  The action $\SSb$ is minimized at $\partial \SSb/\partial \aa_r = 0$ and $\partial \SSb/\partial \aa_i = 0$, by which we get
\begin{subequations} \label{eq:alpha_sigma} % !!!!!!!!!!!!!!!!!!!!!!!!!!!!!!!!
\begin{align}
\omega_{\pi} \aa_r + \rabi_x \braket{\osigma_x} & = 0, \\
\omega_{\pi} \aa_i + \rabi_z \braket{\osigma_z} & = 0,
\end{align}
\end{subequations}
where the expectation values of the Pauli operators are defined, for given $\aa_r$ and $\aa_i$, as
\begin{equation} \label{eq:sigmaa} % !!!!!!!!!!!!!!!!!!!!!!!!!!!!!!!!!!!!!!!!!
\sigmaa{}_{\xi}
\equiv
\braket{\osigma_{\xi}}
\equiv \frac{\Tr[\osigma_{\xi}\ee^{-\oHHa(\aa_r,\aa_i)/(\kB T)}]}{\Tr[\ee^{-\oHHa(\aa_r,\aa_i)/(\kB T)}]}.
\end{equation}
From Eqs.~\eqref{eq:alpha_sigma},  the expectation values of the large spin operators are expressed as
\begin{subequations}
\begin{align}
\braket{\oSigma_{x}}
& = \frac{N}{2}\braket{\osigma_x} = - \frac{N\omega_{\pi}}{2\rabi_x}\aa_r, \\
\braket{\oSigma_{z}}
& = \frac{N}{2}\braket{\osigma_z} = - \frac{N\omega_{\pi}}{2\rabi_z}\aa_i.
\end{align}
\end{subequations}
Substituting these into Eq.~\eqref{eq:oHHa}, we get
\begin{align} \label{eq:oHHa_aa} % !!!!!!!!!!!!!!!!!!!!!!!!!!!!!!!!!!!!!!!!!!!
\frac{\oHHa(\aa_r,\aa_i)}{\hbar}
& = \frac{\wEr}{2} \osigma_{x}
+ \left( 2\rabi_x - \frac{\zEr\JEr\omega_{\pi}}{\hbar\rabi_x} \right)\aa_r\osigma_x
\nonumber \\ & \quad
+ \left( 2\rabi_z + \frac{\zEr\JEr\omega_{\pi}}{\hbar\rabi_z} \right)\aa_i\osigma_z.
\end{align}
By simultaneously solving Eqs.~\eqref{eq:alpha_sigma}, \eqref{eq:sigmaa}, and \eqref{eq:oHHa_aa} for a given temperature $T$, we get the thermal-equilibrium values of the $\Erion$ spins $\sigmaa{}_{x,z}$ and $\Feion$ qAFM magnons $\aa_{r,i}$.  From Eqs.~\eqref{eq:vSa12}, \eqref{eq:delta_ovS_AB}, and \eqref{eq:oTY_a}, the thermal-equilibrium values of the $\Feion$ spins are obtained from those $\aa_{r,i}$ of qAFM magnons as
\begin{subequations} \label{eq:Sa_xyz} % !!!!!!!!!!!!!!!!!!!!!!!!!!!!!!!!!!!!!
\begin{align}
\Sa{}_x
& \equiv \braket{\oS_x^\mathrm{A}} = \braket{\oS_x^\mathrm{B}} \nonumber \\
& = S\sin\beta_0 + \sqrt{2xS} \cos\beta_0 \left( \frac{b+a}{d-c} \right)^{1/4} \aa_r, \\
\Sa{}_y
& \equiv \braket{\oS_y^\mathrm{A}} = - \braket{\oS_y^\mathrm{B}} \nonumber \\
& = - \sqrt{2xS} \left( \frac{d-c}{b+a} \right)^{1/4} \aa_i, \label{eq:Say} \\
\Sa{}_z
& \equiv - \braket{\oS_z^\mathrm{A}} = \braket{\oS_z^\mathrm{B}} \nonumber \\
& = S\cos\beta_0 - \sqrt{2xS} \sin\beta_0 \left( \frac{b+a}{d-c} \right)^{1/4} \aa_r.
\end{align}
\end{subequations}

\begin{figure}[tbp]
\centering
\includegraphics[width=\linewidth]{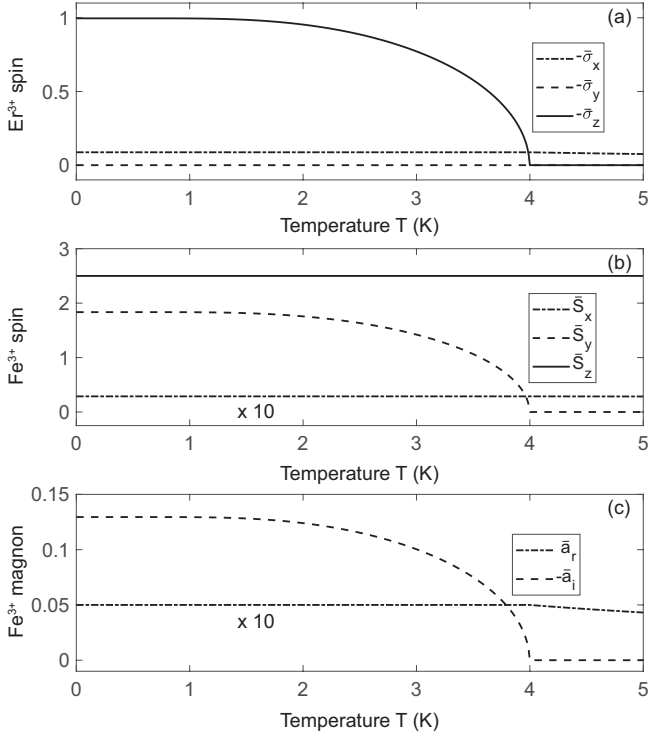}
\caption{Thermal-equilibrium values of (a)~$\Erion$ spins, (b)~$\Feion$ spins, and (c)~$\Feion$ magnon amplitudes are plotted as a function of temperature $T$.  They were calculated by the semiclassical method with the extended Dicke Hamiltonian in the case of zero external DC magnetic field. Figures~\ref{fig:RSa_T}(a) and (b) are almost the same as Figs.~\ref{fig:RS_T}(a) and (b), respectively, except $\Sa{}_z$, which is not largely changed due to bosonization.  The $\Feion$ spins, $\Sa{}_{x,y,z}$, were calculated by Eqs.~\eqref{eq:Sa_xyz} with the thermal-equilibrium value of qAFM magnon annihilation operator $\braket{\oa_{\pi}} = \sqrt{N}(\aa_r + \ii \aa_i)$ plotted in Fig.~\ref{fig:RSa_T}(c).
}
\label{fig:RSa_T}
\end{figure}

In Fig.~\ref{fig:RSa_T}, we plot the thermal-equilibrium values of (a)~$\Erion$ spins, $\sigmaa{}_{x,y,z}$, (b)~$\Feion$ spins, $\Sa{}_{x,y,z}$, and (c)~$\Feion$ qAFM magnons, $\aa_{r,i}$, as a function of temperature in the absence of an external DC magnetic field, $\vBext = \bm{0}$.  We can see that Figs.~\ref{fig:RSa_T}(a) and (b), respectively, well reproduce Figs.~\ref{fig:RS_T}(a) and (b) calculated by the mean-field method, including the critical temperature, $\Tc$, but except $\Sa{}_z$.  In Fig.~\ref{fig:RS_T}(b), $\Sa{}_z$ is seen to decrease, accompanied by the spontaneous appearance of $\Sa{}_y$, as the temperature decreases, while it is almost unchanged in Fig.~\ref{fig:RSa_T}(b).  This is because $\Sa{}_x{}^2 + \Sa{}_y{}^2 + \Sa{}_z{}^2 = S^2$ is no longer satisfied in the extended Dicke Hamiltonian derived through magnon quantization (i.e., bosonization of $\Feion$ spin modulations).  The ultrastrong $\rabi_z$ term causes the spontaneous appearance of $\sigmaa{}_z$ and $\aa_i$, as seen in Fig.~\ref{fig:RSa_T}(a) and (c), respectively, and the latter causes the nonzero $\Sa{}_y$ through Eq.~\eqref{eq:Say}.  The rotation of the $\Feion$ AFM vector occurs in $\ErFeO$ by the spontaneous appearance of nonzero $\Sa{}_y$ when $\Sa{}_x{}^2 + \Sa{}_y{}^2 + \Sa{}_z{}^2 = S^2$ holds.  This is the basic picture of the LTPT in terms of $\Erion$--magnon coupling.

As seen in Eqs.~\eqref{eq:rabi}, the transverse coupling strength, $\rabi_z$, depends on $D_x$, and the longitudinal coupling strength, $\rabi_x$, depends on $J$ and $D_y$.  As seen in Eq.~\eqref{eq:VFeEr}, the $D_x$ antisymmetric $\Erion$--$\Feion$ exchange interaction is essential for the LTPT, because it couples $\osigma^\mathrm{A/B}_z$ and $\oS^\mathrm{A/B}_y$, which appear spontaneously at $T < \Tc$. In contrast, the $J$ and $D_y$ exchange interactions are not directly related to the LTPT.
% because it is characterized as antiferromagnetic ordering of $\Erion$ spins $\vsigma{\text{A/B}}$ along the $c$ axis and the rotation of the $\Feion$ AFM vector ($b$ components of $\Feion$ spins $S^{\text{A/B}}_y$).

In this way,  we can quantitatively reproduce the LTPT as the SRPT in  the extended Dicke Hamiltonian, Eq.~\eqref{eq:extended_Dicke}, which was derived from the spin model of $\ErFeO$.  The essential terms are extracted in Eq.~\eqref{eq:oHH_pi}.  The $\rabi_z$ term (antisymmetric $\Erion$--$\Feion$ exchange interaction with $D_x$) corresponds to the matter--photon coupling and causes the antiferromagnetic ordering of $\Erion$ spins along the $c$ axis and the $b$ component of the $\Feion$ spins through the spontaneous appearance of qAFM magnons.

\subsection{$\Erion$--magnon coupling contribution} \label{sec:quantum}
Although the $\rabi_z$ term causes the spontaneous appearance of both $\sigmaa{}_z$ and $\Sa{}_y$ following the picture of the SRPT, a nonzero $\sigmaa{}_z$ can spontaneously appear also by the $\JEr$ term ($\Erion$--$\Erion$ exchange interactions). While the $\Erion$--magnon coupling is inevitable for the spontaneous rotation of $\Feion$ AFM vector (spontaneous appearance of $\Sa{}_y$), we try to evaluate quantitatively the contributions of the $\Erion$--magnon coupling and $\Erion$--$\Erion$ exchange interactions for the LTPT in this subsection.

\begin{figure}[tbp]
\centering
\includegraphics[width=\linewidth]{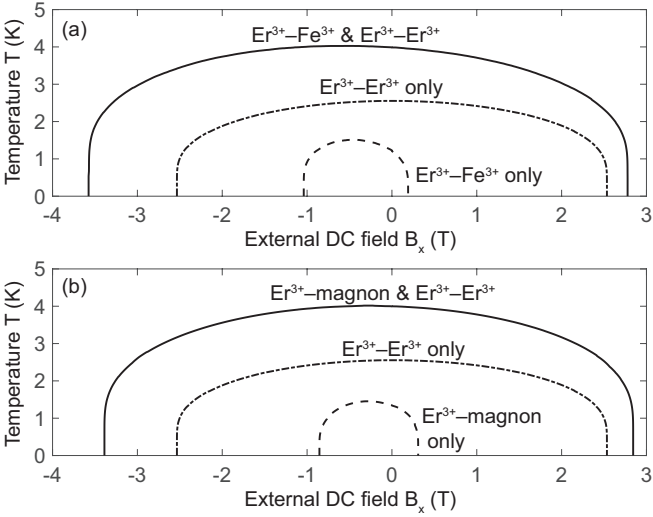}
\caption{Phase boundaries of the LTPT in $\ErFeO$ calculated by (a)~the mean-field method and (c)~the semiclassical method with the extended Dicke Hamiltonian.  An external DC magnetic field is applied along the $a$ axis.  The solid curves are the phase boundaries by the full Hamiltonian, and those in Fig.~\ref{fig:phase_boundary}(a) and \ref{fig:phase_diagram}(a) are equivalent.  The dash-dotted curves are the phase boundaries in the absence of $\Erion$--magnon coupling ($\Erion$--$\Feion$ exchange interactions).  The dashed curves are those obtained in the absence of $\Erion$--$\Erion$ exchange interactions, i.e., the LTPT can be caused solely by the $\Erion$--magnon coupling and thus can be interpreted as a magnonic SRPT.
}
\label{fig:phase_boundary}
\end{figure}
In Fig.~\ref{fig:phase_boundary}, we plot the phase boundaries calculated by the full Hamiltonian (solid lines), in the absence of $\Erion$--$\Feion$ exchange interactions (dash-dotted line; $J = D_x = D_y = \rabi_z = \rabi_x = 0$), and in the absence of $\Erion$--$\Erion$ exchange interactions (dashed line; $\JEr = 0$).  Figures~\ref{fig:phase_boundary}(a) and (b) show results by the mean-field method and by the semiclassical method with the extended Dicke Hamiltonian, respectively.  The solid curve in Fig.~\ref{fig:phase_boundary}(a) is equal to that in Fig.~\ref{fig:phase_diagram}(a).  The small differences between Figs.~\ref{fig:phase_boundary}(a) and (b) are discussed in Appendix \ref{app:boundary}.

As shown by the dashed lines ($\JEr = 0$), the phase transition occurs even in the absence of $\Erion$--$\Erion$ exchange interactions, and the critical temperature $\Tc\sim1.2\;\mathrm{K}$ at $\vBext=\bm{0}$.  This means that the $\Erion$--magnon coupling alone can cause the LTPT.  In this sense, the LTPT can be interpreted as a magnonic SRPT, because the $\Erion$--magnon coupling is strong enough for the phase transition to occur.

On the other hand, in the absence of $\Erion$--magnon coupling, as shown by dash-dotted lines, the critical temperature $\Tc\sim2.6\;\mathrm{K}$ at $\vBext=\bm{0}$.  This result appears to indicate that the contribution of the $\Erion$--$\Erion$ exchange interactions is larger than that of the $\Erion$--magnon coupling.  However, the real critical temperature $\Tc \sim 4\;\mathrm{K}$, meaning that the $\Erion$--magnon coupling enhances the critical temperature of the phase transition.  In the same manner, the critical magnetic field is also enhanced.  These facts are similar to the suggestion of $\Tc$ enhancement through photon--matter coupling by G.~Mazza and A.~Georges  \cite{Mazza2018}, while their phase transition does not occur solely by the photon--matter coupling and their model does not guarantee gauge invariance  \cite{Andolina2019,Nataf2019a}.

In order to quantitatively evaluate their contributions to the LTPT more in detail, we derive the condition for the SRPT in our extended Dicke Hamiltonian, Eq.~\eqref{eq:oHH_pi}, by using the Holstein--Primakoff transformation \cite{Holstein1940,Emary2003PRL,Emary2003PRE}.

We rewrite $\oSigma_{x,y,z}$ by the bosonic annihilation (creation) operator $\ob$ ($\obd$) as
\begin{subequations}
\begin{align}
\oSigma_x & \to \obd\ob - \frac{N}{2}, \\
\oSigma_y & \to \frac{\obd(N-\obd\ob)^{1/2}+(N-\obd\ob)^{1/2}\ob}{2}, \\
\oSigma_z & \to \frac{\obd(N-\obd\ob)^{1/2} - (N-\obd\ob)^{1/2}\ob}{\ii2}.
\end{align}
\end{subequations}
%Then, the Hamiltonian in Eq.~\eqref{eq:oHH_pi} is transformed as
%\begin{align}
%{\oHH}/{\hbar}
%& \to \omega_{\pi}\oad_{\pi}\oa_{\pi} + \wEr\obd\ob
%  + \frac{2\zEr\JEr}{N\hbar} \left[ \left(\obd\ob - \frac{N}{2}\right)^2
%\right. \nonumber \\ & \left.
%- \left( \frac{\obd(N-\obd\ob)^{1/2} - (N-\obd\ob)^{1/2}\ob}{\ii2} \right)^2 \right]
%\nonumber \\ & \quad
%+ \frac{2\rabi_x}{\sqrt{N}}(\oad_\pi+\oa_\pi)\left( \obd\ob - \frac{N}{2} \right)
%+ \frac{\ii2\rabi_z}{\sqrt{N}}(\oad_\pi-\oa_\pi)
%\nonumber \\ & \quad \times
%  \left( \frac{\obd(N-\obd\ob)^{1/2} - (N-\obd\ob)^{1/2}\ob}{\ii2} \right)
%+ \const
%\end{align}
Further, we replace all the operators by c-numbers $\aa_r, \aa_i, \ba \in \mathbb{R}$ as
\begin{subequations}
\begin{align}
\oa & \to \sqrt{N}(\aa_r + \ii \aa_i), \\
\ob & \to \ii \sqrt{N} \ba.
\end{align}
\end{subequations}
Then, the Hamiltonian in Eq.~\eqref{eq:oHH_pi} is transformed to
\begin{align}
\frac{\oHH}{N\hbar}
& \to \omega_{\pi}(\aa_r{}^2+\aa_i{}^2) + \wEr\ba^2
  + \frac{4\zEr\JEr}{\hbar} \ba^2 (\ba^2-1)
\nonumber \\ & \quad
+ 2\rabi_x\aa_r(2\ba^2-1)
- 4\rabi_z\aa_i\ba\sqrt{1-\ba{}^2}
+ \const
\end{align}
The ground state of the system should satisfy
\begin{subequations}
\begin{align}
\frac{1}{2}\frac{1}{\hbar N}\frac{\partial\HH}{\partial\aa_r}
& = \omega_\pi\aa_r + \rabi_x(2\ba{}^2-1) = 0, \\
\frac{1}{2}\frac{1}{\hbar N}\frac{\partial\HH}{\partial\aa_i}
& = \omega_\pi\aa_i - 2\rabi_z\ba\sqrt{1-\ba{}^2} = 0, \\
\frac{1}{2}\frac{1}{\hbar N}\frac{\partial\HH}{\partial\ba}
& = \wEr\ba
    + \frac{4\zEr\JEr}{\hbar}\ba(2\ba{}^2-1)
\nonumber \\ & \quad
    + 4\rabi_x\aa_r\ba
    - 2\rabi_z\aa_i\frac{1-2\ba{}^2}{\sqrt{1-\ba{}^2}} = 0. \label{eq:dH/db}
\end{align}
\end{subequations}
Solving the first two equations, we can express the $\Feion$ qAFM magnon amplitudes as
\begin{subequations}
\begin{align}
\aa_r & = - \frac{\rabi_x}{\omega_\pi}(2\ba{}^2-1), \\
\aa_i & = \frac{2\rabi_z}{\omega_\pi}\ba\sqrt{1-\ba{}^2}.
\end{align}
\end{subequations}
Substituting these into Eq.~\eqref{eq:dH/db}, we get an equation for the $\Erion$ amplitude as
\begin{align} \label{eq:HP_bi} % !!!!!!!!!!!!!!!!!!!!!!!!!!!!!!!!!!!!!!!!!
& \left[
      \wEr - \frac{4\rabi_z{}^2-4\rabi_x{}^2}{\omega_\pi} - \frac{4\zEr\JEr}{\hbar}
\right. \nonumber \\ & \quad \left.
    + \left( \frac{8\rabi_z{}^2-8\rabi_x{}^2}{\omega_\pi} +  \frac{8\zEr\JEr}{\hbar} \right) \ba{}^2
    \right] \ba = 0.
\end{align}
For a real nonzero value of $\ba$ to exist, the parameters must satisfy
\begin{equation} \label{eq:SRPT_extended} % !!!!!!!!!!!!!!!!!!!!!!!!!!!!!!!!!!
\frac{4\rabi_z{}^2}{\omega_\pi\wEr}
- \frac{4\rabi_x{}^2}{\omega_\pi\wEr}
+ \frac{4\zEr\JEr}{\hbar\wEr} > 1.
\end{equation}
For $\JEr = \rabi_x = 0$, this condition is reduced to $4\rabi_z{}^2>\omega_{\pi}\wEr$ for the SRPT in the Dicke model, Eq.~\eqref{eq:Dicke}.

The three terms on the left-hand side of Eq.~\eqref{eq:SRPT_extended} are evaluated as
% coupling criteria, coupling depth, coherence, interaction, mutual
%
\begin{subequations}
\begin{align}
D_{\rabi_z} & \equiv {4\rabi_z{}^2}/({\omega_\pi\wEr}) = 2.65, \\
D_{\rabi_x} & \equiv - {4\rabi_x{}^2}/({\omega_\pi\wEr}) = -0.51, \\
D_{\JEr} & \equiv {4\zEr\JEr}/({\hbar\wEr}) = 9.29.
\end{align}
\end{subequations}
In the following, we call them {\it coupling depths}.
They are dimensionless measures of coupling strengths
and are definitely determined based on the appearance of the SRPT.
As seen in Eq.~\eqref{eq:SRPT_extended},
the SRPT occurs when the sum of these coupling depths is greater than unity:
$D_{\rabi_z} + D_{\rabi_x} + D_{\JEr} > 1$.
The coupling depth $D_{\JEr}$ of the $\JEr$ term is the largest, which is consistent with Fig.~\ref{fig:phase_boundary}.  The $\rabi_x$ term (longitudinal coupling) gives a negative contribution for the SRPT ($D_{\rabi_x}<0$).  Among the three couplings, the contribution of the $\rabi_z$ term is $D_{\rabi_z} / (D_{\rabi_z} + D_{\rabi_x} + D_{\JEr}) = 0.23$, and the contribution of the total $\Erion$--magnon coupling is $(D_{\rabi_z}+D_{\rabi_x}) / (D_{\rabi_z} + D_{\rabi_x} + D_{\JEr}) = 0.19$.  These values are roughly equal to $1.3\;\mathrm{K} / (1.3\;\mathrm{K}+3.4\;\mathrm{K}) = 0.28$ estimated by Kadomtseva, Krynetskii, and Matveev  \cite{Kadomtseva1980}, while they did not consider the longitudinal coupling ($\rabi_x$ term), which is not included in the cooperative Jahn--Teller model  \cite{Gehring1975,Kugel1982,Loos1984,Larson2008PRA}, and the parameters were determined only by the phase boundary for $\vBext//a$.

From the viewpoint of the analogy between the two phase transitions, a remarkable fact is that the coupling depth of the $\rabi_z$ term satisfies $D_{\rabi_z} > 1$ and $D_{\rabi_z}+D_{\rabi_x} > 1$.  This suggests that the transverse $\Erion$--magnon coupling is much stronger than the longitudinal one (giving the negative contribution) and ultrastrong enough to cause the SRPT solely.  Also in this sense, we can conclude that the LTPT in $\ErFeO$ is the magnonic SRPT obtained in the extended Dicke Hamiltonian with the direct atom--atom interaction and the longitudinal coupling ($\rabi_x$ term).

\section{Summary} \label{sec:summary}
From a spin model of $\ErFeO$ that reproduces both the phase diagrams \cite{Zhang2019} and terahertz spectra  \cite{Li2018a}, we derived an extended Dicke model that takes into account $\Erion$--$\Erion$ exchange interactions as well as the cooperative coupling between $\Erion$ spins and $\Feion$ magnon modes.  We found that the LTPT in $\ErFeO$ can be caused solely by the $\Erion$--magnon coupling (in the absence of $\Erion$--$\Erion$ exchange interactions), which demonstrates that the LTPT is a magnonic SRPT in the extended Dicke model.

In the thermodynamic limit, $N\to\infty$, the Dicke model is effectively interpreted as an infinite dimensional system \cite{Larson2017JPA}, because the atoms interact equivalently with each other through the coupling with a single photonic mode. Such a dimensionality is reflected in critical exponents \cite{Larson2017JPA,Shapiro2019} at phase transitions and would differentiate the LTPT in $\ErFeO$ from standard magnetic phase transitions caused by short-range (nearest neighbor, next-nearest-neighbor, \ldots) exchange interactions between spins. Further, the coexistence of the direct (short-range) $\Erion$--$\Erion$ interactions and $\Erion$--magnon couplings (long-range retarded $\Erion$--$\Erion$ interactions) in $\ErFeO$ can lead to rich physics beyond what the normal Dicke model provides. 

The thermal SRPT in $\ErFeO$ would also give us rich physics compared with the quantum or zero-temperature SRPT that has been demonstrated by laser-driven cold atoms \cite{Baumann2010N,Kirton2018a}. In particular, it is known that the thermal and quantum fluctuations of photons and atoms show characteristic behaviors around the SRPT \cite{Shapiro2019}. It is also known that the ground state of an ultrastrongly coupled system is a quantum squeezed vacuum even in the normal phase \cite{Artoni1991,Artoni1989,Schwendimann1992,Schwendimann1992a,quattropani05,Ciuti2005PRB}, and strong two-mode squeezing at the SRPT has been demonstrated numerically \cite{Makihara2020}. Our on-going terahertz magnetospectroscopy experiments of $\ErYFeO$ around the LTPT \cite{Peraca2020} will experimentally examine such characteristic quantum squeezing at the thermal and quantum SRPTs.

% In addition to the non-equilibrium SRPT in cold atoms \cite{Baumann2010N,Kirton2018a}, the thermal SRPT can now be simulated experimentally in $\ErFeO$, 

% Its validity can now be experimentally verified by $\ErFeO$. Especially, critical exponents at the LTPT in $\ErFeO$ will give us rich information about those at the SRPT, because any thermal SRPT have never been observed experimentally, while non-equilibrium one has been demonstrated in cold atom systems 

% In contrast to , the LTPT in $\ErFeO$ is caused also by the long-range $\Erion$--$\Erion$ interactions mediated by the $\Feion$ qAFM magnons.

% The ultrastrong $\Erion$--magnon coupling and the resultant magnonic SRPT will enable us to realize strong intrinsic quantum squeezing of magnons  \cite{Makihara2020}, which can be controlled by the temperature and external DC magnetic field applied to $\ErFeO$.

% While the SRPT has not been observed experimentally in thermal equilibrium since its first proposal in 1973  \cite{Hepp1973AP}, the spins in $\ErFeO$ can be used as an experimental quantum simulator for the SRPT in thermal equilibrium.

\begin{acknowledgments}
This research was supported by JST PRESTO program (grant JPMJPR1767),
National Science Foundation (Cooperative Agreement DMR-1720595),
and U.S. Army Research Office (grant W911NF-17-1-0259).
We thank
Andrey Baydin,
Kenji Hayashida,
Chien-Lung Huang,
Takuma Makihara,
Atsushi Miyake,
Atsuhiko Miyata, and
Fuyang Tay for fruitful discussion.
\end{acknowledgments}

\appendix
\section{Mean-field Calculation} \label{app:meanfield}
Since we simply consider an homogeneous external DC magnetic flux density $\vBext$
in this paper,
the expectation values of $\Erion$ spins
$\vsigma{\text{A/B}} \equiv \braket{\ovsigma^{\text{A/B}}_i}$
and $\Feion$ spins $\vS{\text{A/B}} \equiv \braket{\ovS^{\text{A/B}}_i}$
are independent of the site index $i$.
The bracket represents theoretically the expectation values of operators
at finite temperature in the Heisenberg picture.
It also corresponds to the ensemble average of the spins in each sublattice.
Their equations of motion are obtained from the Heisenberg equations 
derived by the Hamiltonian in Eq.~\eqref{eq:HH_ErYFeO3} as
($s=\text{A,B}$)
\begin{subequations} \label{eq:precession_MF} % !!!!!!!!!!!!!!!!!!!!!!!!!!!!!!
\begin{align}
\hbar\ddtin\vsigma{s}
& = -\vsigma{s} \times \gf\muB\vBMFEr{s}(\{\vsigma{\text{A/B}}\},\{\vS{\text{A/B}}\}),
\label{eq:Sxgrad} \\ % !!!!!!!!!!!!!!!!!!!!!!!!!!!!!!!!!!!!!!!!!!!!!!!!
\hbar\ddtin\vS{s}
& = - \vS{s} \times \gf\muB\vBMFFe{s}(\{\vsigma{\text{A/B}}\},\{\vS{\text{A/B}}\}).
\label{eq:RzRx} % !!!!!!!!!!!!!!!!!!!!!!!!!!!!!!!!!!!!!!!!!!
\end{align}
\end{subequations}
Here, $\vBMFEr{\text{A/B}}$ and $\vBMFFe{\text{A/B}}$ 
are the mean-fields for $\Erion$ and $\Feion$ spins, respectively,
and they are expressed as
\begin{widetext}
\begin{subequations}\label{eq:meanfields} % !!!!!!!!!!!!!!!!!!!!!!!!!!!!!!!!!
\begin{align}
\gf\muB\vBMFEr{\text{A}}(\{\vsigma{\text{A/B}}\},\{\vS{\text{A/B}}\})
& = \muB\mgfEr\cdot\vBext
    + 2\zEr\JEr \vsigma{\text{B}}
    + \sum_{s=\text{A,B}} 2\begin{pmatrix}
      J \SS{s}{x} - (\vD{\text{A},s}\times\vS{s})_x \\
      0 \\
      J \SS{s}{z} - (\vD{\text{A},s}\times\vS{s})_z
    \end{pmatrix}, \label{eq:MF1} \\ % !!!!!!!!!!!!!!!!!!!!!!!!!!!!!!!!!!!!!!!!
\gf\muB\vBMFEr{\text{B}}(\{\vsigma{\text{A/B}}\},\{\vS{\text{A/B}}\})
& = \muB\mgfEr\cdot\vBext
    + 2\zEr\JEr \vsigma{\text{A}}
    + \sum_{s=\text{A,B}} 2\begin{pmatrix}
      J \SS{s}{x} - (\vD{\text{B},s}\times\vS{s})_x \\
      0 \\
      J \SS{s}{z} - (\vD{\text{B},s}\times\vS{s})_z
    \end{pmatrix}, \label{eq:MF2} % !!!!!!!!!!!!!!!!!!!!!!!!!!!!!!!!!!!!!!!!
\end{align}
\begin{align}
\gf\muB\vBMFFe{\text{A}}(\{\vsigma{\text{A/B}}\},\{\vS{\text{A/B}}\})
& = \muB\mgfFe\cdot\vBext
+ \sum_{s=\text{A,B}} x \left( J \vsigma{s} + \vD{s,\text{A}} \times \vsigma{s} \right)
\nonumber \\ & \quad
+ \begin{pmatrix}
\zz\JFe \SS{\text{B}}{x} + \zz\DFe_y \SS{\text{B}}{z} - 2\Ax \SS{\text{A}}{x} - \Axz \SS{\text{A}}{z} \\
\zz\JFe \SS{\text{B}}{y} \\
\zz\JFe \SS{\text{B}}{z} - \zz\DFe_y \SS{\text{B}}{x} - 2\Az \SS{\text{A}}{z} - \Axz \SS{\text{A}}{x}
\end{pmatrix}, \label{eq:MF3} \\ % !!!!!!!!!!!!!!!!!!!!!!!!!!!!!!!!!!!!!!!!
\gf\muB\vBMFFe{\text{B}}(\{\vsigma{\text{A/B}}\},\{\vS{\text{A/B}}\})
& = \muB\mgfFe\cdot\vBext
+ \sum_{s=\text{A,B}} x \left( J \vsigma{s} + \vD{s,\text{B}} \times \vsigma{s} \right)
\nonumber \\ & \quad
+ \begin{pmatrix}
\zz\JFe \SS{\text{A}}{x} - \zz\DFe_y \SS{\text{A}}{z} - 2\Ax \SS{\text{B}}{x} + \Axz \SS{\text{B}}{z} \\
\zz\JFe \SS{\text{A}}{y} \\
\zz\JFe \SS{\text{A}}{z} + \zz\DFe_y \SS{\text{A}}{x} - 2\Az \SS{\text{B}}{z} + \Axz \SS{\text{B}}{x}
\end{pmatrix}. \label{eq:MF4} % !!!!!!!!!!!!!!!!!!!!!!!!!!!!!!!!!!!!!!!!
\end{align}
\end{subequations}
\end{widetext}
In Eqs.~\eqref{eq:MF1} and \eqref{eq:MF2}, the first, second, and third terms
represent the Zeeman effect, \ErEr{} exchange interaction,
and \ErFe{} exchange interaction, respectively. 
In Eqs.~\eqref{eq:MF3} and \eqref{eq:MF4}, the first, second, and third terms
represent the Zeeman effect, \ErFe{} exchange interaction, and \FeFe{} exchange interaction,
respectively.
The dilution of $\Erion$ spins is reflected through
the factors $\zEr = 6x$ and $x$,
i.e., the number of neighboring $\Erion$ is effectively decreased by factor $x$.
Since $(1/2)\ovsigma^{\text{A/B}}$ corresponds to the spin-$\half$ operator,
the factor 2 appears overall in Eqs.~\eqref{eq:MF1} and \eqref{eq:MF2}.
As explained at the end of Sec.~\ref{sec:spin_model}, the $y$ component of the third term
in Eqs.~\eqref{eq:MF1} and \eqref{eq:MF2}
is set to be zero by implicitly considering a high energy potential.

The free energy of the system is minimized
when the thermal-equilibrium values (time-averages) of % the expectation values of 
spins
$\vsigmaa{\text{A/B}}$ and $\vSa{\text{A/B}}$
are parallel to their mean-fields
$\vBMFEra{s} \equiv \vBMFEr{s}(\{\vsigmaa{\text{A/B}}\},\{\vSa{\text{A/B}}\})$
and $\vBMFFea{s} \equiv \vBMFFe{s}(\{\vsigmaa{\text{A/B}}\},\{\vSa{\text{A/B}}\})$
as
\begin{subequations} \label{eq:SzRz} % !!!!!!!!!!!!!!!!!!!!!!!!!!!!!!!!!!!!!!!
\begin{align}
\vsigmaa{s} = \braket{\ovsigma^{s}} & = \braket{\osigma^{s}_{\parallel}}\vuEr{s}, \quad
\osigma^s_{\parallel} \equiv \ovsigma^s\cdot\vuEr{s}, \\
\vSa{s} = \braket{\ovS^s} & = \braket{\oS^s_{\parallel}}\vuFe{s}, \quad
\oS^s_{\parallel} \equiv \ovS^s\cdot\vuFe{s},
\end{align}
\end{subequations}
where we defined unit vectors of the mean-fields as
\begin{subequations}
\begin{align}
\vuEr{s} & \equiv \vBMFEra{s} / |\vBMFEra{s}|, \\
\vuFe{s} & \equiv \vBMFFea{s} / |\vBMFFea{s}|.
\end{align}
\end{subequations}
The thermal-equilibrium values $\vsigmaa{\text{A/B}}$ and $\vSa{\text{A/B}}$ are determined as follows.
For given mean-fields $\vBMFFea{s}$ and $\vBMFEra{s}$,
effective Hamiltonians of each $\Erion$ and $\Feion$ can be defined,
respectively, as
\begin{subequations} \label{eq:oHHMFFeEr} % !!!!!!!!!!!!!!!!!!!!!!!!!!!!!!!!!!
\begin{align}
\oHHMFEr{s}
& = \half\gf\muB\ovsigma^{s}\cdot\vBMFEra{s}
  = \half\gf\muB\osigma^s_{\parallel}|\vBMFEra{s}|, \\
\oHHMFFe{s}
& = \gf\muB\ovS^{s}\cdot\vBMFFea{s}
  = \gf\muB\oS^{s}_{\parallel}|\vBMFFea{s}|.
\end{align}
\end{subequations}
Then, the partition functions are expressed as
\begin{subequations}
\begin{align}
\ZEr{s} & \equiv  \Tr\left[\ee^{-\oHHMFEr{s}/(\kB T)}\right]
= \sum_{m=\pm1} \ee^{-my_s}
\nonumber \\ &
= 2\cosh(y_s), \\
\ZFe{s} & \equiv \Tr\left[\ee^{-\oHHMFFe{s}/(\kB T)}\right]
= \sum_{m=-S}^S \ee^{-mx_s}
\nonumber \\ &
= \frac{\sinh[(S+1/2)x_j]}{\sinh(x_s/2)},
\end{align}
\end{subequations}
where we defined
\begin{subequations}
\begin{align}
y_s & \equiv \gf\muB|\vBMFEra{s}|/(2\kB T), \\
x_s & \equiv \gf\muB|\vBMFFea{s}|/(\kB T).
\end{align}
\end{subequations}
Since $\ovsigma^{\text{A/B}}$ is not a standard spin operator with
an angular momentum of $\hbar$ or $\hbar/2$
but is a vector of the Pauli operators,
the summation is performed for $m = \pm1$.
The free energies are given as
$-\kB T\ln\ZEr{\text{A/B}}$ and $-\kB T\ln\ZFe{\text{A/B}}$,
and the thermal-equilibrium values of the spins are obtained as
\begin{subequations} \label{eq:<oSz><oRz>} % !!!!!!!!!!!!!!!!!!!!!!!!!!!!!!!!!
\begin{align}
\braket{\osigma^s_{\parallel}}
& = - \frac{\partial}{\partial y_s} \ln \ZEr{s}
= -\tanh(y_s), \\
\braket{\oS^s_{\parallel}}
& = - \frac{\partial}{\partial x_s} \ln \ZFe{s}
  = - S B_S(Sx_s),
\end{align}
\end{subequations}
where $B_S(z)$ is the Brillouin function defined as
\begin{equation}
B_J(z) \equiv \frac{2J+1}{2J}\coth\left(\frac{2J+1}{2J}z\right)
- \frac{1}{2J}\coth\left(\frac{z}{2J}\right).
\end{equation}
By consistently solving Eqs.~\eqref{eq:meanfields}, \eqref{eq:SzRz},
and \eqref{eq:<oSz><oRz>},
we can determine $\vsigmaa{\text{A/B}}$ and $\vSa{\text{A/B}}$ at finite temperatures.

\section{Reduction of number of parameters} \label{app:symmetry}
In this appendix, we reduce the number of parameters in our spin model
by considering the spin configuration in the $\Gamma_{12}$ phase of $\ErFeO$
when the external DC magnetic field is zero or along the $a$ axis.
In the ground state ($T=0$), 
the equilibrium values of the spins satisfies Eqs.~\eqref{eq:precession_MF}
with $(\partial/\partial t)\vR{\text{A/B}} = 0$ and $(\partial/\partial t)\vS{\text{A/B}} = 0$.
Here, as depicted in Fig.~\ref{fig:ErFeO3_4.5K},
due to the $\pi$-rotational symmetry about the $a$ axis,
we represent the four spins $\vR{\text{A/B}}$ and $\vS{\text{A/B}}$
(twelve elements) by six values as
\begin{subequations} \label{eq:RSxyz} % !!!!!!!!!!!!!!!!!!!!!!!!!!!!!!!!!!!!!!
\begin{align}
\RR{\text{A}}{x} & = \RR{\text{B}}{x} \equiv \RRR_x, \\
\RR{\text{A}}{y} & = - \RR{\text{B}}{y} = \RRR_y, \\
\RR{\text{A}}{z} & = - \RR{\text{B}}{z} \equiv \RRR_z,
\end{align}
\begin{align}
\SS{\text{A}}{x} & = \SS{\text{B}}{x} \equiv \SSS_x, \\
\SS{\text{A}}{y} & = - \SS{\text{B}}{y} \equiv \SSS_y, \\
- \SS{\text{A}}{z} & = \SS{\text{B}}{z} \equiv \SSS_z.
\end{align}
\end{subequations}
Using these and Eqs.~\eqref{eq:precession_MF} and \eqref{eq:meanfields}, we get
\begin{widetext}
\begin{subequations} \label{eq:RRSS} % !!!!!!!!!!!!!!!!!!!!!!!!!!!!!!!!!!!!!!!
\begin{align}
\begin{pmatrix} \RRR_x \\ \RRR_y \\ \RRR_z \end{pmatrix} \times
    \left[
      \begin{pmatrix} \gfEr_x\muB\Bext \\ 0 \\ 0 \end{pmatrix}
    + 2\zEr\JEr \begin{pmatrix} \RRR_x \\ - \RRR_y \\ - \RRR_z \end{pmatrix}
    + 4 \begin{pmatrix}
      J_{\text{A},+} \SSS_x + D_{\text{A},-,y} \SSS_z + D_{\text{A},-,z} \SSS_y \\
      J_{\text{A},-} \SSS_y - D_{\text{A},+,z} \SSS_x - D_{\text{A},-,x} \SSS_z \\
    - J_{\text{A},-} \SSS_z - D_{\text{A},-,x} \SSS_y + D_{\text{A},+,y} \SSS_x
    \end{pmatrix} \right]
& = \bm{0}, \label{eq:R1} \\ % !!!!!!!!!!!!!!!!!!!!!!!!!!!!!!!!!!!!!!!!!!!!!!!!!!!!!
\begin{pmatrix} \RRR_x \\ - \RRR_y \\ - \RRR_z \end{pmatrix} \times
    \left[
      \begin{pmatrix} \gfEr_x\muB\Bext \\ 0 \\ 0 \end{pmatrix}
    + 2\zEr\JEr \begin{pmatrix} \RRR_x \\ \RRR_y \\ \RRR_z \end{pmatrix}
    + 4 \begin{pmatrix}
      J_{\text{B},+} \SSS_x + D_{\text{B},-,y} \SSS_z + D_{\text{B},-,z} \SSS_y \\
      J_{\text{B},-} \SSS_y - D_{\text{B},+,z} \SSS_x - D_{\text{B},-,x} \SSS_z \\
    - J_{\text{B},-} \SSS_z - D_{\text{B},-,x} \SSS_y + D_{\text{B},+,y} \SSS_x
    \end{pmatrix} \right]
& = \bm{0}, \label{eq:R2} % !!!!!!!!!!!!!!!!!!!!!!!!!!!!!!!!!!!!!!!!!!!!!!!!!!!!!
\end{align}
\begin{align}
\begin{pmatrix} \SSS_x \\ \SSS_y \\ - \SSS_z \end{pmatrix} \times \left\{
      \begin{pmatrix}\gfFe_x\muB\Bext\\0\\0\end{pmatrix}
    + 2x \begin{pmatrix}
        J_{+,\text{A}} \RRR_x + D_{-,\text{A},y} \RRR_z - D_{-,\text{A},z} \RRR_y\\
        J_{-,\text{A}} \RRR_y + D_{+,\text{A},z} \RRR_x - D_{-,\text{A},x} \RRR_z \\
        J_{-,\text{A}} \RRR_z + D_{-,\text{A},x} \RRR_y - D_{+,\text{A},y} \RRR_x
      \end{pmatrix}
\right. \nonumber \quad \\ \left.
    + \begin{bmatrix}
        (\zz\JFe-2\Ax)\SSS_{x} + (\zz\DFe_z-\Axy)\SSS_y + (\zz\DFe_y+\Axz)\SSS_{z} \\
        (\zz\DFe_z-\Axy)\SSS_x - (\zz\JFe+2\Ay)\SSS_{y} \\
       -(\zz\DFe_y+\Axz)\SSS_{x} + (\zz\JFe+2\Az)\SSS_{z}
      \end{bmatrix} \right\}
& = \bm{0}, \label{eq:S1} \\ % !!!!!!!!!!!!!!!!!!!!!!!!!!!!!!!!!!!!!!!!!!!!!!!!!!!!!
\begin{pmatrix} \SSS_x \\ - \SSS_y \\ \SSS_z \end{pmatrix} \times \left\{
      \begin{pmatrix}\gfFe_x\muB\Bext\\0\\0\end{pmatrix}
    + 2x \begin{pmatrix}
        J_{+,\text{B}} \RRR_x + D_{-,\text{B},y} \RRR_z - D_{-,\text{B},z} \RRR_y \\
        J_{-,\text{B}} \RRR_y + D_{+,\text{B},z} \RRR_x - D_{-,\text{B},x} \RRR_z \\
        J_{-,\text{B}} \RRR_z + D_{-,\text{B},x} \RRR_y - D_{+,\text{B},y} \RRR_x
      \end{pmatrix}
\right. \nonumber \quad \\ \left.
    + \begin{bmatrix}
        (\zz\JFe-2\Ax)\SSS_{x} + (\zz\DFe_z-\Axy)\SSS_y + (\zz\DFe_y+\Axz)\SSS_{z} \\
       -(\zz\DFe_z-\Axy)\SSS_x + (\zz\JFe+2\Ay)\SSS_{y} \\
        (\zz\DFe_y+\Axz)\SSS_{x} - (\zz\JFe+2\Az)\SSS_{z}
      \end{bmatrix} \right\}
& = \bm{0}, \label{eq:S2} % !!!!!!!!!!!!!!!!!!!!!!!!!!!!!!!!!!!!!!!!!!!!!!!!!!!!!
\end{align}
\end{subequations}
where we defined
\begin{subequations}
\begin{align}
J_{s,\pm} & \equiv (J_{s,\text{A}} \pm J_{s,\text{B}} ) / 2, \\
J_{\pm,s} & \equiv (J_{\text{A},s} \pm J_{\text{B},s} ) / 2, \\
\vD{s,\pm} & \equiv (\vD{s,\text{A}} \pm \vD{s,\text{B}} ) / 2, \\
\vD{\pm,s} & \equiv (\vD{\text{A},s} \pm \vD{\text{B},s} ) / 2.
\end{align}
\end{subequations}
For the equivalence between Eq.~\eqref{eq:S1} and Eq.~\eqref{eq:S2}, the following equations should be satisfied for any $\RRR_{x,y,z}$:
\begin{subequations}
\begin{align}
  J_{+,\text{A}} \RRR_x + D_{-,\text{A},y} \RRR_z - D_{-,\text{A},z} \RRR_y & = 
  J_{+,\text{B}} \RRR_x + D_{-,\text{B},y} \RRR_z - D_{-,\text{B},z} \RRR_y, \\
  J_{-,\text{A}} \RRR_y + D_{+,\text{A},z} \RRR_x - D_{-,\text{A},x} \RRR_z & = 
- J_{-,\text{B}} \RRR_y - D_{+,\text{B},z} \RRR_x + D_{-,\text{B},x} \RRR_z, \\
  J_{-,\text{A}} \RRR_z + D_{-,\text{A},x} \RRR_y - D_{+,\text{A},y} \RRR_x & =
- J_{-,\text{B}} \RRR_z - D_{-,\text{B},x} \RRR_y + D_{+,\text{B},y} \RRR_x.
\end{align}
\end{subequations}
Then, we get the following relations:
\begin{subequations} \label{eq:S1S2} % !!!!!!!!!!!!!!!!!!!!!!!!!!!!!!!!!!!!!!!!
\begin{align}
J_{+,\text{A}} & = J_{+,\text{B}}, \\
J_{-,\text{A}} & = - J_{-,\text{B}}, \\
D_{-,\text{A},x} & = - D_{-,\text{B},x}, \\
D_{+,\text{A},y} & = - D_{+,\text{B},y}, \\
D_{-,\text{A},y} & = D_{-,\text{B},y}, \\
D_{+,\text{A},z} & = - D_{+,\text{B},z}, \\
D_{-,\text{A},z} & = D_{-,\text{B},z}.
\end{align}
\end{subequations}
On the other hand, for the equivalence between Eq.~\eqref{eq:R1} and Eq.~\eqref{eq:R2} incorporating the consistency with Eqs.~\eqref{eq:S1S2}, the following equations should be satisfied for any $\SSS_{x,y,z}$:
\begin{subequations}
\begin{align}
J_{\text{A},+} \SSS_x + D_{\text{A},-,y} \SSS_z + D_{\text{A},-,z} \SSS_y & =
J_{\text{B},+} \SSS_x + D_{\text{B},-,y} \SSS_z + D_{\text{B},-,z} \SSS_y, \\
J_{\text{A},-} \SSS_y - D_{\text{A},+,z} \SSS_x - D_{\text{A},-,x} \SSS_z & = 
-( J_{\text{B},-} \SSS_y - D_{\text{B},+,z} \SSS_x - D_{\text{B},-,x} \SSS_z ), \\
- J_{\text{A},-} \SSS_z - D_{\text{A},-,x} \SSS_y + D_{\text{A},+,y} \SSS_x & = 
-(- J_{\text{B},-} \SSS_z - D_{\text{B},-,x} \SSS_y + D_{\text{B},+,y} \SSS_x).
\end{align}
\end{subequations}
\end{widetext}
Then, we get the following relations:
\begin{subequations} \label{eq:R1R2} % !!!!!!!!!!!!!!!!!!!!!!!!!!!!!!!!!!!!!!!
\begin{align}
J_{\text{A},+} & = J_{\text{B},+}, \\
J_{\text{A},-} & = - J_{\text{B},-}, \\
D_{\text{A},-,x} & = - D_{\text{B},-,x}, \\
D_{\text{A},+,y} & = - D_{\text{B},+,y}, \\
D_{\text{A},-,y} & = D_{\text{B},-,y}, \\
D_{\text{A},+,z} & = - D_{\text{B},+,z}, \\
D_{\text{A},-,z} & = D_{\text{B},-,z}.
\end{align}
\end{subequations}
A possible choice of parameters for satisfying Eqs.~\eqref{eq:S1S2} and \eqref{eq:R1R2} is
\begin{subequations} \label{eq:Jij_Dij} % !!!!!!!!!!!!!!!!!!!!!!!!!!!!!!!!!!!!
\begin{equation}
J_{\text{A,A}} = J_{\text{B,B}} = J + J', \quad
J_{\text{A,B}} = J_{\text{B,A}} = J - J',
\end{equation}
\begin{align}
& \vD{\text{A,A}} = \begin{pmatrix} D_x + D_x' \\ D_y+D_y' \\ D_z + D_z' \end{pmatrix}, \quad
\vD{\text{A,B}} = \begin{pmatrix} - D_x + D_x' \\ -D_y+D_y' \\ -D_z+D_z' \end{pmatrix},
\nonumber \\ &
\vD{\text{B,A}} = \begin{pmatrix} - D_x + D_x' \\ D_y-D_y' \\ D_z-D_z' \end{pmatrix}, \quad
\vD{\text{B,B}} = \begin{pmatrix} D_x+D_x' \\ -D_y-D_y' \\ -D_z-D_z' \end{pmatrix}.
\end{align}
\end{subequations}
Among these eight parameters $J$, $J'$, $D_{x,y,z}$, and $D_{x,y,z}'$,
we numerically found by the mean-field calculation that $J' = D_y' = D_z = 0$ must be satisfied
in order to make the LTPT a second-order phase transition.
Otherwise, it becomes a crossover between the $\Gamma_{12}$ and $\Gamma_2$ phases.
Further, $D_x'$ and $D_z'$ gives negligible effects on the phase diagrams
under the present parameters.
Therefore, we consider only $J$, $D_x$, and $D_y$ in the spin model discussed in the main text.

\section{Spin resonance frequencies} \label{sec:spin_resonance}
In this Appendix, we discuss spin resonance frequencies
(especially frequency anti-crossing) at $T > \Tc$ ($\Gamma_2$ phase)
in the presence of the external DC magnetic field
along the $a$, $b$, and $c$ axes.
By fitting the calculated resonance frequencies
to the peak positions in THz spectra obtained
in our previous experimental study  \cite{Li2018a},
we determine some parameters ($\JEr$, $J$, $D_y$, and $A_x$) in our spin model
as we explain in Appendix \ref{app:param}.
We show also the consistency between the results by the two approaches:
the mean-field calculation and the extended Dicke Hamiltonian.

In the mean-field approach,
the spin resonance frequencies will be calculated
based on Eqs.~\eqref{eq:precession_MF},
from which equations of motion of the spin fluctuations
$\delta\vsigma{\text{A/B}}(t) \equiv \vsigma{\text{A/B}}(t) - \vsigmaa{\text{A/B}}$
and $\delta\vS{\text{A/B}}(t) \equiv \vS{\text{A/B}}(t) - \vSa{\text{A/B}}$ are obtained as
($s=\text{A,B}$)
\begin{subequations} \label{eq:meanfield_resonance} % !!!!!!!!!!!!!!!!!!!!!!!!!
\begin{align}
\hbar\ddtin\delta\vsigma{s}
& = - \delta\vsigma{s} \times \gf\muB\vBMFEra{s}
\nonumber \\ & \quad
    - \vsigmaa{s} \times \gf\muB\vBMFEr{s}(\{\delta\vsigma{\text{A/B}}\},\{\delta\vS{\text{A/B}}\}), \\
\hbar\ddtin\delta\vS{s}
& = - \delta\vS{s} \times \gf\muB\vBMFFea{s}
\nonumber \\ & \quad
    - \vSa{s} \times \gf\muB\vBMFFe{s}(\{\delta\vsigma{\text{A/B}}\},\{\delta\vS{\text{A/B}}\}).
\end{align}
\end{subequations}
From eigenvalues $E_k$ of the $12\times12$ coefficient matrix for
$\delta\vsigma{\text{A/B}}$ and $\delta\vS{\text{A/B}}$ on the right-hand sides,
we can find four positive eigenfrequencies of the spin resonances
as $\nu_k = \ii E_k / h$.
Another four are negative, and the other four are zero.
The temperature used for determining the equilibrium spins $\vsigmaa{\text{A/B}}$
and $\vSa{\text{A/B}}$ will be assumed as $T = 20\;\mathrm{K} > \Tc$.
While it is higher than the cryostat temperature $10\;\mathrm{K}$
used for measuring the THz spectrum (shown in Fig.~\ref{fig:w_Bz}),
$T = 20\;\mathrm{K}$ is better suited for reproducing the experimental spectrum.
The reason remains as a future problem.

We will also calculate the spin resonance frequencies
from the extended Dicke Hamiltonian, Eq.~\eqref{eq:extended_Dicke}.
We will see that the five $\Erion$--magnon couplings
show a variety of frequency anti-crossings.
It originates from the fact that the $\Feion$ qFM ($K=0$) and qAFM ($K=\pi$) magnon modes
and the $\Erion$ spin resonances in the A and B sublattices
are all coupled in general as seen in the extended Dicke Hamiltonian.

Note that the actual Hamiltonian treated in this Appendix is
\begin{align} \label{eq:extended_Dicke_delta} % !!!!!!!!!!!!!!!!!!!!!!!!!!!!!!
\oHH
& \approx \sum_{K=0,\pi} \hbar\omega_K\oad_K\oa_K
  + E_x \oSigma^+_{x}
  + \sum_{\xi=x,y,z} \gfEr_{\xi}\muB\Bext_{\xi} \oSigma^+_{\xi}
\nonumber \\ & \quad
  + \frac{8\zEr\JEr}{N} \ovSigma^\text{A} \cdot \ovSigma^\text{B}
+ \frac{2\hbar\rabi_x}{\sqrt{N}}(\oad_\pi+\oa_\pi)\delta\oSigma^+_x
\nonumber \\ & \quad
+ \frac{\ii2\hbar\rabi_y}{\sqrt{N}}(\oad_0-\oa_0)\delta\oSigma^+_y
+ \frac{2\hbar\rabi_y'}{\sqrt{N}}(\oad_\pi+\oa_\pi)\delta\oSigma^-_y
\nonumber \\ & \quad
+ \frac{\ii2\hbar\rabi_z}{\sqrt{N}}(\oad_\pi-\oa_\pi)\delta\oSigma^-_z
+ \frac{2\hbar\rabi_z'}{\sqrt{N}}(\oad_0+\oa_0)\delta\oSigma^+_z.
\end{align}
Compared with Eq.~\eqref{eq:extended_Dicke},
the $\Erion$ spin operators $\oSigma^{\pm}_{x,y,z}$ in the coupling terms are replaced
by their fluctuations
$\delta\oSigma^{\pm}_{x,y,z} \equiv \oSigma^{\pm}_{x,y,z} - \Sigmaa{\pm}_{x,y,z}$.
The terms including the equilibrium values $\Sigmaa{\pm}_{x,y,z}$
give shifts of $\Feion$ magnon frequencies.
However, returning to Eq.~\eqref{eq:oHHFeEr_Sigma_S},
we can find that the influence of these terms is smaller
by factor $\NUC{}^{-1/2}$ than the magnon Hamiltonian 
$\sum_{K=0,\pi} \hbar\omega_K\oad_K\oa_K$.
Then, the equilibrium values $\Sigmaa{\pm}_{x,y,z}$ can be omitted
in Eq.~\eqref{eq:extended_Dicke_delta}.

We will calculate the eigenfrequencies of Eq.~\eqref{eq:extended_Dicke_delta}.
However, since we suppose the $\Gamma_2$ phase ($T > \Tc$) in this Appendix,
we do not consider the spontaneous
ordering of $\Erion$ spins nor the rotation of the $\Feion$ spins
in the calculation of the eigenfrequencies.
% The resonance frequencies of Eq.~\eqref{eq:extended_Dicke_delta}
% will be approximately calculated in this condition.
Then, the results are justified only
for relatively high external DC field
that makes the system in the $\Gamma_2$ phase even in the zero-temperature limit.

In the calculation based on the extended Dicke Hamiltonian,
the finite temperature ($T = 20\;\mathrm{K}$) is incorporated
in the following procedure.
We consider the thermal excitation of the $\Erion$ spins
and assume that the $\Erion$ density effectively
depends on the temperature as \cite{Li2018a}
\begin{equation} \label{eq:x_eff} % !!!!!!!!!!!!!!!!!!!!!!!!!!!!!!!!!!!!!!!!!!
x = \tanh\left(\frac{\EEr}{2\kB T}\right),
\end{equation}
where the $\Erion$ excitation energy $E_{\mathrm{Er}}$
(excluding the $\Erion$--$\Erion$ exchange interaction) is represented as
\begin{equation}
\EEr
\equiv
\sqrt{(E_x+\gfEr_{x}\muB\Bext_{x})^2 + \sum_{\xi=y,z}(\gfEr_{\xi}\muB\Bext_{\xi})^2}.
\end{equation}
The temperature dependence appears through this effective $x$
and $\zEr = 6x$.

Note that, in this Appendix, the results by the mean-field approach
is more reliable than those by the extended Dicke Hamiltonian,
which are derived under some approximations.
However, the spin resonance frequencies and anti-crossing on them
will be better clarified by the extended Dicke Hamiltonian.

In the following subsections, we discuss how the five $\Erion$-magnon couplings
are reflected in three configurations:
$\vBext//a$ (Appendix \ref{sec:B//a}),
$\vBext//b$ (Appendix \ref{sec:B//b}),
and $\vBext//c$ (Appendix \ref{sec:B//c}).
We compare them with our experimental results \cite{Li2018a}
in Appendix \ref{sec:experiment}.

\subsection{$\vBext//a$} \label{sec:B//a}
If the external DC magnetic field is along the $a$ axis,
the $\Erion$ subsystem is most stable
when the $\Erion$ spins are along the $a$ axis.
For calculating the spin resonance frequencies
from the extended Dicke Hamiltonian
in the weak excitation limit (linear optical response),
we here bosonize the spin operators.
By the lowest-order Holstein--Primakoff transformation,
the spin-$\frac{N}{4}$ operators are transformed as ($s=\text{A,B}$)
\begin{subequations}
\begin{align}
\oSigma^s_x & \to \obd_s\ob_s - \frac{N}{4}, \\
\delta\oSigma^s_x & \to \obd_s\ob_s\\
\oSigma^s_y = \delta\oSigma^s_y & \to \sqrt{\frac{N}{2}}\frac{\obd_s+\ob_s}{2}, \\
\oSigma^s_z = \delta\oSigma^s_z & \to \sqrt{\frac{N}{2}}\frac{\obd_s-\ob_s}{\ii2}.
\end{align}
\end{subequations}
Then, the total Hamiltonian in Eq.~\eqref{eq:extended_Dicke_delta} is transformed as
\begin{align} \label{eq:Dicke_x} % !!!!!!!!!!!!!!!!!!!!!!!!!!!!!!!!!!!!!!!!!!!
\oHH
& \approx \sum_{K=0,\pi} \hbar\omega_K\oad_K\oa_K
  + ( E_x + \gfEr_{x}\muB\Bext_{x} ) (\obd_+\ob_+ + \obd_-\ob_-)
\nonumber \\ & \quad
  - 4\zEr\JEr \obd_-\ob_-
  + \hbar\rabi_x(\oad_{\pi}+\oa_{\pi})(\obd_+\ob_+ + \obd_-\ob_-)
\nonumber \\ & \quad
  + \ii\hbar\rabi_y(\oad_0-\oa_0)(\obd_++\ob_+)
  + \hbar\rabi_y'(\oad_\pi+\oa_\pi)(\obd_-+\ob_-)
\nonumber \\ & \quad
  + \hbar\rabi_z(\oad_\pi-\oa_\pi)(\obd_--\ob_-)
  - \ii\hbar\rabi_z'(\oad_0+\oa_0)(\obd_+-\ob_+)
\nonumber \\ & \quad
  + \const
\end{align}
Here, we defined operators of the in-phase oscillation $\ob_+$
and out-of-phase one $\ob_-$
of the two $\Erion$ spins $\ob_{\text{A/B}}$ as
\begin{equation}
\ob_{\pm} = \frac{\ob_\text{A}\pm\ob_\text{B}}{\sqrt{2}}.
\end{equation}
In the weak excitation limit, the $\rabi_x$ term can be neglected,
since it is involved with the number of $\Erion$ excitations
$\obd_{\pm}\ob_{\pm}$.
Then, the Hamiltonian can be divided into two parts as
\begin{align} \label{eq:oHH=oHH0++oHHpi-} % !!!!!!!!!!!!!!!!!!!!!!!!!!!!!!!!!!
\oHH
& \approx \oHH_{0+} + \oHH_{\pi-} + \const
\end{align}
The first term consists of the $\Feion$ qFM magnon mode
and $\Erion$ in-phase mode,
and it is expressed as
\begin{widetext}
\begin{align} \label{eq:oHH0+} % !!!!!!!!!!!!!!!!!!!!!!!!!!!!!!!!!!!!!!!!!!!!!
\oHH_{0+}
& \equiv \hbar\omega_0\oad_0\oa_0
  + |E_x + \gfEr_{x}\muB\Bext_{x}| \obd_+\ob_+
  + \ii\hbar\rabi_y(\oad_0-\oa_0)(\obd_++\ob_+)
\nonumber \\ & \quad
  - \ii\hbar\rabi_z'(\oad_0+\oa_0) \times
  \begin{cases}
  (\obd_+-\ob_+) & \gfEr_{x}\muB\Bext_{x} > - E_x \\
  (\ob_+-\obd_+) & \gfEr_{x}\muB\Bext_{x} < - E_x
  \end{cases}
\end{align}
If the coefficient $(E_x + \gfEr_{x}\muB\Bext_{x})$ of the second term
in Eq.~\eqref{eq:Dicke_x}
is negative for negative $\Bext_x$,
the roles of the annihilation operator $\ob_+$ and creation one $\obd_+$
are flipped.
As a result of it, the sign of the last term in Eq.~\eqref{eq:oHH0+} was flipped.
On the other hand, the second term in Eq.~\eqref{eq:oHH=oHH0++oHHpi-}
consists of the $\Feion$ qAFM magnon mode
and $\Erion$ out-of-phase mode, and it is expressed as
\begin{align} \label{eq:oHHpi-} % !!!!!!!!!!!!!!!!!!!!!!!!!!!!!!!!!!!!!!!!!!!!!
\oHH_{\pi-}
& \equiv \hbar\omega_\pi\oad_\pi\oa_\pi
  + \hbar\rabi_y'(\oad_\pi+\oa_\pi)(\obd_-+\ob_-)
\nonumber \\ & \quad
+ \begin{cases}
    (E_x + \gfEr_{x}\muB\Bext_{x}-4\zEr\JEr) \obd_-\ob_-
  + \hbar\rabi_z(\oad_\pi-\oa_\pi)(\obd_--\ob_-) &
    \gfEr_{x}\muB\Bext_{x} >  - E_x + 4\zEr\JEr \\
    (-E_x - \gfEr_{x}\muB\Bext_{x}-4\zEr\JEr) \obd_-\ob_-
  + \hbar\rabi_z(\oad_\pi-\oa_\pi)(\ob_--\obd_-) &
    \gfEr_{x}\muB\Bext_{x} < -E_x - 4\zEr\JEr
  \end{cases}
\end{align}
\end{widetext}
The $\Erion$--$\Erion$ exchange interaction,
the third term in Eq.~\eqref{eq:Dicke_x},
gives a negative frequency shift to the $\Erion$ out-of-phase mode.
Since it is always negative,
this calculation cannot be used in the case of
$-4\zEr\JEr < E_x + \gfEr_x\muB\Bext_x < 4\zEr\JEr$.
Such a situation corresponds to the $\Gamma_{12}$ phase,
and the present expression cannot be used.

\begin{figure}[tbp]
\centering
\includegraphics[scale=.45]{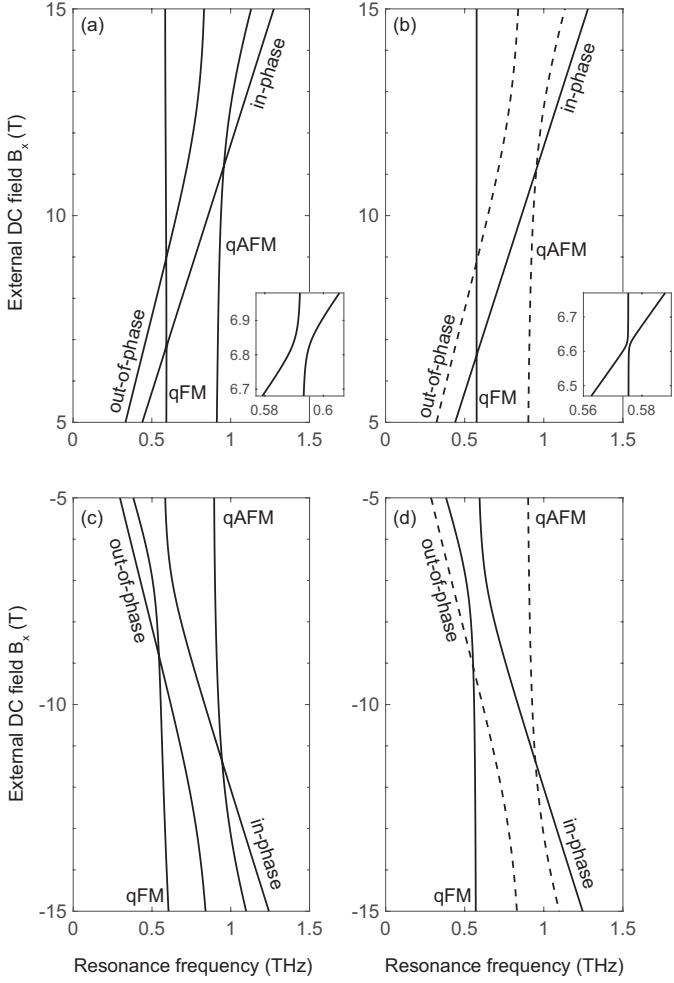}
\caption{Resonance frequencies of $\Erion$ and $\Feion$ spins in $\ErFeO$
at $T = 20\;\mathrm{K}$ under an external DC magnetic field along the $a$ axis
for positive (a,b) and negative (c,d) directions.
Figures \ref{fig:w_Bx}(a,c) and (b,d) are calculated, respectively,
by the mean-field approach,
Eqs.~\eqref{eq:meanfield_resonance},
and by the extended Dicke Hamiltonian,
Eqs.~\eqref{eq:oHH0+} (solid lines) and \eqref{eq:oHHpi-} (dashed lines).
The $\Erion$ out-of-phase and $\Feion$ qAFM modes
show frequency anti-crossing around $B_x = 13\;\mathrm{T}$
and $B_x = - 14\;\mathrm{T}$
obeying Eq.~\eqref{eq:oHHpi-} (dashed lines in Figs.~\ref{fig:w_Bx}(b,d)).
The anti-crossing between $\Erion$ in-phase and $\Feion$ qFM modes
is very small for positive $B_x$ (solid lines in Fig.~\ref{fig:w_Bx}(b))
but is relatively large for negative $B_x$
obeying Eq.~\eqref{eq:oHH0+} (solid lines in Fig.~\ref{fig:w_Bx}(d)).
The frequency splitting between the $\Erion$ in-phase and out-of-phase modes
is narrowed at small $|B_x|$ due to the thermal excitation of $\Erion$ spins.
The two approaches show almost the same resonance frequencies
except the slight frequency blue-shift of the $\Feion$ qFM mode
at large $-B_x$.
It is not obtained by the extended Dicke Hamiltonian
due to the approximations used to derive it.}
\label{fig:w_Bx}
\end{figure}
In Figs.~\ref{fig:w_Bx}(a,c) and (b,d), 
we plot the spin resonance frequencies calculated by the mean-field approach,
Eqs.~\eqref{eq:meanfield_resonance},
and by Eqs.~\eqref{eq:oHH0+} (solid lines) and \eqref{eq:oHHpi-} (dashed lines),
respectively.
Due to the broken mirror symmetry of spins about the $bc$ plane
even in the absence of the DC field,
the resonance frequencies depend on the sign of the DC field $\Bext_x$.

As shown by the dashed lines in Figs.~\ref{fig:w_Bx}(b,d),
the $\Erion$ out-of-phase and $\Feion$ qAFM modes
show frequency anti-crossing around $\Bext_x = 13\;\mathrm{T}$
and $\Bext_x = -14\;\mathrm{T}$
obeying Eq.~\eqref{eq:oHHpi-}.
As shown by solid lines in Figs.~\ref{fig:w_Bx}(b,d),
the anti-crossing between $\Erion$ in-phase and $\Feion$ qFM modes
clearly appears around $\Bext_x \sim -7\;\mathrm{T}$
obeying Eq.~\eqref{eq:oHH0+},
while it is very small around $\Bext_x \sim 7\;\mathrm{T}$
as shown in the insets.
This is because the coupling strength
$\rabi_y+\rabi_z' = 2\pi\times7\times10^{-4}\;\THz$
for the rotating terms $(\oad_0\ob_+-\obd_+\oa_0)$
is small for $\Bext_x > 0$.

The frequency splitting between the in-phase and out-of-phase $\Erion$ resonances
originates from the $\Erion$--$\Erion$ exchange interaction as explained above.
At a fixed temperature $T = 20\;\mathrm{K}$,
as we discussed also in our previous study  \cite{Li2018a},
the effective density (ratio $x$) of $\Erion$ spins
(involved with coherent dynamics such as spin precession) is decreased
by decreasing the $\Erion$ excitation energy $\EEr$
following Eq.~\eqref{eq:x_eff}.
Then, the splitting frequency $4\zEr\JEr$ is decreased
by the decrease in $|B_x|$.

The two approaches (mean-field method and extended Dicke Hamiltonian)
show almost the same resonance frequencies
except the slight frequency blue-shift of the $\Feion$ qFM mode
at large $-B_x$.
It is obtained by the mean-field approach but are not 
by the extended Dicke Hamiltonian.
This shift of the $\Feion$ magnon mode is due to
the Zeeman effect (external DC field) in the $\Feion$ subsystem
and the influence from the macroscopic paramagnetic $\Erion$ spins.
They are not considered in the present calculation with the extended Dicke Hamiltonian.

\subsection{$\vBext//b$} \label{sec:B//b}
When the external DC magnetic field along the $b$ axis
is large enough ($|\gfEr_y\muB\Bext_y| \gg E_x$),
the $E_x\oSigma^+_x$ term in Eq.~\eqref{eq:extended_Dicke_delta} can be neglected.
In the same manner as the previous subsection,
we transform the $\Erion$ spins as
\begin{subequations}
\begin{align}
\oSigma^s_y & \to \obd_s\ob_s - \frac{N}{4}, \\
\delta\oSigma^s_y & \to \obd_s\ob_s, \\
\oSigma^s_z = \delta\oSigma^s_z & \to \sqrt{\frac{N}{2}}\frac{\obd_s+\ob_s}{2}, \\
\oSigma^s_x = \delta\oSigma^s_x & \to \sqrt{\frac{N}{2}}\frac{\obd_s-\ob_s}{\ii2}.
\end{align}
\end{subequations}
Then, in the weak excitation limit,
the total Hamiltonian in Eq.~\eqref{eq:extended_Dicke_delta} is transformed to
\begin{align} \label{eq:Dicke_y} % !!!!!!!!!!!!!!!!!!!!!!!!!!!!!!!!!!!!!!!!!!!
\oHH
& \approx \sum_{K=0,\pi} \hbar\omega_K\oad_K\oa_K
  + |\gfEr_{y}\muB\Bext_{y}| (\obd_+\ob_+ + \obd_-\ob_-)
\nonumber \\ & \quad
  - 4\zEr\JEr \obd_-\ob_-
  - \ii\hbar\rabi_x(\oad_\pi+\oa_\pi)(\obd_+-\ob_+)
\nonumber \\ & \quad
  + \ii\hbar\rabi_z(\oad_\pi-\oa_\pi)(\obd_-+\ob_-)
  + \hbar\rabi_z'(\oad_0+\oa_0)(\obd_++\ob_+)
\nonumber \\ & \quad
  + \const
\end{align}
\begin{subequations}
This Hamiltonian can be used for $|\gfEr_y\muB\Bext_y| > 4\zEr\JEr$
similarly as the previous subsection.
In this configuration,
the two $\Feion$ magnon modes and two $\Erion$ modes are all coupled
in general.
However, when we focus around the $\Feion$ qFM magnon frequency,
the Hamiltonian can be simplified as
\begin{align}
\oHH
& \approx \hbar\omega_0\oad_0\oa_0
  + |\gfEr_{y}\muB\Bext_{y}| \obd_+\ob_+
\nonumber \\ & \quad
  + \hbar\rabi_z'(\oad_0+\oa_0)(\obd_++\ob_+) + \const
\end{align}
In this way, the $\Feion$ qFM mode shows anti-crossing with
the $\Erion$ in-phase mode.
On the other hand, when we focus on the $\Feion$ qAFM magnon mode,
the Hamiltonian is simplified as
\begin{align}
\oHH_{\pi-}
& \approx \hbar\omega_\pi\oad_\pi\oa_\pi
  + (|\gfEr_{y}\muB\Bext_{y}|-4\zEr\JEr)\obd_-\ob_-
\nonumber \\ & \quad
  + |\gfEr_{y}\muB\Bext_{y}| \obd_+\ob_+
  - \ii\hbar\rabi_x(\oad_\pi+\oa_\pi)(\obd_+-\ob_+)
\nonumber \\ & \quad
  + \ii\hbar\rabi_z(\oad_\pi-\oa_\pi)(\obd_-+\ob_-)
  + \const
\end{align}
\end{subequations}
In this way, the $\Feion$ qAFM mode shows anti-crossing
with both $\Erion$ in-phase and out-of-phase modes.

\begin{figure}[tbp]
\centering
\includegraphics[scale=.45]{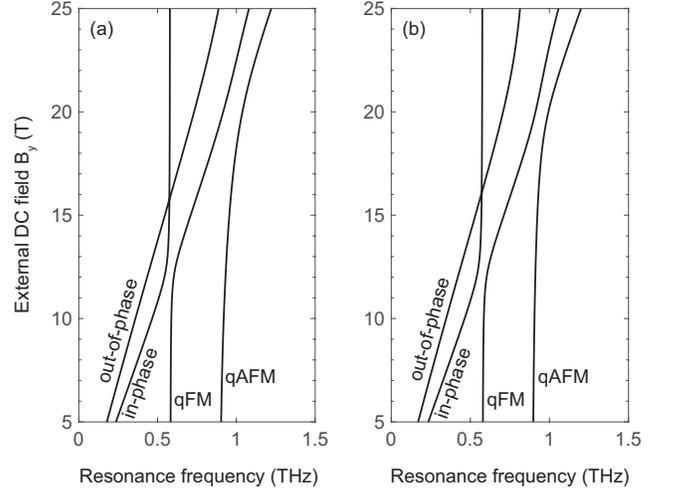}
\caption{Resonance frequencies of $\Erion$ and $\Feion$ spins in $\ErFeO$
at $T = 20\;\mathrm{K}$
under an external DC magnetic field along the $b$ axis.
Figures \ref{fig:w_By}(a) and (b) are calculated, respectively,
by the mean-field approach,
Eqs.~\eqref{eq:meanfield_resonance},
and by the extended Dicke Hamiltonian,
Eq.~\eqref{eq:Dicke_y}.
The $\Erion$ in-phase and $\Feion$ qFM modes
show frequency anti-crossing around $B_y = 12\;\mathrm{T}$
and the $\Feion$ qAFM mode shows anti-crossing
with the two $\Erion$ modes around $B_y = 20\;\mathrm{T}$.
The frequency splitting between the $\Erion$ in-phase and out-of-phase modes
is narrowed at small $|B_y|$ due to the thermal excitation of $\Erion$ spins.
The two approaches show almost the same resonance frequencies.}
\label{fig:w_By}
\end{figure}

In Figs.~\ref{fig:w_By}(a) and (b), 
we plot the spin resonance frequencies calculated by the mean-field approach,
Eqs.~\eqref{eq:meanfield_resonance},
and by Eq.~\eqref{eq:Dicke_y}, respectively.
The $\Erion$ in-phase and $\Feion$ qFM modes
show frequency anti-crossing around $B_y = 12\;\mathrm{T}$.
The $\Feion$ qAFM mode shows anti-crossing
with the two $\Erion$ modes around $B_y = 20\;\mathrm{T}$.
The two approaches show almost the same resonance frequencies
in the present case.

\subsection{$\vBext//c$} \label{sec:B//c}
Finally, when the external DC magnetic field along the $c$ axis
is large enough ($|\gfEr_z\muB\Bext_z| \gg E_x$),
the $E_x\oSigma^+_x$ term in Eq.~\eqref{eq:extended_Dicke_delta} can be neglected.
In the same manner as the previous subsections,
we transform the $\Erion$ spins as
\begin{subequations}
\begin{align}
\oSigma^s_z & = \obd_s\ob_s - \frac{N}{4}, \\
\delta\oSigma^s_z & = \obd_s\ob_s, \\
\oSigma^s_x = \delta\oSigma^s_x & = \sqrt{\frac{N}{2}}\frac{\obd_s+\ob_s}{2}, \\
\oSigma^s_y = \delta\oSigma^s_y & = \sqrt{\frac{N}{2}}\frac{\obd_s-\ob_s}{\ii2}.
\end{align}
\end{subequations}
In the weak excitation limit,
the total Hamiltonian in Eq.~\eqref{eq:extended_Dicke_delta} is transformed to
\begin{align}
\oHH
& \approx \sum_{K=0,\pi} \hbar\omega_K\oad_K\oa_K
  + |\gfEr_{z}\muB\Bext_{z}| (\obd_+\ob_+ + \obd_-\ob_-)
\nonumber \\ & \quad
  - 4\zEr\JEr \obd_-\ob_-
\nonumber \\ & \quad
  + \hbar\rabi_x(\oad_\pi+\oa_\pi)(\obd_++\ob_+)
  + \hbar\rabi_y(\oad_0-\oa_0)(\obd_+-\ob_+)
\nonumber \\ & \quad
  - \ii\hbar\rabi_y'(\oad_\pi+\oa_\pi)(\obd_--\ob_-)
  + \const
\end{align}
This Hamiltonian can be used for $|\gfEr_z\muB\Bext_z| > 4\zEr\JEr$
similarly as the previous subsections.
Also in this configuration,
the two $\Feion$ magnon modes and two $\Erion$ modes are all coupled.
However, since $g_y' \ll g_x, g_y$,
we can neglect the $g_y'$ term.
Then, the Hamiltonian is simplified as
\begin{equation} \label{eq:Dicke_Bz} % !!!!!!!!!!!!!!!!!!!!!!!!!!!!!!!!!!!!!!!
\oHH \approx \oHH_{0\pi+} + \oHH_{-} + \const
\end{equation}
The first term consists of the two $\Feion$ magnon modes
and the $\Erion$ in-phase mode as
\begin{align} \label{eq:oHH0pi+} % !!!!!!!!!!!!!!!!!!!!!!!!!!!!!!!!!!!!!!!!!!!!
\oHH_{0\pi+}
& \equiv \sum_{K=0,\pi}\hbar\omega_K\oad_0\oa_K
  + |\gfEr_{z}\muB\Bext_{z}| \obd_+\ob_+
\nonumber \\ & \quad
  + \hbar\rabi_y(\oad_0-\oa_0)(\obd_+-\ob_+)
\nonumber \\ & \quad
  + \hbar\rabi_x(\oad_\pi+\oa_\pi)(\obd_++\ob_+).
\end{align}
In this way, the $\Erion$ in-phase mode shows anti-crossing
with both the two $\Feion$ magnon modes.
The second term in Eq.~\eqref{eq:Dicke_Bz}
represents only the $\Erion$ out-of-phase mode as
\begin{align} \label{eq:oHH-} % !!!!!!!!!!!!!!!!!!!!!!!!!!!!!!!!!!!!!!!!!!!!!!!
\oHH_{-}
\equiv (|\gfEr_{z}\muB\Bext_{z}|-4\zEr\JEr)\obd_-\ob_-.
\end{align}
This mode is coupled only with the qAFM mode by the strength of
$\rabi_y'\ll\rabi_x,\rabi_y$
under the approximation used for deriving the extended Dicke Hamiltonian.

\begin{figure}[tbp]
\centering
\includegraphics[scale=.45]{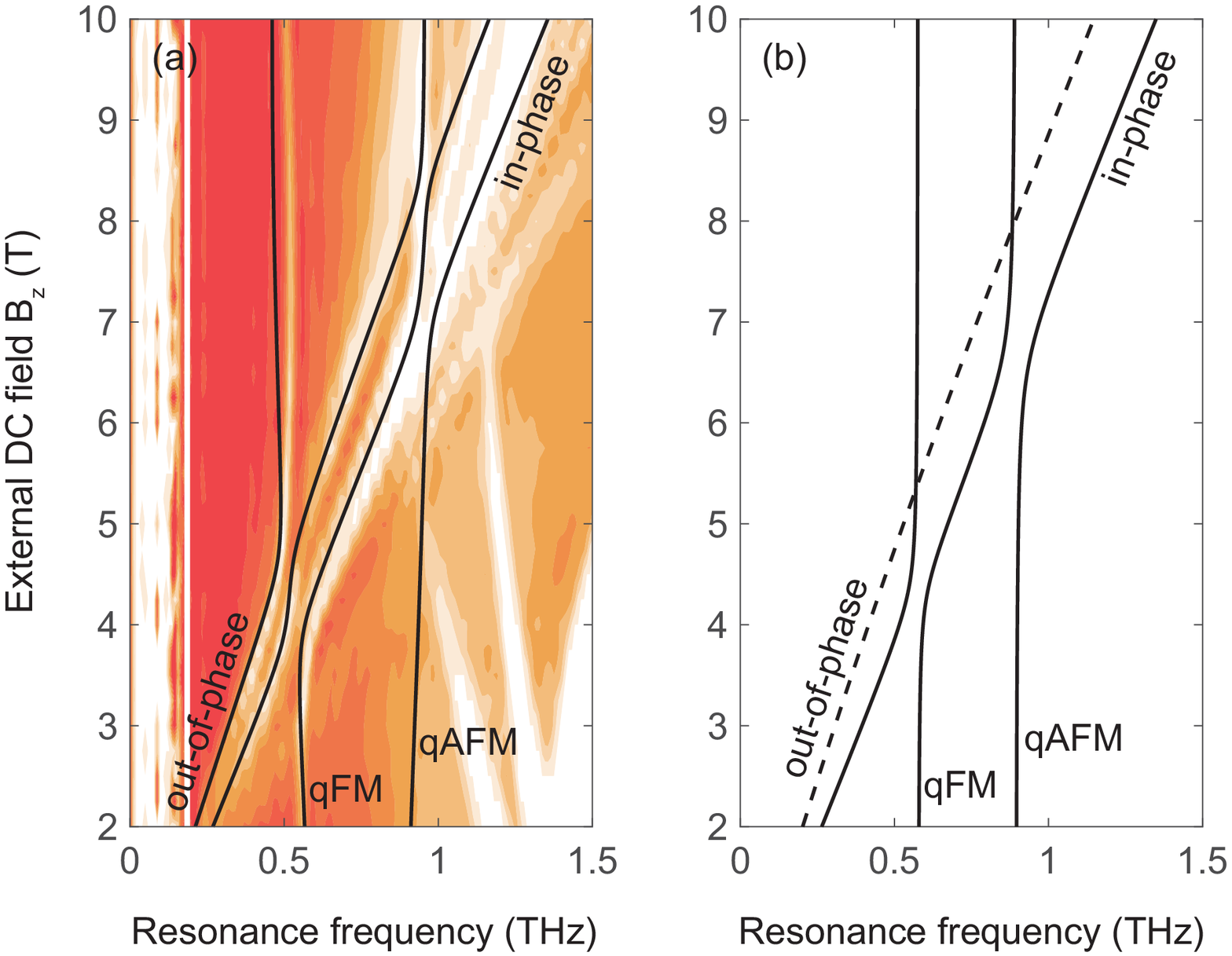}
\caption{Resonance frequencies of $\Erion$ and $\Feion$ spins in $\ErFeO$
at $T = 20\;\mathrm{K}$
under an external DC magnetic field along the $c$ axis.
Figures \ref{fig:w_Bz}(a) and (b) are calculated, respectively,
by the mean-field approach,
Eqs.~\eqref{eq:meanfield_resonance},
and by the extended Dicke Hamiltonian,
Eqs.~\eqref{eq:oHH0pi+} (solid lines) and \eqref{eq:oHH-} (dashed lines).
The experimentally observed absorption spectrum \cite{Li2018a} is plotted by red color
in Fig.~\ref{fig:w_Bz}(a).
The $\Erion$ in-phase mode shows frequency anti-crossing
with $\Feion$ qFM mode around $B_z = 4\;\mathrm{T}$
and with qAFM mode around $B_z = 7\;\mathrm{T}$
obeying Eq.~\eqref{eq:oHH0pi+} (solid lines in Fig.~\ref{fig:w_Bz}(b)).
The frequency splitting between the $\Erion$ in-phase and out-of-phase modes
is narrowed at small $|B_y|$ due to the thermal excitation of $\Erion$ spins.
In contrast to Figs.~\ref{fig:w_Bx} ($\vBext//a$)
and \ref{fig:w_By} ($\vBext//b$),
the two results show an inconsistency concerning
the anti-crossing between the $\Erion$ out-of-phase and the two $\Feion$ magnon modes
(around $B_z=4.5\;\mathrm{T}$ and 8.5\;T).
The reason is discussed at the end of Appendix \ref{sec:B//c}.
The frequency shifts of the $\Feion$ qFM and qAFM modes
at high external DC field are also not obtained by the extended Dicke Hamiltonian
due to the approximations used to derive it.}
\label{fig:w_Bz}
\end{figure}
In Figs.~\ref{fig:w_Bz}(a) and (b), 
we plot the spin resonance frequencies calculated by the mean-field approach,
Eqs.~\eqref{eq:meanfield_resonance},
and by the extended Dicke Hamiltonian,
Eqs.~\eqref{eq:oHH0pi+} (solid lines) and \eqref{eq:oHH-} (dashed lines),
respectively.
As shown by solid lines in Fig.~\ref{fig:w_Bz}(b),
obeying Eq.~\eqref{eq:oHH0pi+},
the $\Erion$ in-phase mode shows frequency anti-crossing
with $\Feion$ qFM mode around $B_z = 4\;\mathrm{T}$
and with qAFM mode around $B_z = 7\;\mathrm{T}$.
The frequency shifts of the $\Feion$ magnon modes at large $B_z$
are not reproduced in Fig.~\ref{fig:w_Bz}(b)
due to the approximations explained at the end of Appendix \ref{sec:B//a}.

As shown in Fig.~\ref{fig:w_Bz}(a),
the $\Erion$ out-of-phase mode shows frequency anti-crossing
with the $\Feion$ qFM mode around $B_z=4.5\;\mathrm{T}$
and with the qAFM mode around $B_z=8.5\;\mathrm{T}$.
They are not obtained by the present calculation with the extended Dicke Hamiltonian
as shown in Fig.~\ref{fig:w_Bz}(b).
Such an inconsistency does not appear in the previous cases ($\vBext//a,b$).
We checked that the inconsistency cannot be resolved even by considering
the equilibrium contribution $\Sigmaa{\pm}_{x,y,z}$
in the $\Erion$--magnon couplings in Eq.~\eqref{eq:extended_Dicke_delta}.
The $\rabi_y'$ term also cannot resolve it,
since it induces only the coupling between the $\Erion$ out-of-phase and $\Feion$ qAFM modes.

This inconsistency originates from the fact
that we did not properly consider
the change of the equilibrium values of $\Erion$ and $\Feion$ spins
by the presence of the external DC field $\vBext$
in the derivation of the extended Dicke Hamiltonian.
In fact, in the presence of $\vBext//c$,
we can find by the mean-field method that
the $\Feion$ spins become strongly asymmetric about the $ab$ plane
due to the large $z$ component of the macroscopic $\Erion$ spins
induced by $\vBext//c$.
Such an asymmetry causes the coupling between the $\Erion$ out-of-phase mode
and the two $\Feion$ magnon modes.
Then, the anti-crossing appears in Fig.~\ref{fig:w_Bz}(a).

The reproduction of these anti-crossing by the extended Dicke Hamiltonian
is beyond the scope of the present paper
and it remains as a future task.
% We have to consider the correct equilibrium configuration
% of the spins in the derivation of the extended Dicke Hamiltonian.

\subsection{Comparison with experimental results} \label{sec:experiment}
Since the maximum external DC magnetic flux density was
limited by around $10\;\mathrm{T}$
in our previous study  \cite{Li2018a},
the $\Erion$--magnon anti-crossing was experimentally observed
mainly for $\vBext//c$.
The anti-crossing around $\Bext_z = 4\;\mathrm{T}$ ($7\;\mathrm{T}$)
was clearly (slightly) observed.
If we apply the external DC field in the anti-parallel direction
to the magnetization along the $a$ axis,
we could observe anti-crossing around $\Bext_x = -7\;\mathrm{T}$
as shown in Fig.~\ref{fig:w_Bx}.
If we can apply a stronger DC magnetic field
and the linewidth is narrow enough,
we could observe the anti-crossing around $\Bext_y = 20\;\mathrm{T}$
for $\vBext//b$ as shown in Fig.~\ref{fig:w_By}.
In our previous study  \cite{Li2018a},
the anti-crossing was slightly observed around $\Bext_y = 7\;\mathrm{T}$.
It corresponds to the one around $\Bext_y = 12\;\mathrm{T}$ in Fig.~\ref{fig:w_By}.
The difference between the theoretical and experimental
external DC fields is due to the red-shift of $\Feion$ qFM mode
caused by the DC-field-induced structural change,
which is not considered in the present calculation.
For $\vBext//a$, in order to observe the large anti-crossing
around $\Bext_x = 13\;\mathrm{T}$ and $-14\;\mathrm{T}$ in Fig.~\ref{fig:w_Bx},
the probe THz wave should be irradiated along the $b$ or $c$ axis,
since the $\Feion$ qAFM modes can be excited
by the oscillating magnetic field only along the $a$ axis.

% The neutron diffraction measurement can also gives a similar information as the THz absorption spectroscopy. Further, the neutron diffraction can give the spin resonance energies at finite wavenumbers, while the THz spectroscopy basically gives the ones almost in the zero wavenumber limit. However, the selection rule (oscillator strength) for the electromagnetic wave can be obtained in the THz absorption spectroscopy. This selection rule is the information that cannot be obtained by the neutron diffraction measurement. We will discuss the influence of the LTPT on the THz spectroscopy and the analogy with the SRPT incorporating that selection rule in our future papers.

As shown in Fig.~\ref{fig:phase_diagram},
the phase diagrams around the LTPT of $\ErFeO$
are well reproduced by the mean-field method with our spin model.
Concerning other phase transitions at higher temperature and stronger DC field,
the present spin model can reproduce
the transition between the $\Gamma_2$ phase and the $\Gamma_4$ one,
where the $\Feion$ spins are ordered antiferromagnetically along the $a$ axis
with a slight canting to the $c$ axis, in the case of $\vBext//c$.
It occurs around $\Bext_z \sim 20\;\mathrm{T}$  \cite{Zhang2019}.
However, the temperature-induced $\Gamma_2$--$\Gamma_4$
spin-reorientation phase transition
around $90\;\mathrm{K} \lesssim T \lesssim 100\;\mathrm{K}$
 \cite{Gorodetsky1973,Klochan1975,Zhang2019}
cannot be reproduced in the present model.
We need a more complicated spin model for the $\Feion$ subsystem,
 \cite{Shane1968PRLe,Levinson1969PR,Yamaguchi1974JPCS,Balbashov1995,Zubov2019}
while it is beyond the scope of this paper.
Further, the phase transitions around $\Bext_x = 15\;\mathrm{T}$ for $\vBext//a$
and $\Bext_y = 20\;\mathrm{T}$ for $\vBext//b$
reported by Zhang {\it et al.} \cite{Zhang2019}
cannot also be reproduced in the present spin model.
The reproduction of these phase transitions remains a future task.
Existence of these transitions are the reason
why we restrict $|\Bext_x| < 15\;\mathrm{T}$
in Fig.~\ref{fig:w_Bx} and $\Bext_y < 25\;\mathrm{T}$ in Fig.~\ref{fig:w_By},
while we enlarged the latter range for clearly showing
the anti-crossing around $\Bext_y = 20\;\mathrm{T}$.

% We theoretically reproduced the phase diagrams around the LTPT of $\ErFeO$ by the meanfield calculation with the spin model of $\ErYFeO$ including the $\Erion$--$\Erion$ exchange interaction. The spin resonance frequencies obtained by the meanfield calculation well agree with the experimentally observed THz spectra in our previous study  \cite{Li2018a}, especially the fine structure at the anti-crossing between the two $\Erion$ resonances and the $\Feion$ magnon mode for $\vBext//c$ as shown in Fig.~\ref{fig:w_Bz}(a). By the analysis based on the extended Dicke model derived from the spin model, the origin of the spin resonances and the anti-crossing are well clarified. Especially, we found that the fine structure (three-mode anti-crossing) is caused not only by the $\Erion$--magnon coupling but also by the $\Erion$--$\Erion$ exchange interaction. Owing to the experimental THz spectrum and the present calculation, all the $\Erion$--$\Erion$ and $\Erion$--$\Feion$ exchange interaction strengths are determined consistently. It indicates a capability of the THz spectroscopy for estimating the strengths of spin--spin interactions in magnetic materials as a complementary method to the magnetization and neutron diffraction measurements.

\section{Parameters} \label{app:param}
Following our previous study  \cite{Li2018a},
we used the following values for the $\Feion$ subsystem
in our numerical calculations,
except $\Ax$, which was determined for fitting
the spin resonance frequencies in Fig.~\ref{fig:w_Bz}
to the corresponding THz absorption spectrum in our experiments: \cite{Li2018a}
\begin{subequations} \label{eq:param_Fe} % !!!!!!!!!!!!!!!!!!!!!!!!!!!!!!!!!!!
\begin{align}
\JFe & = 4.96\;\meV, \\
\DFe_y & = -0.107\;\meV, \\
\Ax  & = 0.0073\;\meV, \\
\Az  & = 0.0150\;\meV, \\
\Axz  & = 0.
\end{align}
\end{subequations}

The anisotropic $g$-factors for $\Erion$ spins were assumed to be
\begin{subequations}
\begin{align}
\gfEr_x & = 6, \\
\gfEr_y & = 3.4, \\
\gfEr_z & = 9.6.
\end{align}
\end{subequations}
They were determined for fitting the $\Erion$ spin resonance frequencies
in Figs.~\ref{fig:w_Bx}, \ref{fig:w_By}, and \ref{fig:w_Bz}
to their absorption peak positions observed in our experiments  \cite{Li2018a}.
They are basically multiplied by factor 2 from the values
estimated in our previous study \cite{Li2018a}
due to the additional factor $1/2$ in Eq.~\eqref{eq:ovmu}.

The anisotropic $g$-factors for $\Feion$ spins were assumed to be
\begin{subequations}
\begin{align}
\gfFe_x & = 2, \\
\gfFe_y & = 2, \\
\gfFe_z & = 0.6.
\end{align}
\end{subequations}
Here, $\gfFe_z$ was determined for reproducing
the critical magnetic flux density $\Bext_z \sim 20\;\mathrm{T}$
 \cite{Zhang2019} of the transition between
the $\Gamma_2$ phase and the $\Gamma_4$ one,
where the $\Feion$ spins are ordered antiferromagnetically along the $a$ axis
with a slight canting to the $c$ axis, in the case of $\vBext//c$.
On the other hand, $\gfFe_x$ and $\gfFe_y$ were simply set to be that of the free electron spin,
since the results in the present paper is very insensitive to these values.
% Note that our spin model (especially our simple $\Feion$ spin model) does not reproduce the spin-reorientation transition at $90\;\mathrm{K}\lesssim T \lesssim 100\;\mathrm{K}$ \cite{Gorodetsky1973,Klochan1975,Zhang2019} nor the phase transitions at $\Bext_x \sim 15\;\mathrm{T}$ for $\vBext//a$ and at $\Bext_x \sim 20\;\mathrm{T}$ for $\vBext//b$ reported by Zhang {it et al.} \cite{Zhang2019}

Concerning the $\Erion$--$\Erion$ and $\Erion$--$\Feion$ exchange interactions,
we used the following values:
\begin{subequations}
\begin{align}
\JEr & = 0.037\;\meV \\
J & = 0.60\;\meV \\
D_x & = 0.034\;\meV \\
D_y & = 0.003\;\meV
\end{align}
\end{subequations}
They were roughly determined for fitting Figs.~\ref{fig:phase_diagram}
to the phase diagrams reported by Zhang {it et al.} \cite{Zhang2019}.
The precise values of $\JEr$, $J$, and $D_y$ were mainly determined for fitting
our calculated spin resonance frequencies for $\vBext//c$
to the corresponding THz absorption spectrum in our experiments  \cite{Li2018a},
which are both shown in Fig.~\ref{fig:w_Bz}(a).
On the other hand, $D_x$ was determined for reproducing the critical temperature
$\Tc = 4.0\;\mathrm{K}$.

% As shown in Fig.~\ref{fig:w_Bz}(a), by fitting the spin resonance frequencies (solid lines) ($\vBext//c$) to the THz absorption spectrum (color plot) observed in our experiments  \cite{Li2018a}, we determined the $\Erion$--$\Erion$ exchange interaction strength $\JEr = 0.037\;\meV$, the $\Erion$--$\Feion$ ones $J = 0.60\;\meV$ and $D_y = 0.003\;\meV$. Of course, the fitting of the phase diagrams in Fig.~\ref{fig:phase_diagram} to the experimental ones by Zhang {\it et al.} \cite{Zhang2019} was also important, and actually we determined $D_x = 0.034\;\meV$ for reproducing $\Tc = 4.0\;\mathrm{K}$.
%Especially, the anti-crossing between the in-phase, out-of-phase $\Erion$ resonances,
%and the $\Feion$ qFM magnon mode around $\Bext_z \sim 4\;\mathrm{T}$
%was the most fruitful information for determining them.

Although the ratio between the \ErEr{} and \ErFe{} interaction strengths
was theoretically investigated by the phase boundary for $\vBext//a$  \cite{Kadomtseva1980},
the phase diagrams (critical temperature and DC fields) themselves
were not enough at least for determining all our parameters,
while we do not intend to scientifically claim its impossibility in this paper.
As far as we tried, the phase diagrams give only some ranges of the parameters.
Since the LTPT is caused
not only by the $\Erion$--$\Erion$ exchange interaction
but also by the $\Erion$--$\Feion$ ones,
there are at least four parameters $\JEr$, $J$, $D_x$, and $D_y$
even if we reduce the number of parameters by the analysis
in Appendix \ref{app:symmetry}.
Further, the anisotropic $g$-factors $\gfEr_x$, $\gfEr_y$, and $\gfEr_z$
of $\Erion$ spins were also free parameters,
and they can easily change the critical DC fields.
The critical temperature and the three critical DC fields
obtained by the magnetization measurements were not enough
for determining the above parameters.

In order to determine all of them,
the spin resonance frequencies are informative.
Especially, as we discussed in Appendix \ref{sec:spin_resonance}
by the extended Dicke Hamiltonian,
the $\Erion$--$\Erion$ exchange interaction strength $\JEr$
clearly appears as the frequency splitting between
the $\Erion$ in-phase and out-of-phase resonances.
The out-of-phase mode cannot be excited by the THz wave
unless it couples with the $\Feion$ magnon modes.
In that sense,
the anti-crossing between the $\Erion$ in-phase, out-of-phase resonances,
and the $\Feion$ qFM magnon mode around $\Bext_z \sim 4\;\mathrm{T}$
in Fig.~\ref{fig:w_Bz}
gave the most fruitful information for determining $\JEr$ and other parameters.

\section{Magnon quantization} \label{app:magnon}
\begin{figure}[tbp]
\centering
\includegraphics[scale=.4]{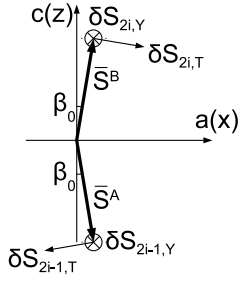}
\caption{Definitions of spin fluctuations $\delta\oS_{\ell,T/Y}$.
The index $\ell=2i-1$ and $2i$ correspond to the spins at the $i$-th site
in the A and B sublattices, respectively.}
\label{fig:STY}
\end{figure}
Here, we rewrite the Hamiltonian of $\Feion$ spins described by $\oHHFe$
in Eq.~\eqref{eq:VFe}
in terms of the annihilation and creation operators of a magnon.
As shown in Fig.~\ref{fig:STY},
we define the modulations $\{\delta\oS_{\ell,T}, \delta\oS_{\ell,Y}\}$
of $\Feion$ spins from their most stable values $\vSa{\text{A/B}}$ in its subsystem.
The index $\ell=2i-1$ and $2i$ correspond to the spins at the $i$-th site
in the A and B sublattices, respectively.
The spin modulations are expressed in the original axes as
\begin{subequations}
\begin{align}
\delta\ovS^\text{A}_i
& = \ovS^\text{A}_i - \vSa{\text{A}}_i
  = \sqrt{S}\begin{pmatrix}
    - \delta\oS_{2i-1,T}\cos\beta_0 \\
      \delta\oS_{2i-1,Y} \\
    - \delta\oS_{2i-1,T}\sin\beta_0
    \end{pmatrix}, \\
\delta\ovS^\text{B}_i
& = \ovS^\text{B}_i - \vSa{\text{B}}_i
  = \sqrt{S}\begin{pmatrix}
      \delta\oS_{2i,T}\cos\beta_0 \\
      \delta\oS_{2i,Y} \\
    - \delta\oS_{2i,T}\sin\beta_0
    \end{pmatrix}.
\end{align}
\end{subequations}
The factor $\sqrt{S}$ appears by considering the consistency
with the Holstein--Primakoff transformation,
while this factor did not appear in our previous studies,
 \cite{Li2018a,Bamba2019SPIE}
since the $\Feion$ spins are normalized
based on Herrmann's calculation  \cite{Herrmann1963JPCS}.

In the weak excitation limit, the spin fluctuations can be approximated as bosons
(magnons), and we define a bosonic commutation relations
for these modulations as
\begin{subequations}
\begin{align}
[\delta\oS_{\ell,T}, \delta\oS_{\ell',Y}] & = \ii \delta_{\ell,\ell'},\\
[\delta\oS_{\ell,T}, \delta\oS_{\ell',T}] & = 
[\delta\oS_{\ell,Y}, \delta\oS_{\ell',Y}] = 0.
\end{align}
\end{subequations}
Extending Herrmann's calculation into a magnon model propagating in the $z$ direction (under averaging in the $x$--$y$ plane)  \cite{Tsang1978}, we can derive the equations of motion for these modulations as
\begin{subequations} \label{eq:motion_magnon} % !!!!!!!!!!!!!!!!!!!!!!!!!!!!
\begin{align}
\frac{1}{\gamma}\ddt{}\delta\oS_{\ell,T}
& = - a \delta\oS_{\ell,Y}
    + \frac{b}{2}\left( \delta\oS_{\ell-1,Y} + \delta\oS_{\ell+1,Y} \right), \\
\frac{1}{\gamma}\ddt{}\delta\oS_{\ell,Y}
& = - c \delta\oS_{\ell,T}
    - \frac{d}{2}\left( \delta\oS_{\ell-1,T} + \delta\oS_{\ell+1,T} \right).
\end{align}
\end{subequations}
Here, $\gamma = \gf \muB /\hbar$ is the gyromagnetic ratio for the free electron $g$-factor $\gf$ and the Bohr magneton $\muB$.
The coefficients $a$, $b$, $c$, and $d$
are defined in Eqs.~\eqref{eq:abcd}  \cite{Herrmann1963JPCS}.
Then, the Hamiltonian of the $\Feion$ spins is approximated (bosonized) as
\begin{align} \label{eq:H_magnon} % !!!!!!!!!!!!!!!!!!!!!!!!!!!!!!!!!!!!!!!
\oHHFe
& \approx \hbar\gamma \sum_{\ell=1}^{2N_z} \left(
  - \frac{a}{2} \delta\oS_{\ell,Y}{}^2
  + \frac{c}{2} \delta\oS_{\ell,T}{}^2
  + \frac{b}{2} \delta\oS_{\ell,Y} \delta\oS_{\ell+1,Y}
\right. \nonumber \\ & \quad \left.
  + \frac{d}{2} \delta\oS_{\ell,T} \delta\oS_{\ell+1,T}
  \right) + \const
\end{align}
Here, $N_z$ and $2N_z$ are the number of unit cells and of $\Feion$ spins, respectively, in the $z$ direction. In terms of the annihilation operator $\oa_K$ of a magnon with a dimensionless wavenumber $K$, satisfying $[\oa_K, \oad_{K'}] = \delta_{K,K'}$, the modulation operators are expressed as
\begin{subequations}
\begin{align}
\delta\oS_{\ell,T}
& = \frac{1}{\sqrt{2N_z}} \sum_{K=-\pi}^\pi \ee^{\ii K \ell} \oT_K, \\
\delta\oS_{\ell,Y}
& = \frac{1}{\sqrt{2N_z}} \sum_{K=-\pi}^\pi \ee^{\ii K \ell} \oY_K,
\end{align}
\end{subequations}
\begin{subequations}
\begin{align}
\oT_K
& = \left( \frac{b\cos K-a}{d\cos K+c} \right)^{1/4}
    \frac{(\oad_{-K}+\oa_K)}{\sqrt{2}}, \\
\oY_K
& = \left( \frac{d\cos K+c}{b\cos K-a} \right)^{1/4}
    \frac{\ii(\oad_{-K}-\oa_K)}{\sqrt{2}}.
\end{align}
\end{subequations}
The Hamiltonian in Eq.~\eqref{eq:H_magnon} is rewritten as
\begin{align}
\oHHFe
& \approx \sum_{K} \hbar\omega_K \left(
    \oad_{K}\oa_{K} + \frac{1}{2}
  \right) + \const
\end{align}

Since we want to discuss a phase transition
where spin configurations are changed homogeneously in space
(we set the same assumption in the mean-field calculation),
we focus on only the two modes with $K = 0$ and $\pi$.
Then, the above Hamiltonian is approximated to Eq.~\eqref{eq:oHHFe2}.
The fluctuations are approximated as
\begin{subequations}
\begin{align}
\delta\oS_{2\ell-1,T}
& \approx \frac{1}{\sqrt{2N_z}} ( \oT_0 - \oT_\pi ), \\
\delta\oS_{2\ell-1,Y}
& \approx \frac{1}{\sqrt{2N_z}} ( \oY_0 - \oY_\pi ), \\
\delta\oS_{2\ell,T}
& \approx \frac{1}{\sqrt{2N_z}} ( \oT_0 + \oT_\pi ), \\
\delta\oS_{2\ell,Y}
& \approx \frac{1}{\sqrt{2N_z}} ( \oY_0 + \oY_\pi ).
\end{align}
\end{subequations}
Under this approximation,
the fluctuations do not depend on the index $\ell$ of unit cell.
In the original $xyz$-axes shown in Fig.~\ref{fig:STY},
the fluctuation vectors are
expressed in Eqs.~\eqref{eq:delta_ovS_AB}.

\section{Aspects of phase boundaries} \label{app:boundary}
In Fig.~\ref{fig:phase_boundary},
the phase boundaries obtained by the two approaches
show small differences.
The dash-dotted curves (phase transition only by the $\Erion$--$\Erion$ exchange interaction)
are almost the same.
However, the solid and dashed curves by the extended Dicke Hamiltonian
are shifted to the positive side
from those obtained by the mean-field approach.
These shifts of the critical magnetic fields
are mainly due to the neglect of $\vBext$-dependence of $\Feion$ spins
in the derivation of the extended Dicke Hamiltonian.
Then, a more sophisticated derivation of the extended Dicke Hamiltonian
will resolve these differences,
while it is beyond the scope of the present paper.

Note also that, in both approaches,
the absolute values of the negative critical fields
are larger than the positive ones
for the solid and dash-dotted curves,
while they are almost the same (symmetric about the origin) for the dashed curves.
The symmetric phase boundary is obtained
because the $\Erion$ spins are not influenced by the weak magnetization
of $\Feion$ spins in the absence of the $\Erion$--$\Feion$ exchange interactions
($\Erion$--magnon couplings).
In contrast, the phase boundaries become asymmetric about the origin
in the presence of the $\Erion$--$\Feion$ exchange interactions
($\Erion$--magnon couplings).
It is for compensating the magnetization along the $a$ axis.

% \bibliographystyle{bamba_prl}
% \bibliography{../../bib/library}

\begin{thebibliography}{10}
\expandafter\ifx\csname href\endcsname\relax\def\href#1#2{#2}\fi

\bibitem{Hepp1973AP}
K.~Hepp and E.~H. Lieb,
\newblock {On the superradiant phase transition for molecules in a quantized
  radiation field: the dicke maser model},
\newblock \href{https://doi.org/10.1016/0003-4916(73)90039-0}{Ann. Phys. (N.
  Y.) {\bf 76}, 360 (1973)}.

\bibitem{Wang1973PRA}
Y.~K. Wang and F.~T. Hioe,
\newblock {Phase transition in the dicke model of superradiance},
\newblock \href{https://doi.org/10.1103/PhysRevA.7.831}{Phys. Rev. A {\bf 7},
  831 (1973)}.

\bibitem{Ciuti2005PRB}
C.~Ciuti, G.~Bastard, and I.~Carusotto,
\newblock {Quantum vacuum properties of the intersubband cavity polariton
  field},
\newblock \href{https://doi.org/10.1103/PhysRevB.72.115303}{Phys. Rev. B {\bf
  72}, 115303 (2005)}.

\bibitem{Forn-Diaz2018}
P.~Forn-D{\'{i}}az, L.~Lamata, E.~Rico, J.~Kono, and E.~Solano,
\newblock {Ultrastrong coupling regimes of light-matter interaction},
\newblock \href{https://doi.org/10.1103/RevModPhys.91.025005}{Rev. Mod. Phys.
  {\bf 91}, 025005 (2019)}.

\bibitem{Kockum2018}
A.~{Frisk Kockum}, A.~Miranowicz, S.~{De Liberato}, S.~Savasta, and F.~Nori,
\newblock {Ultrastrong coupling between light and matter},
\newblock \href{https://doi.org/10.1038/s42254-018-0006-2}{Nat. Rev. Phys. {\bf
  1}, 19 (2019)}.

\bibitem{Dicke1954PR}
R.~H. Dicke,
\newblock {Coherence in spontaneous radiation processes},
\newblock \href{https://doi.org/10.1103/PhysRev.93.99}{Phys. Rev. {\bf 93}, 99
  (1954)}.

\bibitem{Baumann2010N}
K.~Baumann, C.~Guerlin, F.~Brennecke, and T.~Esslinger,
\newblock {Dicke quantum phase transition with a superfluid gas in an optical
  cavity},
\newblock \href{https://doi.org/10.1038/nature09009}{Nature {\bf 464}, 1301
  (2010)}.

\bibitem{Kirton2018a}
P.~Kirton, M.~M. Roses, J.~Keeling, and E.~G. {Dalla Torre},
\newblock {Introduction to the Dicke Model: From Equilibrium to Nonequilibrium,
  and Vice Versa},
\newblock \href{https://doi.org/10.1002/qute.201800043}{Adv. Quantum Technol.
  {\bf 2}, 1800043 (2019)}.

\bibitem{Bamba2016circuitSRPT}
M.~Bamba, K.~Inomata, and Y.~Nakamura,
\newblock {Superradiant Phase Transition in a Superconducting Circuit in
  Thermal Equilibrium},
\newblock \href{https://doi.org/10.1103/PhysRevLett.117.173601}{Phys. Rev.
  Lett. {\bf 117}, 173601 (2016)}.

\bibitem{Rzazewski1975PRL}
K.~Rz{\c{a}}{\.{z}}ewski, K.~W{\'{o}}dkiewicz, and W.~{\.{Z}}akowicz,
\newblock {Phase Transitions, Two-Level Atoms, and the {\$}A{\^{}}2{\$} Term},
\newblock \href{https://doi.org/10.1103/PhysRevLett.35.432}{Phys. Rev. Lett.
  {\bf 35}, 432 (1975)}.

\bibitem{Knight1978}
J.~M. Knight, Y.~Aharonov, and G.~T.~C. Hsieh,
\newblock {Are super-radiant phase transitions possible?},
\newblock \href{https://doi.org/10.1103/PhysRevA.17.1454}{Phys. Rev. A {\bf
  17}, 1454 (1978)}.

\bibitem{Bialynicki-Birula1979}
I.~Bialynicki-Birula and K.~Rz{\c{a}}{\.{z}}ewski,
\newblock {No-go theorem concerning the superradiant phase transition in atomic
  systems},
\newblock \href{https://doi.org/10.1103/PhysRevA.19.301}{Phys. Rev. A {\bf 19},
  301 (1979)}.

\bibitem{Gawedzki1981PRA}
K.~Gawedzki and K.~Rz{\c{a}}{\.{z}}ewski,
\newblock {No-go theorem for the superradiant phase transition without dipole
  approximation},
\newblock \href{https://doi.org/10.1103/PhysRevA.23.2134}{Phys. Rev. A {\bf
  23}, 2134 (1981)}.

\bibitem{Hepp1973PRA}
K.~Hepp, E.~H. Lieb, R.~Field, and K.~Etudes,
\newblock {Equilibrium Statistical Mechanics of Matter Interacting with the
  Quantized Radiation Field},
\newblock \href{https://doi.org/10.1103/PhysRevA.8.2517}{Phys. Rev. A {\bf 8},
  2517 (1973)}.

\bibitem{Hemmen1980PLA}
J.~L. van Hemmen and K.~Rz{\c{a}}{\.{z}}ewski,
\newblock {On the thermodynamic equivalence of the Dicke maser model and a
  certain spin system},
\newblock \href{https://doi.org/10.1016/0375-9601(80)90645-3}{Phys. Lett. A
  {\bf 77}, 211 (1980)}.

\bibitem{Bamba2017NogoCircuit}
M.~Bamba and N.~Imoto,
\newblock {Circuit configurations which may or may not show superradiant phase
  transitions},
\newblock \href{https://doi.org/10.1103/PhysRevA.96.053857}{Phys. Rev. A {\bf
  96}, 053857 (2017)}.

\bibitem{Keeling2007JPCM}
J.~Keeling,
\newblock {Coulomb interactions, gauge invariance, and phase transitions of the
  Dicke model},
\newblock \href{https://doi.org/10.1088/0953-8984/19/29/295213}{J. Phys.
  Condens. Matter {\bf 19}, 295213 (2007)}.

\bibitem{Vukics2012PRA}
A.~Vukics and P.~Domokos,
\newblock {Adequacy of the Dicke model in cavity QED: A counter-no-go
  statement},
\newblock \href{https://doi.org/10.1103/PhysRevA.86.053807}{Phys. Rev. A {\bf
  86}, 53807 (2012)}.

\bibitem{Vukics2014PRL}
A.~Vukics, T.~Grie{\ss}er, and P.~Domokos,
\newblock {Elimination of the A-square problem from cavity QED},
\newblock \href{https://doi.org/10.1103/PhysRevLett.112.073601}{Phys. Rev.
  Lett. {\bf 112}, 73601 (2014)}.

\bibitem{Bamba2014SPT}
M.~Bamba and T.~Ogawa,
\newblock {Stability of polarizable materials against superradiant phase
  transition},
\newblock \href{https://doi.org/10.1103/PhysRevA.90.063825}{Phys. Rev. A {\bf
  90}, 063825 (2014)}.

\bibitem{Vukics2015PRA}
A.~Vukics, T.~Grie{\ss}er, and P.~Domokos,
\newblock {Fundamental limitation of ultrastrong coupling between light and
  atoms},
\newblock \href{https://doi.org/10.1103/PhysRevA.92.043835}{Phys. Rev. A {\bf
  92}, 43835 (2015)}.

\bibitem{Griesser2016PRA}
T.~Grie{\ss}er, A.~Vukics, and P.~Domokos,
\newblock {Depolarization shift of the superradiant phase transition},
\newblock \href{https://doi.org/10.1103/PhysRevA.94.033815}{Phys. Rev. A {\bf
  94}, 033815 (2016)}.

\bibitem{Hagenmuller2012PRL}
D.~Hagenm{\"{u}}ller and C.~Ciuti,
\newblock {Cavity QED of the graphene cyclotron transition},
\newblock \href{https://doi.org/10.1103/PhysRevLett.109.267403}{Phys. Rev.
  Lett. {\bf 109}, 267403 (2012)}.

\bibitem{Chirolli2012PRL}
L.~Chirolli, M.~Polini, V.~Giovannetti, and A.~H. MacDonald,
\newblock {Drude weight, cyclotron resonance, and the dicke model of graphene
  cavity QED},
\newblock \href{https://doi.org/10.1103/PhysRevLett.109.267404}{Phys. Rev.
  Lett. {\bf 109}, 267404 (2012)}.

\bibitem{Mazza2018}
G.~Mazza and A.~Georges,
\newblock {Superradiant Quantum Materials},
\newblock \href{https://doi.org/10.1103/PhysRevLett.122.017401}{Phys. Rev.
  Lett. {\bf 122}, 017401 (2019)}.

\bibitem{Andolina2019}
G.~M. Andolina, F.~M.~D. Pellegrino, V.~Giovannetti, A.~H. MacDonald, and
  M.~Polini,
\newblock {Cavity quantum electrodynamics of strongly correlated electron
  systems: A no-go theorem for photon condensation},
\newblock \href{https://doi.org/10.1103/PhysRevB.100.121109}{Phys. Rev. B {\bf
  100}, 121109 (2019)}.

\bibitem{Nataf2019a}
P.~Nataf, T.~Champel, G.~Blatter, and D.~M. Basko,
\newblock {Rashba Cavity QED: A Route Towards the Superradiant Quantum Phase
  Transition},
\newblock \href{https://doi.org/10.1103/PhysRevLett.123.207402}{Phys. Rev.
  Lett. {\bf 123}, 207402 (2019)}.

\bibitem{Zhang2014PRLa}
X.~Zhang, C.~L. Zou, L.~Jiang, and H.~X. Tang,
\newblock {Strongly coupled magnons and cavity microwave photons},
\newblock \href{https://doi.org/10.1103/PhysRevLett.113.156401}{Phys. Rev.
  Lett. {\bf 113}, 156401 (2014)}.

\bibitem{Goryachev2014PRA}
M.~Goryachev, W.~G. Farr, D.~L. Creedon, Y.~Fan, M.~Kostylev, and M.~E. Tobar,
\newblock {High-cooperativity cavity QED with magnons at microwave
  frequencies},
\newblock \href{https://doi.org/10.1103/PhysRevApplied.2.054002}{Phys. Rev.
  Appl. {\bf 2}, 54002 (2014)}.

\bibitem{Bourhill2016}
J.~Bourhill, N.~Kostylev, M.~Goryachev, D.~L. Creedon, and M.~E. Tobar,
\newblock {Ultrahigh cooperativity interactions between magnons and resonant
  photons in a YIG sphere},
\newblock \href{https://doi.org/10.1103/PhysRevB.93.144420}{Phys. Rev. B {\bf
  93}, 1 (2016)}.

\bibitem{Kostylev2016}
N.~Kostylev, M.~Goryachev, and M.~E. Tobar,
\newblock {Superstrong coupling of a microwave cavity to yttrium iron garnet
  magnons},
\newblock \href{https://doi.org/10.1063/1.4941730}{Appl. Phys. Lett. {\bf 108}
  (2016)}.

\bibitem{Flower2019}
G.~Flower, M.~Goryachev, J.~Bourhill, and M.~E. Tobar,
\newblock {Experimental implementations of cavity-magnon systems: From ultra
  strong coupling to applications in precision measurement},
\newblock \href{https://doi.org/10.1088/1367-2630/ab3e1c}{New J. Phys. {\bf
  21}, 095004 (2019)}.

\bibitem{Li2018a}
X.~Li, M.~Bamba, N.~Yuan, Q.~Zhang, Y.~Zhao, M.~Xiang, K.~Xu, Z.~Jin, W.~Ren,
  G.~Ma, S.~Cao, D.~Turchinovich, and J.~Kono,
\newblock {Observation of Dicke cooperativity in magnetic interactions},
\newblock \href{https://doi.org/10.1126/science.aat5162}{Science (80-. ). {\bf
  361}, 794 (2018)}.

\bibitem{Zhang2019}
X.~X. Zhang, Z.~C. Xia, Y.~J. Ke, X.~Q. Zhang, Z.~H. Cheng, Z.~W. Ouyang, J.~F.
  Wang, S.~Huang, F.~Yang, Y.~J. Song, G.~L. Xiao, H.~Deng, and D.~Q. Jiang,
\newblock {Magnetic behavior and complete high-field magnetic phase diagram of
  the orthoferrite ErFeO3},
\newblock \href{https://doi.org/10.1103/PhysRevB.100.054418}{Phys. Rev. B {\bf
  100}, 054418 (2019)}.

\bibitem{Tabuchi2014PRL}
Y.~Tabuchi, S.~Ishino, T.~Ishikawa, R.~Yamazaki, K.~Usami, and Y.~Nakamura,
\newblock {Hybridizing ferromagnetic magnons and microwave photons in the
  quantum limit},
\newblock \href{https://doi.org/10.1103/PhysRevLett.113.083603}{Phys. Rev.
  Lett. {\bf 113}, 83603 (2014)}.

\bibitem{Tabuchi2015}
Y.~Tabuchi, S.~Ishino, A.~Noguchi, T.~Ishikawa, R.~Yamazaki, K.~Usami, and
  Y.~Nakamura,
\newblock {Coherent coupling between a ferromagnetic magnon and a
  superconducting qubit},
\newblock \href{https://doi.org/10.1126/science.aaa3693}{Science (80-. ). {\bf
  349}, 405 (2015)}.

\bibitem{Tabuchi2016}
Y.~Tabuchi, S.~Ishino, A.~Noguchi, T.~Ishikawa, R.~Yamazaki, K.~Usami, and
  Y.~Nakamura,
\newblock {La magnonique des quanta: Le magnon rencontre le qubit
  supraconducteur},
\newblock \href{https://doi.org/10.1016/j.crhy.2016.07.009}{Comptes Rendus
  Phys. {\bf 17}, 729 (2016)}.

\bibitem{Morris2017}
R.~G. Morris, A.~F. {Van Loo}, S.~Kosen, and A.~D. Karenowska,
\newblock {Strong coupling of magnons in a YIG sphere to photons in a planar
  superconducting resonator in the quantum limit},
\newblock \href{https://doi.org/10.1038/s41598-017-11835-4}{Sci. Rep. {\bf 7},
  1 (2017)}.

\bibitem{Flower2019a}
G.~Flower, J.~Bourhill, M.~Goryachev, and M.~E. Tobar,
\newblock {Broadening frequency range of a ferromagnetic axion haloscope with
  strongly coupled cavity-magnon polaritons},
\newblock \href{https://doi.org/10.1016/j.dark.2019.100306}{Phys. Dark Universe
  {\bf 25}, 100306 (2019)}.

\bibitem{Macneill2019}
D.~Macneill, J.~T. Hou, D.~R. Klein, P.~Zhang, P.~Jarillo-Herrero, and L.~Liu,
\newblock {Gigahertz Frequency Antiferromagnetic Resonance and Strong
  Magnon-Magnon Coupling in the Layered Crystal CrCl3},
\newblock \href{https://doi.org/10.1103/PhysRevLett.123.047204}{Phys. Rev.
  Lett. {\bf 123}, 47204 (2019)}.

\bibitem{Liensberger2019}
L.~Liensberger, A.~Kamra, H.~Maier-Flaig, S.~Gepr{\"{a}}gs, A.~Erb, S.~T.~B.
  Goennenwein, R.~Gross, W.~Belzig, H.~Huebl, and M.~Weiler,
\newblock {Exchange-enhanced Ultrastrong Magnon-Magnon Coupling in a
  Compensated Ferrimagnet},
\newblock \href{https://doi.org/10.1103/PhysRevLett.123.117204}{Phys. Rev.
  Lett. {\bf 123}, 117204 (2019)}.

\bibitem{Lachance-Quirion2019}
D.~Lachance-Quirion, S.~P. Wolski, Y.~Tabuchi, S.~Kono, K.~Usami, and
  Y.~Nakamura,
\newblock {Entanglement-based single-shot detection of a single magnon with a
  superconducting qubit},
\newblock \href{https://doi.org/10.1126/science.aaz9236}{Science (80-. ). {\bf
  367}, 425 (2020)}.

\bibitem{Gorodetsky1973}
G.~Gorodetsky, R.~M. Hornreich, I.~Yaeger, H.~Pinto, G.~Shachar, and H.~Shaked,
\newblock {Magnetic Structure of ErFeO3 below 4.5 K},
\newblock \href{https://doi.org/10.1103/PhysRevB.8.3398}{Phys. Rev. B {\bf 8},
  3398 (1973)}.

\bibitem{Klochan1975}
V.~A. Klochan, N.~M. Kovtun, and V.~M. Khmara,
\newblock {Low-temperature spin configuration of iron ions in erbium
  orthoferrite},
\newblock Zh. Eksp. Teor. Fiz. {\bf 68}, 721 (1975).

\bibitem{Vitebskii1978}
I.~M. Vitebskii and D.~A. Yablonskii,
\newblock {Theory of Low-Temperature Spin Reorientation in ErFeO3},
\newblock Sov. Phys. Solid State {\bf 20}, 1327 (1978).

\bibitem{Kadomtseva1980}
A.~M. Kadomtseva, I.~B. Krynetskil, and V.~M. Matveev,
\newblock {Nature of the spontaneous and field-induced low-temperature
  orientational transitions in erbium orthoferrite},
\newblock Sov. Phys. JETP {\bf 52}, 732 (1980).

\bibitem{Gehring1975}
G.~A. Gehring and K.~A. Gehring,
\newblock {Co-operative Jahn-Teller effects},
\newblock \href{https://doi.org/10.1088/0034-4885/38/1/001}{Reports Prog. Phys.
  {\bf 38}, 1 (1975)}.

\bibitem{Kugel1982}
K.~I. Kugel' and D.~I. Khomski,
\newblock {The Jahn-Teller effect and magnetism: transition metal compounds},
\newblock \href{https://doi.org/10.1070/PU1982v025n04ABEH004537}{Sov. Phys.
  Uspekhi {\bf 25}, 231 (1982)}.

\bibitem{Loos1984}
J.~Loos,
\newblock {On the Fluctuations and Phase Transitions in Dicke-Like Models},
\newblock \href{https://doi.org/10.1002/pssb.2221230223}{Phys. status solidi
  {\bf 123}, 595 (1984)}.

\bibitem{Larson2008PRA}
J.~Larson,
\newblock {Jahn-Teller systems from a cavity QED perspective},
\newblock \href{https://doi.org/10.1103/PhysRevA.78.033833}{Phys. Rev. A {\bf
  78}, 33833 (2008)}.

\bibitem{Herrmann1964}
G.~F. Herrmann,
\newblock {Magnetic Resonances and Susceptibility in Orthoferrites},
\newblock \href{https://doi.org/10.1103/PhysRev.133.A1334}{Phys. Rev. {\bf
  133}, A1334 (1964)}.

\bibitem{Wood1969}
D.~L. Wood, J.~P. Remeika, L.~M. Holmes, and E.~M. Gyorgy,
\newblock {Effect of Y and Bi Substitution on Spin Reorientation and Optical
  Absorption in ErFeO 3},
\newblock \href{https://doi.org/10.1063/1.1657613}{J. Appl. Phys. {\bf 40},
  1245 (1969)}.

\bibitem{Herrmann1963JPCS}
G.~Herrmann,
\newblock {Resonance and high frequency susceptibility in canted
  antiferromagnetic substances},
\newblock \href{https://doi.org/10.1016/S0022-3697(63)80001-3}{J. Phys. Chem.
  Solids {\bf 24}, 597 (1963)}.

\bibitem{Bamba2019SPIE}
M.~Bamba, X.~Li, and J.~Kono,
\newblock {Terahertz strong-field physics without a strong external terahertz
  field},
\newblock in {\em Ultrafast Phenom. Nanophotonics XXIII}, edited by M.~Betz and
  A.~Y. Elezzabi, SPIE, 2019, p.~5.

\bibitem{Holstein1940}
T.~Holstein and H.~Primakoff,
\newblock {Field Dependence of the Intrinsic Domain Magnetization of a
  Ferromagnet},
\newblock \href{https://doi.org/10.1103/PhysRev.58.1098}{Phys. Rev. {\bf 58},
  1098 (1940)}.

\bibitem{Emary2003PRL}
C.~Emary and T.~Brandes,
\newblock {Quantum Chaos Triggered by Precursors of a Quantum Phase Transition:
  The Dicke Model},
\newblock \href{https://doi.org/10.1103/PhysRevLett.90.044101}{Phys. Rev. Lett.
  {\bf 90}, 044101 (2003)}.

\bibitem{Emary2003PRE}
C.~Emary and T.~Brandes,
\newblock {Chaos and the quantum phase transition in the Dicke model},
\newblock \href{https://doi.org/10.1103/PhysRevE.67.066203}{Phys. Rev. E {\bf
  67}, 066203 (2003)}.

\bibitem{Larson2017JPA}
J.~Larson and E.~K. Irish,
\newblock {Some remarks on `superradiant' phase transitions in light-matter
  systems},
\newblock \href{https://doi.org/10.1088/1751-8121/aa65dc}{J. Phys. A Math.
  Theor. {\bf 50}, 174002 (2017)}.

\bibitem{Shapiro2019}
D.~S. Shapiro, W.~V. Pogosov, and Y.~E. Lozovik,
\newblock {Hierarchy of universal behaviors in generalized Dicke model near the
  superradiant phase transition} , 1 (2019).

\bibitem{Artoni1991}
M.~Artoni and J.~L. Birman,
\newblock {Quantum-optical properties of polariton waves},
\newblock \href{https://doi.org/10.1103/PhysRevB.44.3736}{Phys. Rev. B {\bf
  44}, 3736 (1991)}.

\bibitem{Artoni1989}
M.~Artoni and J.~L. Birman,
\newblock {Polariton squeezing: theory and proposed experiment},
\newblock \href{https://doi.org/10.1088/0954-8998/1/2/002}{Quantum Opt. J. Eur.
  Opt. Soc. Part B {\bf 1}, 91 (1989)}.

\bibitem{Schwendimann1992}
P.~Schwendimann and A.~Quattropani,
\newblock {Nonclassical Properties of Polariton States},
\newblock \href{https://doi.org/10.1209/0295-5075/17/4/013}{Europhys. Lett.
  {\bf 17}, 355 (1992)}.

\bibitem{Schwendimann1992a}
P.~Schwendimann and A.~Quattropani,
\newblock {Nonclassical Properties of Polariton States},
\newblock \href{https://doi.org/10.1209/0295-5075/18/3/016}{Europhys. Lett.
  {\bf 18}, 281 (1992)}.

\bibitem{quattropani05}
A.~Quattropani and P.~Schwendimann,
\newblock {Polariton squeezing in microcavities},
\newblock \href{https://doi.org/10.1002/pssb.200560963}{Phys. status solidi
  {\bf 242}, 2302 (2005)}.

\bibitem{Makihara2020}
T.~Makihara, K.~Hayashida, G.~T. {Noe II}, X.~Li, N.~M. Peraca, X.~Ma, Z.~Jin,
  W.~Ren, G.~Ma, I.~Katayama, J.~Takeda, H.~Nojiri, D.~Turchinovich, S.~Cao,
  M.~Bamba, and J.~Kono,
\newblock under reviewing (2020).

\bibitem{Peraca2020}
N.~Marquez Peraca, X.~Li, M.~Bamba, C.-L.~Huang, N.~Yuan, X.~Ma, G.~T.~Noe II,
E.~Morosan, S.~Cao, and J. Kono,
\newblock {Terahertz Magnon Spectroscopy Mapping of the Low-Temperature Phases of $\mathrm{Er}_x\mathrm{Y}_{1-x}\mathrm{FeO_3}$},
\newblock Proceedings of 2020 Conference on Lasers and Electro-Optics (CLEO), FM4D.5.

\bibitem{Shane1968PRLe}
J.~R. Shane,
\newblock {Resonance frequencies of the orthoferrites in the spin reorientation
  region},
\newblock \href{https://doi.org/10.1103/PhysRevLett.20.728}{Phys. Rev. Lett.
  {\bf 20}, 728 (1968)}.

\bibitem{Levinson1969PR}
L.~M. Levinson, M.~Luban, and S.~Shtrikman,
\newblock {Microscopic model for reorientation of the easy axis of
  magnetization},
\newblock \href{https://doi.org/10.1103/PhysRev.187.715}{Phys. Rev. {\bf 187},
  715 (1969)}.

\bibitem{Yamaguchi1974JPCS}
T.~Yamaguchi,
\newblock {Theory of spin reorientation in rare-earth orthochromites and
  orthoferrites},
\newblock \href{https://doi.org/10.1016/S0022-3697(74)80003-X}{J. Phys. Chem.
  Solids {\bf 35}, 479 (1974)}.

\bibitem{Balbashov1995}
A.~M. Balbashov, G.~V. Kozlov, A.~A. Mukhin, and A.~S. Prokhorov,
\newblock {Submillimeter Spectroscopy of Antiferromagnetic Dielectrics:
  Rare-Earth Orthoferrites},
\newblock in {\em High Freq. Process. Magn. Mater.} (World Scientific, 1995),
  pp. 56--98.

\bibitem{Zubov2019}
E.~E. Zubov, V.~Markovich, I.~Fita, A.~Wisniewski, and R.~Puzniak,
\newblock {Magnetic order in ErFeO3 single crystals studied by mean-field
  theory},
\newblock \href{https://doi.org/10.1103/PhysRevB.99.184419}{Phys. Rev. B {\bf
  99}, 1 (2019)}.

\bibitem{Tsang1978}
C.~H. Tsang, R.~L. White, and R.~M. White,
\newblock {Spin-wave damping of domain walls in YFeO3},
\newblock \href{https://doi.org/10.1063/1.324577}{J. Appl. Phys. {\bf 49}, 6063
  (1978)}.

\end{thebibliography}

\end{document}